\documentclass[aps,prd,floatfix,groupedaddress,nofootinbib,superscriptaddress,twocolumn]{revtex4-2}

\usepackage{amsmath,amssymb,bm,xcolor,float,graphicx,mathrsfs}
\usepackage[hidelinks]{hyperref}
\hypersetup{colorlinks=true,linkcolor=blue,citecolor=blue,urlcolor=blue}
\usepackage{Macro}
\usepackage{orcidlink}

\def\eprint#1{\href{https://arxiv.org/abs/#1}{arXiv:~\nolinkurl{#1}}}

\def\estk{\widehat{\kappa}}

\def\T{\Theta}
\def\hE{\widehat{E}}
\def\hB{\widehat{B}}
\def\hX{\widehat{X}}

\def\wE{\widehat{E}^{\rm WF}}
\def\Blens{B^{\rm lens}}

\begin{document}

% Title %
\title{The Simons Observatory: Constraining inflationary gravitational waves with multi-tracer $B$-mode delensing}

%
%\iffalse
\author{Toshiya Namikawa\,\orcidlink{0000-0003-3070-9240}}
\affiliation{Kavli Institute for the Physics and Mathematics of the Universe (WPI), UTIAS, The University of Tokyo, Kashiwa, Chiba 277-8583, Japan}
\affiliation{Department of Applied Mathematics and Theoretical Physics, University of Cambridge, Wilberforce Road, Cambridge CB3 0WA, UK}

\author{Anton Baleato Lizancos\,\orcidlink{0000-0002-0232-6480}}
\affiliation{Institute of Astronomy, Madingley Road, Cambridge, CB3 0HA, UK}
\affiliation{Kavli Institute for Cosmology Cambridge, Madingley Road, Cambridge, CB3 0HA, UK}
\affiliation{Berkeley Center for Cosmological Physics, Department of Physics, University of California, Berkeley, CA 94720, USA}
\affiliation{Lawrence Berkeley National Laboratory, One Cyclotron Road, Berkeley, CA 94720, USA}

\author{Naomi Robertson\,\orcidlink{0000-0001-5519-9158}}
\affiliation{Institute of Astronomy, Madingley Road, Cambridge, CB3 0HA, UK}

\author{Blake D. Sherwin\,\orcidlink{0000-0002-4495-1356}}
\affiliation{Department of Applied Mathematics and Theoretical Physics, University of Cambridge, Wilberforce Road, Cambridge CB3 0WA, UK}
\affiliation{Kavli Institute for Cosmology Cambridge, Madingley Road, Cambridge, CB3 0HA, UK}

\author{Anthony Challinor\,\orcidlink{0000-0003-3479-7823}}
\affiliation{Institute of Astronomy, Madingley Road, Cambridge, CB3 0HA, UK}
\affiliation{Department of Applied Mathematics and Theoretical Physics, University of Cambridge, Wilberforce Road, Cambridge CB3 0WA, UK}
\affiliation{Kavli Institute for Cosmology Cambridge, Madingley Road, Cambridge, CB3 0HA, UK}

\author{David Alonso\,\orcidlink{0000-0002-4598-9719}}
\affiliation{Department of Physics, University of Oxford, Denys Wilkinson Building, Keble Road, Oxford OX1 3RH, UK}

\author{Susanna Azzoni\,\orcidlink{0000-0002-8132-4896}}
\affiliation{Department of Physics, University of Oxford, Denys Wilkinson Building, Keble Road, Oxford OX1 3RH, UK}
\affiliation{Kavli Institute for the Physics and Mathematics of the Universe (WPI), UTIAS, The University of Tokyo, Kashiwa, Chiba 277-8583, Japan}

\author{Carlo Baccigalupi\,\orcidlink{0000-0002-8211-1630}}
\affiliation{International School for Advanced Studies (SISSA), Via Bonomea 265, 34136, Trieste, Italy}
\affiliation{Institute for Fundamental Physics of the Universe (IFPU), Via Beirut 2, 34151, Grignano (TS), Italy}
\affiliation{National Institute for Nuclear Physics (INFN), Sezione di Trieste Via Valerio 2, I-34127, Trieste, Italy}

\author{Erminia Calabrese\,\orcidlink{0000-0003-0837-0068}}
\affiliation{School of Physics and Astronomy, Cardiff University, The Parade, Cardiff, Wales CF24 3AA, UK}

\author{Julien Carron\,\orcidlink{0000-0002-5751-1392}}
\affiliation{Universit\'e de Gen\`eve, D\'epartement de Physique Th\'eorique et CAP, 24 Quai Ansermet, CH-1211 Gen\`eve 4, Switzerland}

\author{Yuji Chinone\,\orcidlink{0000-0002-3266-857X}}
\affiliation{Research Center for the Early Universe, School of Science, The University of Tokyo, Tokyo 113-0033, Japan}
\affiliation{Kavli Institute for the Physics and Mathematics of the Universe (WPI), UTIAS, The University of Tokyo, Kashiwa, Chiba 277-8583, Japan}

\author{Jens Chluba\,\orcidlink{0000-0003-3725-6096}}
\affiliation{Jodrell Bank Centre for Astrophysics, Department of Physics and Astronomy, University of Manchester, Alan Turing Building, Oxford Road, Manchester, M13 9PL, UK}

\author{Gabriele Coppi\orcidlink{0000-0002-6362-6524}}
\affiliation{Department of Physics, University of Milano-Bicocca, Piazza della Scienza 3, 20126 Milano, Italy}

\author{Josquin Errard\,\orcidlink{0000-0002-1419-0031}}
\affiliation{Universit\'{e} de Paris, CNRS, Astroparticule et Cosmologie, F-75013 Paris, France}

\author{Giulio Fabbian\,\orcidlink{0000-0002-3255-4695}}
\affiliation{Center for Computational Astrophysics, Flatiron Institute, 162 5th Avenue, New York, NY 10010, USA}
\affiliation{School of Physics and Astronomy, Cardiff University, The Parade, Cardiff, CF24 3AA, UK}

\author{Simone Ferraro\,\orcidlink{0000-0003-4992-7854}}
\affiliation{Lawrence Berkeley National Laboratory, One Cyclotron Road, Berkeley, CA 94720, USA}

\author{Alba Kalaja\,\orcidlink{0000-0001-9342-9399}}
\affiliation{Van Swinderen Institute for Particle Physics and Gravity, University of Groningen, Nijen- borgh 4, 9747 AG Groningen, The Netherlands}

\author{Antony Lewis\,\orcidlink{0000-0001-5927-6667}}
\affiliation{Department of Physics \& Astronomy, University of Sussex, Brighton BN1 9QH, UK}

\author{Mathew S. Madhavacheril\,\orcidlink{0000-0001-6740-5350}}
\affiliation{Centre for the Universe, Perimeter Institute, Waterloo, ON N2L 2Y5, Canada}
\affiliation{Department of Physics and Astronomy, University of Southern California, Los Angeles, CA 90007, USA}

\author{P. Daniel Meerburg\,\orcidlink{0000-0002-6080-6845}}
\affiliation{Van Swinderen Institute for Particle Physics and Gravity, University of Groningen, Nijen- borgh 4, 9747 AG Groningen, The Netherlands}

\author{Joel Meyers\,\orcidlink{0000-0001-8510-2812}}
\affiliation{Department of Physics, Southern Methodist University, 3215 Daniel Ave, Dallas, TX 75275, USA}

\author{Federico Nati\orcidlink{0000-0002-8307-5088}}
\affiliation{Department of Physics, University of Milano-Bicocca, Piazza della Scienza 3, 20126 Milano, Italy}

\author{Giorgio Orlando\,\orcidlink{0000-0002-4497-4910}}
\affiliation{Van Swinderen Institute for Particle Physics and Gravity, University of Groningen, Nijen- borgh 4, 9747 AG Groningen, The Netherlands}

\author{Davide Poletti\,\orcidlink{0000-0001-9807-3758}}
\affiliation{Universit\`a di Milano - Bicocca, 20126, Milano, Italy}
\affiliation{INFN sezione di Milano - Bicocca, 20216 Milano, Italy}

\author{Giuseppe Puglisi\,\orcidlink{0000-0002-0689-4290}}
\affiliation{Dipartimento di Fisica, Universit\`a di Roma “Tor Vergata”, Via della Ricerca Scientifica 1, 00133 Roma, Italy}
\affiliation{Computational Cosmology Center, Lawrence Berkeley National Laboratory, Berkeley, CA 94720, USA}

\author{Mathieu Remazeilles\,\orcidlink{0000-0001-9126-6266}}
\affiliation{Jodrell Bank Centre for Astrophysics, Department of Physics and Astronomy, University of Manchester, Alan Turing Building, Oxford Road, Manchester, M13 9PL, UK}

\author{Neelima Sehgal\,\orcidlink{0000-0002-9674-4527}}
\affiliation{Physics and Astronomy Department, Stony Brook University, Stony Brook, NY 11794, USA}

\author{Osamu Tajima}
\affiliation{Department of Physics, Kyoto University, Kitashirakawa Oiwake-cho, Sakyo-ku, Kyoto 606-8502, Japan}

\author{Grant Teply}
\affiliation{Department of Physics, University of California San Diego, La Jolla, CA 92093, USA}

\author{Alexander van Engelen\,\orcidlink{0000-0002-3495-158X}}
\affiliation{School of Earth and Space Exploration, Arizona State University, Tempe, AZ 85287, USA}

\author{Edward J. Wollack\,\orcidlink{0000-0002-7567-4451}}
\affiliation{NASA Goddard Space Flight Center, Greenbelt, MD 20771, USA}

\author{Zhilei Xu\,\orcidlink{0000-0001-5112-2567}}
\affiliation{MIT Kavli Institute, Massachusetts Institute of Technology, 77 Massachusetts Avenue, Cambridge, MA 02139, USA}

\author{Byeonghee Yu\,\orcidlink{0000-0003-3698-426X}}
\affiliation{Berkeley Center for Cosmological Physics, Department of Physics, University of California, Berkeley, CA 94720, USA}

\author{Ningfeng Zhu\,\orcidlink{0000-0002-3037-2003}}
\affiliation{Department of Physics and Astronomy, University of Pennsylvania, Philadelphia, PA 19104, USA}

\author{Andrea Zonca\,\orcidlink{0000-0001-6841-1058}}
\affiliation{San Diego Supercomputer Center, University of California San Diego, La Jolla, CA 92093, USA}

%\fi
%

% Date %
\date{\today}

% Abstract %
%----------------------------------------------------------------------------------------------------%
\begin{abstract}
We introduce and validate a delensing framework for the Simons Observatory (SO), which will be used to improve constraints on inflationary gravitational waves (IGWs) by reducing the lensing noise in measurements of the $B$-modes in CMB polarization. SO will initially observe CMB by using three small aperture telescopes and one large-aperture telescope. While polarization maps from small-aperture telescopes will be used to constrain IGWs, the internal CMB lensing maps used to delens will be reconstructed from data from the large-aperture telescope. Since lensing maps obtained from the SO data will be noise-dominated on sub-degree scales, the SO lensing framework constructs a template for lensing-induced $B$-modes by combining internal CMB lensing maps with maps of the cosmic infrared background from Planck as well as galaxy density maps from the LSST survey. We construct a likelihood for constraining the tensor-to-scalar ratio $r$ that contains auto- and cross-spectra between observed $B$-modes and the lensing $B$-mode template. We test our delensing analysis pipeline on map-based simulations containing survey non-idealities, but that, for this initial exploration, does not include contamination from Galactic and extragalactic foregrounds. 
We find that the SO survey masking and inhomogeneous and atmospheric noise have very little impact on the delensing performance, and the $r$ constraint becomes $\sigma(r)\approx 0.0015$ which is close to that obtained from the idealized forecasts in the absence of the Galactic foreground and is nearly a factor of two tighter than without delensing. 
We also find that uncertainties in the external large-scale structure tracers used in our multi-tracer delensing pipeline lead to bias much smaller than the $1\,\sigma$ statistical uncertainties. 
\end{abstract} 
%----------------------------------------------------------------------------------------------------%

% Output Title/Abstract etc %

%//////////////////////////////////////////////////////////////////////////////////////////%
% MAIN MATTER 
%//////////////////////////////////////////////////////////////////////////////////////////%

\maketitle

% Contents %

%//////////////////////////////////////////////////////////////////////////////////////////%
\section{Introduction}
%//////////////////////////////////////////////////////////////////////////////////////////%

Measuring the polarization of the cosmic microwave background (CMB) anisotropies will be at the forefront of observational cosmology in the next decade. In particular, measurements of the curl component ($B$-modes) in the CMB polarization will be of great importance, as these provide us with a unique window to probe inflationary gravitational waves (IGWs) and gain new insights into the early Universe~\cite{Polnarev:1985,Kamionkowski:1996:GW,Seljak:1996:GW}. 
CMB observations have not yet confirmed the presence of these IGWs but have placed upper bounds on the IGW amplitude. The best current constraints on the IGW background, parameterized by the tensor-to-scalar ratio $r$ (at a pivot scale of $0.05\,$Mpc$^{-1}$), are from the combination of BICEP/Keck Array measurements and Planck and WMAP: $r<0.036$ ($2\,\sigma$)~\cite{BK13,BK15:data}.
Several ongoing and upcoming CMB experiments, including the BICEP Array \cite{BICEPArray}, Simons Array \cite{SimonsArray}, Simons Observatory (SO) \cite{SimonsObservatory}, LiteBIRD \cite{LiteBIRD}, and CMB-S4 \cite{CMBS4}, are targeting a detection of IGW $B$-modes over the next decade.

A high-precision measurement of the large-scale $B$-modes can tightly constrain $r$ \cite{CMBS4:r-forecast}. The precision of the IGW $B$-mode measurement is, however, limited by other sources of $B$-modes. 
In addition to Galactic foregrounds \cite{B2I}, gravitational lensing leads to $B$-modes from conversion of part of the $E$-mode polarization \cite{Zaldarriaga:1998:LensB}; these lensing $B$-modes behave as an additional noise component when constraining $r$. 
Indeed, the current best constraint on $r$ is limited by the lensing $B$-modes more than by Galactic foregrounds at the low dust region \cite{BK13}. 
Reducing statistical uncertainties by subtracting off the lensing-induced $B$-modes (or equivalent methods) -- a process usually referred to as delensing -- will hence be of critical importance for improving the constraints on $r$ \cite{Kesden:2002ku,Seljak:2003pn}. 
To estimate the lensing-induced $B$-modes in the survey region, known as a $B$-mode template, the simplest method is to combine the measured $E$-modes with a reconstructed lensing map derived from CMB data, and multiple works have studied this technique (e.g. \cite{Teng:2011xc,Namikawa:2014:patchwork,Sehgal:2016,Namikawa:2017:delens,Carron:2017,PB:2019:delens,ref:baleato_20}).

In addition to the lensing map measured internally with CMB data \cite{Smith:2010gu,Carron:2018:GW,ref:baleato_20}, we can also use external mass tracers that correlate with the CMB lensing signal efficiently, such as the cosmic infrared background (CIB) \cite{Sherwin:2015,Simard:2015}, radio and optical galaxies \cite{Namikawa:2015:delens,Manzotti:2018}, galaxy weak lensing \cite{Marian:2007}, and intensity mapping signals \cite{Sigurdson:2005cp,Karkare:2019:delens}. In the last few years, several analyses, beginning with Ref.~\cite{Larsen:2016}, have demonstrated delensing using real small-scale CMB temperature and polarization data  \cite{SPTpol:delens,Carron:2017,PB:2019:delens,DW:2020:delens}. Recently, Ref.~\cite{ref:bkspt_cib_delensing} (hereafter, BKSPT) demonstrated for the first time $B$-mode delensing on the large scales relevant for constraining IGWs using the CIB as a mass tracer. 

SO, which we focus on throughout this paper, is targeting a measurement of $r$ with the $1\,\sigma$ uncertainty, $\sigma(r)=0.002$. 
SO will measure the large-scale $B$-modes with three small-aperture telescope (SAT) and delens the large-scale $B$-modes by combining the lensing map measured from the large-aperture telescope (LAT) and external large-scale structure (LSS) tracers. 
Achieving $\sigma(r)=0.002$ will require removing approximately $70$\% of the lensing $B$-mode power spectrum, based on idealized forecasts building on Ref.~\cite{SO:2018:forecast}. 
However, it is an open question whether a practical delensing method can match this somewhat idealized forecast performance.

In real analyses, the delensing efficiency may be degraded by, e.g., the presence of survey boundaries, inhomogeneous instrumental noise and atmospheric noise. For example, the efficiency of the SPTpol $B$-mode delensing at high multipoles is $19.7$\% while the idealized analytic estimate is $27$\% \cite{SPTpol:delens}. 

Estimating the actual delensing performance including realistic survey effects in SO is important given the significant improvement in $\sigma(r)$ we can hope to achieve with delensing. 
For SO, the noise properties of the LAT used for measuring the lensing map will be significantly different from those of the SAT, the $B$-modes from which will be delensed and used to constrain $r$. This difference further complicates the situation.
Other significant practical concerns include the astrophysical uncertainties inherent in our use of the external mass tracers. The SO baseline delensing strategy utilizes external mass tracers, i.e., LSS tracers, such as galaxies and CIB to enhance the delensing performance; it is therefore important to evaluate the impact of the astrophysical uncertainties (e.g., redshift or bias uncertainties) associated with mass tracers on $\sigma(r)$ and mitigate the relevant uncertainties if necessary \cite{Sherwin:2015,ref:bkspt_cib_delensing}. 

In this paper, we present a delensing framework for SO (which relies on multi-tracer delensing), test it on simulations, and address the practical concerns listed above. Although accurate removal of Galactic foreground emission is of critical concern for IGW $B$-mode searches, we shall not consider this issue here. Our aim is to validate the delensing framework in the presence of realistic survey effects, the impacts of which would be difficult to isolate if (residual) foregrounds were also included. The impact of Galactic foreground on the SO large-scale $B$-mode analysis has been explored in the SO overview paper \cite{SO:2018:forecast}. The integration of Galactic-foreground cleaning and delensing has already been demonstrated by BKSPT, and in future work we will explore this issue within the context of SO.

This paper is organized as follows. 
In Sec.~\ref{sec:method}, after briefly reviewing the lensing effect on the CMB, we present the baseline multi-tracer strategy for SO delensing. 
In Sec.~\ref{sec:results}, we test our method with SO simulations including realistic survey effects and show the expected constraints on $r$. 
In Sec.~\ref{sec:systematics}, we discuss how to incorporate astrophysical uncertainties in mass tracers for delensing. 
We conclude in Sec.~\ref{sec:summary}. Appendix~\ref{appendix:covariances} contains technical details of the covariances of the auto- and cross-power spectra of the $B$-mode template and the observed $B$-modes, which are used in the likelihood to constrain $r$.

%//////////////////////////////////////////////////////////////////////////////////////////%
\section{Large-Scale $B$-mode Delensing} 
\label{sec:method}
%//////////////////////////////////////////////////////////////////////////////////////////%

%<><><><><><><><><><><><><><><><><><><><><><><><><><><><><><><><><><><><><><><><><><><><><><><><><><>%
\begin{figure}[t]
\bc
\includegraphics[width=8.5cm,clip]{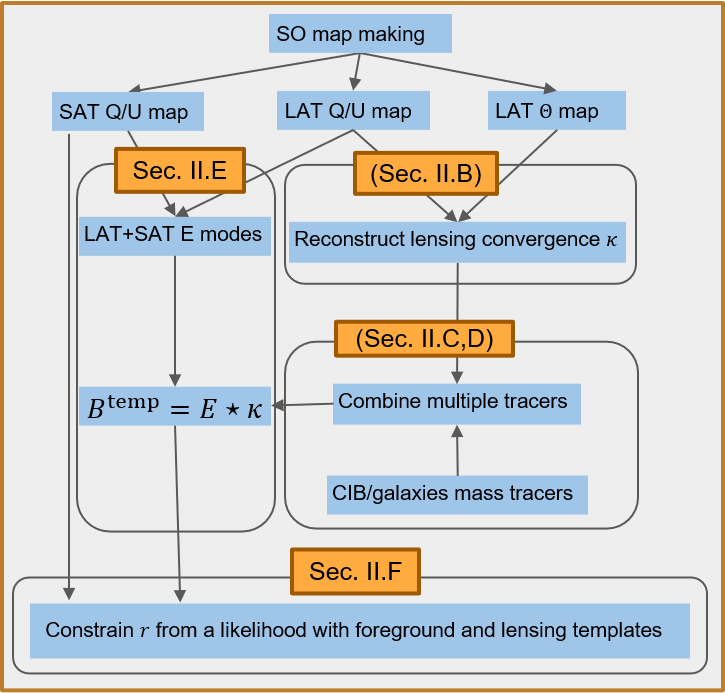}
\caption{
Flowchart of the SO delensing pipeline. 
}
\label{Fig:flowchart}
\ec
\end{figure}
%<><><><><><><><><><><><><><><><><><><><><><><><><><><><><><><><><><><><><><><><><><><><><><><><><><>%

In this section, we first briefly review CMB $B$-mode delensing and introduce our notation. Then, we describe our method for delensing SO data. Figure \ref{Fig:flowchart} shows our flowchart of the delensing pipeline. 

\subsection{CMB lensing and lensing $B$-modes} 

The distortion effect of lensing on the primary CMB temperature and polarization anisotropies is expressed by a remapping. Denoting the primary temperature and polarization anisotropies from the last-scattering surface as $\T$ and $Q\pm\iu U$, respectively, the lensed temperature and polarization anisotropies in the sky direction $\hatn$, are given by (see, e.g., Ref.~\cite{Lewis:2006fu})
%----------------------------------------------------------------------------------------------------%
\al{
    \tT(\hatn) &= \T(\hatn+\bm{d}(\hatn)) \,, \\
	[\tQ\pm\iu\tU](\hatn) &= [Q\pm\iu U](\hatn + \bm{d}(\hatn)) 
	\,, \label{Eq:remap}
}
%----------------------------------------------------------------------------------------------------%
where tildes indicate lensed quantities and where $\bm{d}$ is the deflection angle.\footnote{See Ref.~\cite{Challinor:2002cd} for the detailed form that these lensing displacements take on the spherical sky.}
In the Born approximation, $\bm{d}$ is given by the gradient of the lensing potential $\bn\grad$ and is related to the lensing convergence as $\bn\cdot\bm{d}=-2\kappa$. (Here we ignore all curl modes.) It is generally more convenient to work with the scalar-valued $E$ and $B$-modes rather than the spin-$2$ Stokes parameters, $Q$ and $U$. In harmonic space, these are related by
%----------------------------------------------------------------------------------------------------%
\al{
	E_{\l m} \pm \iu B_{\l m} = -\Int{2}{\hatn}{} {}_{\pm 2}Y_{\l m}^*(\hatn) [Q\pm\iu U](\hatn) 
	\,, \label{Eq:EB-def}
}
%----------------------------------------------------------------------------------------------------% 
where we denote the spin-$2$ spherical harmonics as ${}_{\pm 2}Y_{\l m}(\hatn)$. Similarly with the spin-$0$ (scalar) spherical harmonics, $Y_{\l m}(\hatn)$, the temperature and lensing potential maps are transformed into harmonic space as
%----------------------------------------------------------------------------------------------------% 
\al{
	\T_{\l m} &= \Int{2}{\hatn}{} Y_{\l m}^*(\hatn) \T (\hatn) 
	\,, \\
	\kappa_{LM} &= \Int{2}{\hatn}{} Y_{LM}^*(\hatn) \kappa (\hatn) 
	\,.
}
%----------------------------------------------------------------------------------------------------% 
Expanding the right-hand side of \eq{Eq:remap} up to first order in the lensing potential, and then transforming the Stokes $Q/U$ parameters to $E$/$B$-modes with \eq{Eq:EB-def}, the $B$-modes of the lensed polarization field at linear order in $\phi$ are given by~\cite{Smith:2010gu}
%----------------------------------------------------------------------------------------------------%
\al{
	\Blens_{\l m} &= \iu\sum_{\l'm'}\sum_{LM} 
	\Wjm{\l}{\l'}{L}{m}{m'}{M} p^- F^{(2)}_{\l L\l'} E_{\l'm'}^* \kappa_{LM}^* 
	\label{Eq:Lensed-B} \,, 
} 
%----------------------------------------------------------------------------------------------------%
where we ignore the primary $B$-modes. The quantity in round brackets is the Wigner-3$j$ symbol, $p^+$ ($p^-$) is unity
if $\l+L+\l'$ is an even (odd) integer and zero otherwise, and $F^{(2)}_{\l L\l'}$ represents the mode coupling induced by the lensing \cite{OkamotoHu:quad,Smith:2010gu}: 
%----------------------------------------------------------------------------------------------------%
\al{ 
	F^{(s)}_{\l L\l'} &= \frac{2}{L(L+1)}
	\sqrt{\frac{(2\l+1)(2\l'+1)(2L+1)}{16\pi}}
	\notag \\
	&\times 
	[-\l(\l+1)+\l'(\l'+1)+L(L+1)]\Wjm{\l}{\l'}{L}{-s}{s}{0} 
	\,. \label{Eq:Spm}
} 
%----------------------------------------------------------------------------------------------------%
Equation~\eqref{Eq:Lensed-B} is known to be a good analytic approximation to the lensing $B$-modes on large scales~\citep{Challinor:2005jy, BaleatoLizancos:2020b}. From \eq{Eq:Lensed-B}, the lensing $B$-modes are simply expressed in terms of a convolution between the unlensed $E$-modes and lensing potential. 
Once we obtain an estimate for the lensing map, we simply approximate the lensing $B$-modes as a convolution of the Wiener-filtered $E$ and lensing maps as described below.

\subsection{Internal CMB lensing map} 

From the lensed temperature and polarization maps, we reconstruct the lensing potential $\grad$ using the quadratic-estimator approach of Ref.~\cite{OkamotoHu:quad}.\footnote{For the expected noise levels from SO, the improvements in precision of lensing reconstruction and delensing efficiency from applying more optimal, maximum-likelihood methods are negligible~\cite{Carron:2017mqf}.}
Lensing induces off-diagonal elements of the covariance ($\l\not=\l'$ or $m\not=m'$) between two lensed CMB anisotropy fields ($X,Y=\T,E,B$) as 
%----------------------------------------------------------------------------------------------------%
\al{
	\ave{\tX_{\l m}\tY_{\l'm'}}\rom{CMB} 
		&= \sum_{LM}\Wjm{\l}{\l'}{L}{m}{m'}{M}f^{XY}_{\l L\l'}\kappa^*_{LM}
	\,, \label{Eq:weight} 
}
%----------------------------------------------------------------------------------------------------%
where the operation $\ave{\ldots}\rom{CMB}$ denotes the ensemble average over the primary unlensed CMB anisotropies. The response functions $f_{\l L\l'}^{XY}$ in \eq{Eq:weight} are defined as~\cite{OkamotoHu:quad}\footnote{We ignore quadratic combinations with $XY=\T B$ and $BB$ since the signal-to-noise of the associated estimators is much lower than that of the other quadratic estimators for SO noise levels.}
%----------------------------------------------------------------------------------------------------%
\al{
	f^{\T\T}_{\l L\l'} &= 
	F^{(0)}_{\l L\l'}\CTT_{\l'} + F^{(0)}_{\l'L\l}\CTT_\l 
	\,, \\
	f^{\T E}_{\l L\l'} &= p^+ F^{(0)}_{\l L\l'}\CTE_{\l'} 
	+ p^+ F^{(2)}_{\l'L\l}\CTE_\l
	\,, \\
	f^{EE}_{\l L\l'} &= p^+ F^{(2)}_{\l L\l'}\CEE_{\l'} 
	+ p^+ F^{(2)}_{\l'L\l}\CEE_\l 
	\,, \\
	f^{EB}_{\l L\l'} &= \iu p^- F^{(2)}_{\l'L\l}\CEE_\l 
	\,. \label{Eq:f}
}
%----------------------------------------------------------------------------------------------------%
Here, $F^{(s)}$ is defined in \eq{Eq:Spm}, and $C^{XY}_\l$ is the angular power spectrum of the unlensed CMB anisotropies. In our analysis, we replace the unlensed CMB spectra with their lensed counterparts, $\tC^{XY}_\l$, giving a good approximation to the non-perturbative response functions~\cite{Lewis:2011fk} and mitigating higher-order biases in the power spectrum of the lens reconstruction~\cite{Hanson:2010:N2}. 
Equation~\eqref{Eq:weight} motivates the following form for a quadratic lensing estimator \cite{OkamotoHu:quad}:
%----------------------------------------------------------------------------------------------------%
\al{
	(\estk^{XY}_{LM})^* = A^{XY}_L\sum_{\l\l'mm'}\Wjm{\l}{\l'}{L}{m}{m'}{M}
		\frac{(f^{XY}_{\l L\l'})^*}{\Delta^{XY}}\ol{X}_{\l m}\ol{Y}_{\l'm'}
	\,, \label{Eq:estg-XY}
}
%----------------------------------------------------------------------------------------------------%
where we introduce $\Delta^{XY}$ which is $2$ if $X=Y$ and $1$ otherwise. Here, $\ol{X}$ and $\ol{Y}$ are observed anisotropies filtered by their inverse variance. In the idealistic case, the inverse-variance filtering is diagonal:
%----------------------------------------------------------------------------------------------------%
\al{
    \ol{X}_{\l m} = (\hC^{XX}_{\l})^{-1}\hX_{\l m} 
    \,, 
    \label{Eq:FilterDiag}
}
%----------------------------------------------------------------------------------------------------%
where $\hX_{\l m}$ are the observed CMB anisotropies and $\hC^{XX}_\l$ is their angular power spectra. 
We ignore the correlation between $\T$ and $E$ in the above filtering, making the inverse-variance filtering diagonal in CMB anisotropies as well. The normalization $A^{XY}_L$ is then given by 
%----------------------------------------------------------------------------------------------------%
\al{
	A^{XY}_L &= \left\{\frac{1}{2L+1}\sum_{\l\l'} 
		\frac{|f^{XY}_{\l\l'L}|^2}{\Delta^{XY}\hC^{XX}_\l\hC^{YY}_{\l'}}\right\}^{-1}
	\,. \label{Eq:Rec:N0}
}
%----------------------------------------------------------------------------------------------------%

For a realistic (anisotropic) survey, the diagonal filtering approximation (in $\l$) of Eq.~\eqref{Eq:FilterDiag} generally makes the reconstruction sub-optimal. 
However, it also makes the computational cost very low.
As we show later, the reconstruction and delensing performances for the SO surveys are not degraded significantly compared to an isotropic case (i.e., for the same total integration time, but distributed evenly over the survey region), even if we use the diagonal approximation. Therefore, we choose the diagonal filtering for our baseline analysis due to its low computational cost.

In practice, it is necessary to subtract a mean-field correction from the reconstruction since $\ave{\estk^{XY}_{LM}}$ becomes non-zero due to, e.g., the survey boundary and inhomogeneous noise~\cite{Namikawa:2012:bhe,Namikawa:2013:bhepol}. In this paper, we estimate the mean-field biases, $\ave{\estk^{XY}_{LM}}$, by averaging over simulation realizations and subtract these estimates from the $\estk^{XY}_{LM}$.

It is possible to combine the quadratic estimators together to improve the precision of the reconstruction.
In this paper, we construct a minimum variance (MV) estimator following Ref.~\cite{OkamotoHu:quad}, i.e., the linear combination of the individual estimators, $\estk^{\rm MV}_{LM}=\sum_{XY}\alpha^{XY}_L\estk^{XY}_{LM}$, where $\alpha_L$ is determined so that the reconstruction noise of $\estk^{\rm MV}_{LM}$ is minimized. 
Note that Ref.~\cite{Maniyar:2021:GMV} showed that the use of more optimal weights originally derived by Ref.~\cite{Hirata:2003:mle:pol} can improve the signal-to-noise by around $10$\% at $L\alt 100$ compared to the use of the MV estimator developed by Ref.~\cite{OkamotoHu:quad}. 
However, for delensing, the improvement is not significant; delensing requires a lensing mass map at intermediate scales, $L\sim 200$--$800$~\cite{Simard:2015,Sherwin:2015}, where the increase in signal-to-noise from the more optimal weights is only a few percent \cite{Maniyar:2021:GMV}. The impact of the sub-optimal weights on the delensing performance is reduced further since we combine with other LSS tracers, which are significant contributors on these delensing scales.
Therefore, in this paper, we use the linear combination of the estimators of Ref.~\cite{OkamotoHu:quad} to construct the CMB lensing map.

\subsection{External mass-tracer map}
\label{subsec:tracer}

In addition to being reconstructed internally from the CMB fields themselves, the lensing convergence field can be estimated from observations of the LSS tracers such as the spatial distribution of galaxies or the CIB~\cite{Smith:2010gu,Sherwin:2015,Simard:2015,Namikawa:2015:delens,Karkare:2019:delens}.

\begin{figure}
\includegraphics[width=1.0\columnwidth]{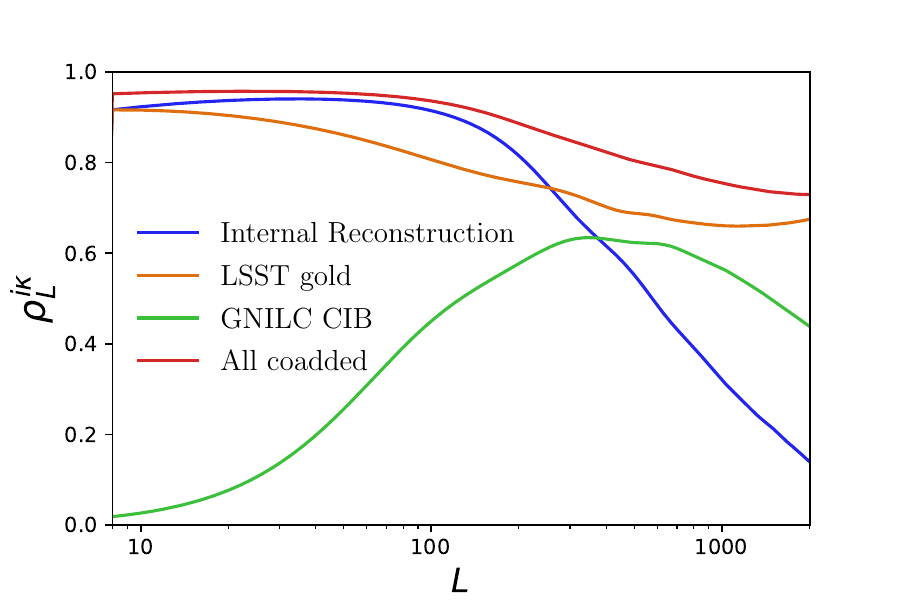}
\caption{
Correlation coefficients of the true CMB lensing field with several LSS tracers, and with a co-added tracer. On large angular scales, correlation between the CIB map extracted from Planck data using the GNILC algorithm drops due to the presence of residual CIB in the dust maps (which, in turn, gets filtered out of the CIB maps). Fortunately, on those scales internal techniques can very accurately reconstruct lensing, as shown here for a projected minimum-variance quadratic estimator reconstruction with SO (goal) noise levels~\cite{SO:2018:forecast} and standard internal-linear-combination (ILC) foreground cleaning. On the other hand, the relevance of shot noise on small scales means that the correlation with the CIB decreases for large $L$. The forecasted curves involving LSST galaxies correspond to the case where tomographic observations of galaxies in the ``gold'' sample are divided into six redshift bins. The auto- and cross-spectra of mass tracers for this plot are taken from Ref.~\cite{ref:yu_17}.
}
\label{fig:corrcoeffs}
\end{figure}

As proposed in Refs.~\cite{Sherwin:2015,ref:yu_17}, different tracers can also be linearly combined using weights designed to maximize the cross-correlation between the co-added tracer and the true convergence. Reference~\cite{Sherwin:2015} determined that the weights that achieve this are
\begin{equation}\label{eqn:multitracer_weights}
    c^i_L = \sum_{j}(\rho^{-1})^{ij}_L \rho^{j\kappa}_L\sqrt{\frac{C_L^{\kappa\kappa}}{C_L^{\widehat{\kappa}^i\widehat{\kappa}^i}}}
    \,, 
\end{equation}
where the linearly combined tracer is $\widehat{\kappa}^{\rm comb}_{LM}=\sum_i c^i_L \widehat{\kappa}^i_{LM}$. Here, $\rho^{i\kappa}_L$ is the cross-correlation coefficient, at multipole $L$, between tracer $\widehat{\kappa}^i$ and the true convergence; $\rho^{ij}_L$ is the cross-correlation between tracers $\widehat{\kappa}^i$ and $\widehat{\kappa}^j$; and $C_L^{\widehat{\kappa}^i\widehat{\kappa}^i}$ is the angular power spectrum of tracer $\widehat{\kappa}^i$. Qualitatively, on a given angular scale, this scheme brings to the fore the tracers that best correlate with the underlying truth. In practice, this means that internal reconstructions, which accurately reconstruct lensing on the largest angular scales, can be supplemented with external tracers on the small scales where they are dominated by reconstruction noise. Figure~\ref{fig:corrcoeffs} illustrates this for an experiment with the characteristics of the Simons Observatory. Notice that information gleaned from Planck CIB data (extracted using the GNILC algorithm~\cite{Remazeilles:2011:GNILC,ref:gnilc}), and from a galaxy survey with the characteristics expected of the Vera Rubin Observatory Legacy Survey of Space and Time (LSST) ``gold'' sample (approximately 40 galaxies per arcmin$^2$)~\cite{ref:lsst} enables the co-added tracer to maintain a high degree of correlation with the true lensing convergence on scales of $250<L<1000$. This is of particular importance for delensing, since it is those intermediate and small-scale lenses located primarily at high redshifts (see Fig.~3 of Ref.~\cite{Lewis:2006fu}) that are most relevant~\cite{Smith:2011we}. 
The recent Planck lensing analysis demonstrates delensing by combining the CMB lensing map with the GNILC CIB map \cite{P18:phi}.

\subsection{Optimal combination of mass tracers}
\begin{figure*}[t]
\bc
\includegraphics[width=8.5cm,clip]{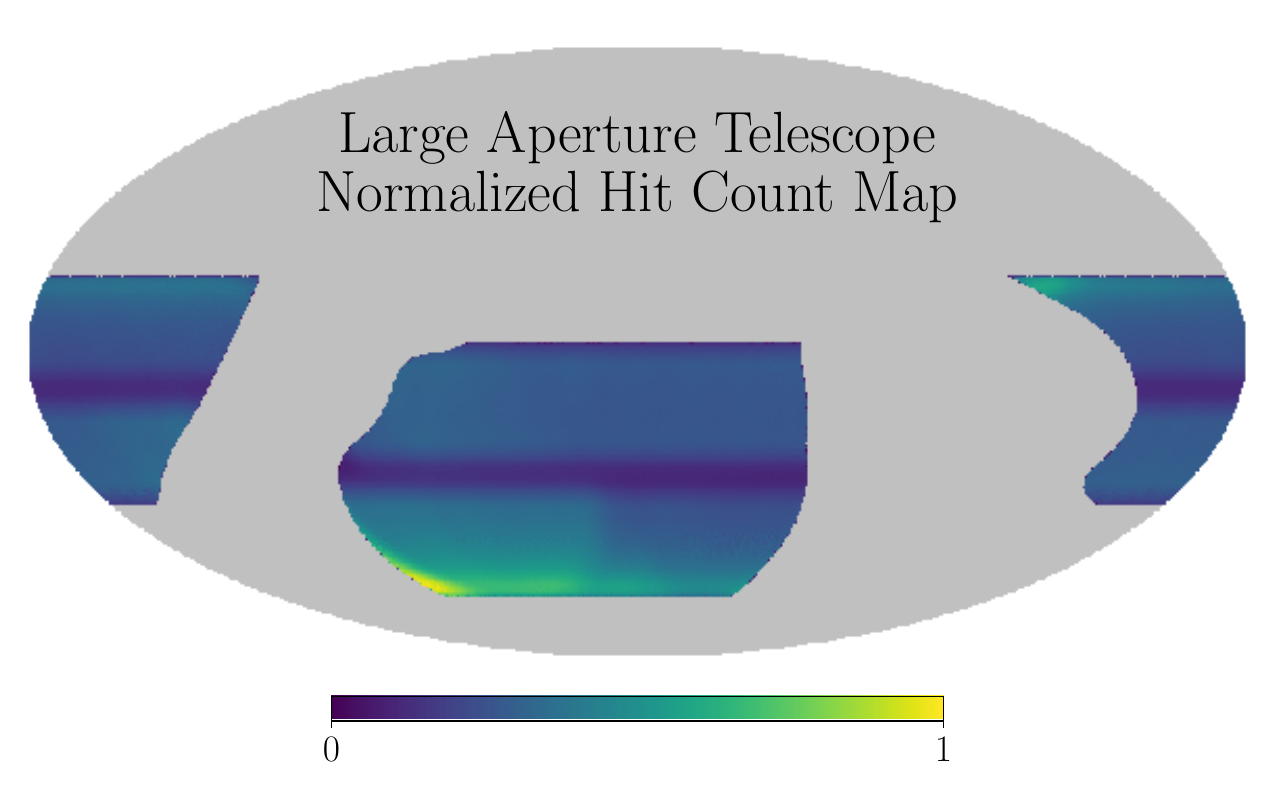}
\includegraphics[width=8.5cm,clip]{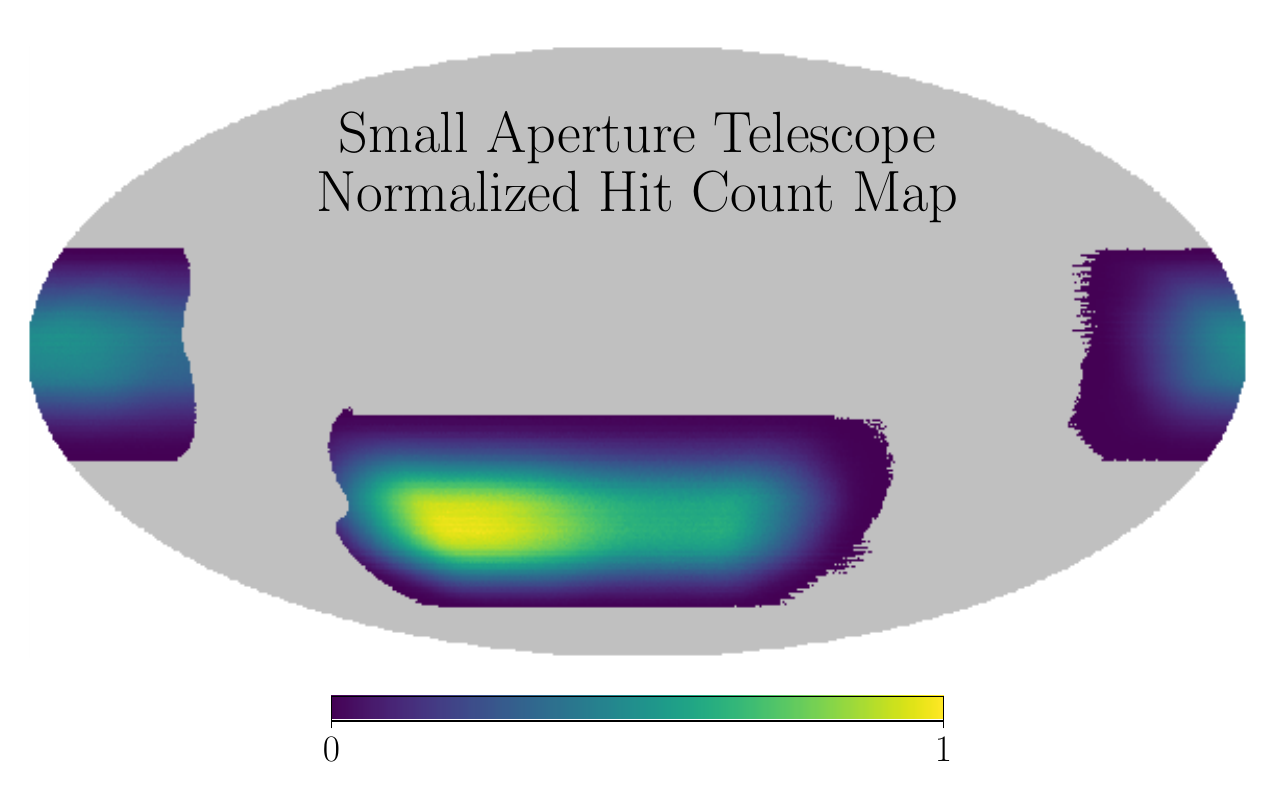}
\caption{%
Normalized SO hit-count maps multiplied by the nominal Galactic binary masks for the LAT (left) and SAT (right) regions. The LAT and SAT Galactic masks coincide with those currently used in the preparations for the LAT lensing and SAT $B$-mode analyses by the SO Collaboration, respectively. 
%The sky fraction of the LAT and SAT regions are $f_{\rm sky}=0.29$ and $0.27$, respectively, although these values do not take account of the variation in the hit counts across the regions.
The hit count maps are obtained from the map-based SO simulation package\footnote{\url{https://github.com/simonsobs/map_based_simulations}} which are one of the possible scan strategies for SO, although work is ongoing to optimize the strategy further for a range of SO science goals.
}
\label{Fig:window}
\ec
\end{figure*}

The optimal estimate of the CMB lensing potential is obtained as a linear combination of the quadratic estimators and external mass tracers. 
In practice, the analytic weights in~\eq{eqn:multitracer_weights} could be no longer optimal due to, e.g., an analysis mask and inhomogeneous noise and residual foregrounds.
Instead of using the analytic optimal weights, our pipeline empirically evaluates the weights, $c^i_L$, from smoothed auto- and cross-spectra determined from simulations to mimic the actual procedure that would likely be applied with new SO and LSST data. 
We compute $c^i_L$ from the covariance of mass tracers and the input $\kappa$. The Wiener-filtered mass map, $\estk^{\rm comb}$, is then obtained as defined in the previous subsection. 
Here, the indices of the mass tracers, $i$, include the $\T\T$, $\T E$, $EE$ and $EB$ quadratic estimators for CMB lensing reconstruction, the galaxy overdensity at six tomographic redshift bins with edges at $z = [0, 0.5, 1, 2, 3, 4, 7]$, and the CIB. 
When combining mass tracers, we restrict the full-sky mass-tracer maps (galaxies at each photo-$z$ bin and the CIB) to the region surveyed by the LAT (see Fig.~\ref{Fig:window}). We do not take into account correlations between different $L$. 

%<><><><><><><><><><><><><><><><><><><><><><><><><><><><><><><><><><><><><><><><><><><><><><><><><><>%
%<><><><><><><><><><><><><><><><><><><><><><><><><><><><><><><><><><><><><><><><><><><><><><><><><><>%

\subsection{Lensing $B$-mode template construction}

On large scales, we estimate the lensing $B$-modes as
%----------------------------------------------------------------------------------------------------%
\al{
	B^{\mathrm{temp}}_{\l m} = \iu\sum_{\l'm'}\sum_{LM} 
	\Wjm{\l}{\l'}{L}{m}{m'}{M}p^- F^{(2)}_{\l L\l'} 
	(\wE_{\l'm'})^*(\estk^{\rm comb}_{LM})^*
	\,, \label{Eq:Quad-LensB}
}
%----------------------------------------------------------------------------------------------------%
where $\wE_{\l m}$ are the Wiener-filtered, observed $E$-modes. 
This first-order lensing template built from lensed $E$-modes is indistinguishable for our purposes from
an optimal `remapping' method, and will continue to be so until the fidelity of $\estk$ and $\wE$ allow for residuals to be as low as $O(1\%)$ of the original lensing $B$-mode power~\cite{BaleatoLizancos:2020b}. 

To construct the optimal lensing $B$-mode template, we compute the Wiener-filtered $E$-modes, $\hE^{\rm WF}_{\l m}$, which are obtained by solving the following equation~\cite{Eriksen:2004:wiener}: 
\al{
	&\left[1+\sum_{t,\nu}\bR{C}^{1/2}b_{t,\nu}\bR{Y}^\dagger\bR{N}^{-1}_{t,\nu}\bR{Y}b_{t,\nu}\bR{C}^{1/2}\right] (\bR{C}^{-1/2}\bm{x}^w) 
	\notag \\ 
	&\quad = \sum_{t,\nu}\bR{C}^{1/2}b_{t,\nu}\bR{Y}^\dagger \bR{N}^{-1}_{t,\nu} \bm{d}_{t,\nu}
	\,. \label{Eq:Cinv}
}
Here, $t$ and $\nu$ are indices for the input maps specifying telescope type (LAT or SAT) and frequency ($93$, $145$, or $225$\,GHz), respectively. 
The vector $\bm{x}^w$ has as its components the harmonic coefficients of the Wiener-filtered $E$- and $B$-modes, $\bR{C}$ is the diagonal signal covariance of the lensed $E$- and $B$-modes in spherical-harmonic space, and $b_{t,\nu}$ is the beam function, 
The matrix $\bR{C}^{1/2}$ is defined so that its square is equal to $\bR{C}$. The real-space vector $\bm{d}_{t,\nu}$ contains the Stokes $Q$ and $U$ maps observed by telescope $t$ at frequency $\nu$, and 
$\bR{N}_{t,\nu}$ is the covariance matrix of the instrumental noise in these maps. The matrix $\bR{Y}$ is defined so that it transforms the multipoles of the $E$- and $B$-modes into real-space maps of the Stokes parameters $Q$ and $U$. 
Solving \eq{Eq:Cinv} for $\bm{x}^w$ is computationally demanding and we adopt the conjugate gradient inversion algorithm \cite{NumericalRecipes}.
At unobserved pixels, we assign infinite noise in the noise covariance. In this paper, we do not include any extragalactic foregrounds, but in practice, we should include masks for extragalactic contaminants as unobserved pixels. 
Note that constructing $\hE^{\rm WF}_{\l m}$ in this way naturally combines the SAT and LAT polarization measurements optimally. It also combines the maps across frequencies optimally under the assumption that foreground emission is negligible. In practice, it may be necessary to work with foreground-cleaned maps from the SAT and LAT rather than individual frequency maps. In this case, the instrument noise entering in Eq.~\eqref{Eq:Cinv} should be generalized to describe the noise in the foreground-cleaned maps.

When constructing $\hE^{\rm WF}_{\l m}$, we assume that the noise covariance matrix in real space, $\bR{N}_{t,\nu}$, is diagonal, although the actual simulations have noise correlations between different pixels due to atmospheric noise. The diagonal elements of the noise covariance are taken to be
of the form $\sigma_{t,\nu}^2/H_t(\hatn)$ where $H_t(\hatn)$ is the normalized hit count and $\sigma_{t,\nu}$ is the white-noise level of each map $t,\nu$. The hit count is assumed to be the same for all frequency maps of a given telescope. The above filtering naturally takes into account the large difference between SAT and LAT hit-count maps as shown in Fig.~\ref{Fig:window}. We show below that this assumption of diagonal noise covariance is sufficient to achieve good delensing performance. 

\subsection{Likelihood for constraining IGWs} \label{sec:likelihood}

Here we describe the approach we take to implement delensing and constrain the tensor-to-scalar ratio using the lensing $B$-mode template. The choice of delensing approach depends on how the observed $B$-mode maps across frequencies are to be combined to clean foregrounds. Several schemes for such cleaning are being pursued within SO (see, e.g.,~\cite{SO:2018:forecast}), but here we focus on the cross-spectral approach. This compresses the frequency maps into their auto- and cross-spectra and models these as the sum of CMB signal, instrumental noise and parametrized foreground spectra. An approximate likelihood for these spectra is constructed, which is combined with priors on the foreground parameters to obtain parameter constraints. The cross-spectral approach has been demonstrated on $B$-mode data from BICEP/Keck Array (e.g.,~\cite{BKP}) and on simulated SO data in~\cite{Azzoni:2020hpw}. Delensing is simply incorporated in this framework by viewing the template as an additional ``frequency channel''. The auto-spectrum of the template, and its cross-spectra with the frequency maps, are included in the likelihood along with the cross-frequency spectra. This spectral approach for combined foreground cleaning and delensing has recently been demonstrated on data in BKSPT.

Since we do not consider foreground cleaning in this work, we work with a single $B$-mode map from the SAT and the lensing $B$-mode template. For the former, we adopt noise levels appropriate to a co-addition of the 93, 145 and 225\,GHz frequency channels assuming that the remaining frequency channels are used to clean foregrounds (with the noise level in the cleaned map being similar to the co-addition we consider).
We construct the auto- and cross-spectra between the observed $B$-mode map and the lensing template over the region common to the SAT and LAT surveys. To minimize the additional variance from leakage of $E$-modes due to the survey boundary, we use the pure-$B$-mode formalism~\cite{Smith:2005:chi-estimator} as implemented in the NaMaster code\footnote{\url{https://namaster.readthedocs.io/en/latest/index.html}}. We use an apodization length of $8^\circ$, but do not otherwise weight the data to account for noise inhomogeneities. In practice, the noise varies significantly (see Fig.~\ref{Fig:window})
as the SAT scan strategy concentrates integration time on around 10\% of the sky in a region of low Galactic foreground emission. Given this, adopting a more optimal weighting in the construction of the pure-$B$-modes might improve performance somewhat and also lessen the demands on foreground cleaning in analysis of the real data.
For example, combining with Planck data to get the larger-scale $E$-modes \cite{Ghosh:2020}, or adopting the optimal Wiener filtering of Eq.~\eqref{Eq:Cinv}, would provide a more optimal measurement of the SAT $B$-modes. 
Binned estimates of the auto- and cross-spectra are used in an approximate likelihood following Ref.~\cite{Hamimeche:2008}, 
which accounts for the non-Gaussian form of the likelihood on large scales where there are few modes. The likelihood requires model auto- and cross-spectra, and the covariance of the measured spectra in a fiducial model. We now describe how these are calculated.

\subsubsection{Model spectra} \label{ref:model_spectra}

For the likelihood analysis, we need to model the auto-spectrum of the $B$-mode lensing template and its cross-spectrum with the observed $B$-modes. For an isotropic survey, these can be calculated simply (see Sec.~\ref{sec:systematics}). However, in a realistic setup they are not straightforward to model because of, e.g., the inhomogeneous Wiener-filtering applied to $E$-modes. In the likelihood analysis, therefore, we model these spectra with the means of simulation realizations. This approach is also convenient if the analysis includes more complicated realistic effects in the mass tracers. We similarly use the mean of simulations of noisy, lensing $B$-modes to model the observed $B$-mode auto-spectrum, to capture properly complications due to noise inhomogeneity. Note that we can also avoid the noise complications in the observed $B$-mode auto-spectrum by using the cross-correlations between split data, and our choice of modeling the observed $B$-mode spectrum does not undermine any of our results. We add to this spectrum a theory tensor spectrum, parameterized by the tensor-to-scalar ratio $r$. These mean spectra (with $r=0$) are used for the fiducial spectra that are also needed in the likelihood.

\subsubsection{Covariances}

As noted earlier, the approximate likelihood that we use requires a set of fiducial angular power spectra and their covariances~\cite{Hamimeche:2008}. These can be obtained either from simulations or analytically. In our analysis, we use the covariance derived from simulations. 
Simulated covariances, which fully capture the effects of masking and inhomogeouneous noise, are expensive to compute given that we require the Monte Carlo error to be small in order to resolve the small correlations between band powers \cite{ref:smith_2004, BenoitLevy:2012va, ref:baleato_20}. Hence, as a cross-check, we also compare our results based on simulated covariances with those based on the analytic covariances presented in Appendix \ref{appendix:covariances}.

\subsection{Summary of delensing procedure and likelihood}

For convenience, we summarize the steps we take to produce a $B$-mode lensing template and how this is used to implement delensing within the likelihood framework.

\bi
\item We first prepare the lensing mass map. We combine the CMB lensing map reconstructed from the foreground-cleaned and Wiener-filtered LAT temperature and polarization maps, with the external mass tracer data from galaxy counts and the CIB. 
\item We also prepare the Wiener-filtered $E$-modes by combining LAT and SAT polarization maps as in \eq{Eq:Cinv}. 
\item The above two maps are combined to form the lensing $B$-mode template as in \eq{Eq:Quad-LensB}. The multipoles $B_{lm}^{\text{temp}}$ are projected to a real-space polarization map.
\item The polarization observed with the SAT, and the lensing template map, are projected onto pure $B$-modes over the region common to the LAT and SAT surveys. Their auto- and cross-spectra are used in an approximate, cross-spectral likelihood.
Note that this approach is the same as the BKSPT analysis. 
\item Generally, we would include SAT $B$-modes at all observing frequencies in the cross-spectral likelihood, and constrain simultaneously the tensor-to-scalar ratio, $r$, Galactic foreground-related parameters, and (potentially) parameters describing uncertainties in the expected $B$-mode lensing power and the auto- and cross-power of the lensing template (to incorporate the uncertainty of the mass tracer). In this paper, however, we ignore Galactic foregrounds and only constrain $r$ since our purpose is to see the impact of the practical effects in the lensing template construction on the $r$ constraint. The impact of uncertainties in the mass tracer are discussed in Sec.~\ref{sec:systematics}.
\ei

%//////////////////////////////////////////////////////////////////////////////////////////%
\section{Simulated delensing performance: constraints on inflationary gravitational waves}
\label{sec:results}
%//////////////////////////////////////////////////////////////////////////////////////////%

%<><><><><><><><><><><><><><><><><><><><><><><><><><><><><><><><><><><><><><><><><><><><><><><><><><>%
\begin{figure*}[t]
\bc
\includegraphics[width=5.8cm,clip]{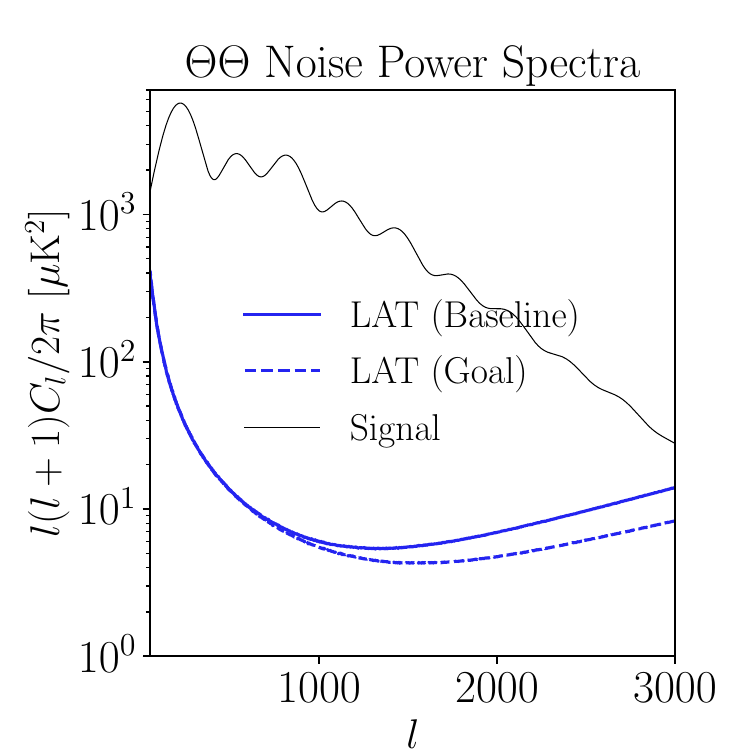}
\includegraphics[width=5.8cm,clip]{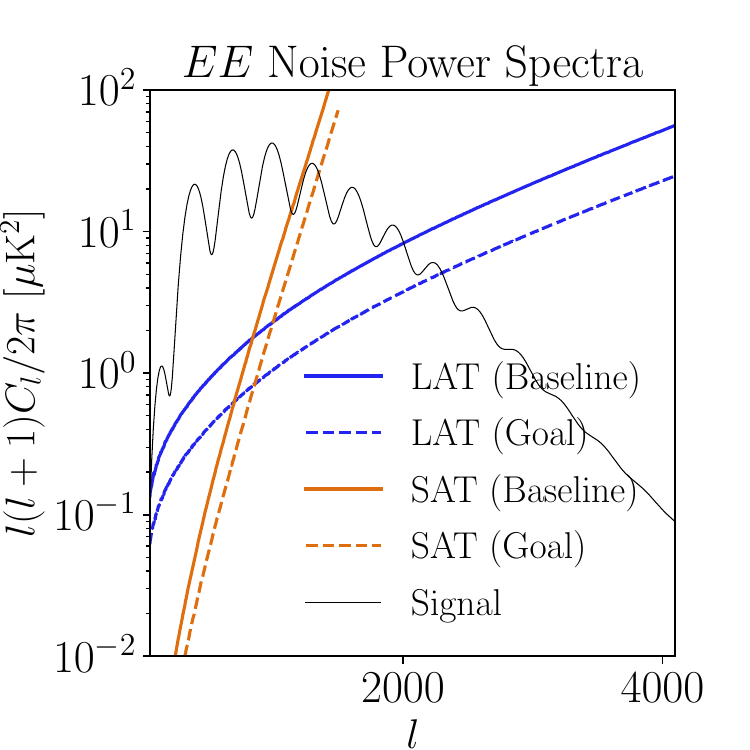}
\includegraphics[width=5.8cm,clip]{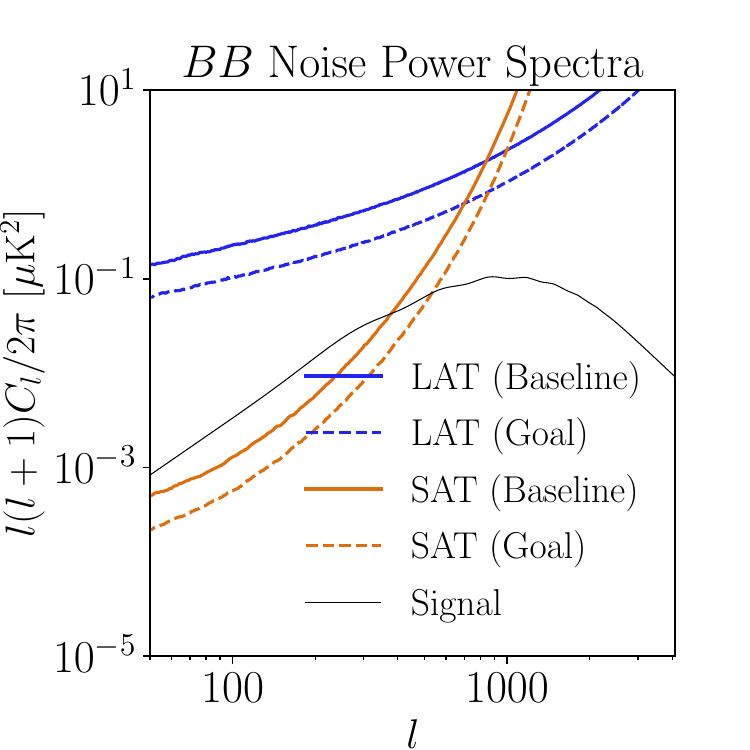}
\caption{
Noise angular power spectra for temperature anisotropies (left), $E$-mode polarization (middle) and $B$-mode polarization (right) computed from the SO map-based simulations after a pixel weighting with the square root of the hit count map and beam deconvolution. The blue and orange lines show the spectra obtained from the LAT and SAT maps, respectively.
The solid lines are for the baseline noise case and the dashed for the goal (see Ref.~\cite{SO:2018:forecast}).
We show the optimally co-added noise spectra from $93$, $145$ and $225$\,GHz.
The black solid lines are the model, lensed CMB spectra (with $r=0$). Note that we only use polarization from the SAT.
}
\label{Fig:cmb-aps}
\ec
\end{figure*}
%<><><><><><><><><><><><><><><><><><><><><><><><><><><><><><><><><><><><><><><><><><><><><><><><><><>%

\subsection{Simulations}\label{sec:sims}

For CMB maps, we use the precomputed data obtained from the map-based SO simulation package.\footnote{ \url{https://github.com/simonsobs/map_based_simulations}} 
The signal maps are convolved with a symmetric Gaussian beam at each frequency whose FWHM is given by Ref.~\cite{SO:2018:forecast}. 
The noise maps are generated using the map-based simulation package at each frequency for the LAT and SAT using a model of the instrumental and atmospheric noise spectra and hit-count maps. This allows efficient production of a large number of simulations, which would otherwise be computationally expensive if relying on simulated time-ordered data. The knee frequency for the $1/f$ polarization noise for the SAT, due primarily to atmospheric leakage and electronic noise, is chosen to be the ``pessimistic case'' of Ref.~\cite{SO:2018:forecast}. This choice means that the results of the $r$ constraint we obtain below are actually conservative. 
The model implements a damping of the large-scale power at $\l<\l_{\rm roll}$ to account for the filtering process applied in an actual analysis to reduce the atmospheric noise contamination of the large-scale modes. We set $\l_{\rm roll}=50$ for both SAT and LAT which equals to the knee multipole of the $1/f$ noise in the pessimistic case and do not use CMB multipoles at $\l<\l_{\rm roll}$ in the following analysis. 
We do not include Galactic and extragalactic foregrounds in the simulations. 
Thus, we also do not include the point-source masks. 
We generate $100$ realizations of the lensed CMB and noise at $93$, $145$, and $225$\,GHz for this work. 
The effective white noise levels in temperature at each frequency are $8.0$ ($93$), $10$ ($145$), and $22\,\mu$K-arcmin ($225$\,GHz) for the LAT, and $2.6$ ($93$), $3.3$ ($145$), and $6.3\,\mu$K-arcmin ($225$\,GHz) for the SAT \cite{SO:2018:forecast}. Note that for the LAT, which does not employ a rotating half-wave plate, the dominant source of the noise is the $1/f$ component at $\l\alt 1000$. 
We do not use other frequencies since these have much larger instrumental noise and do not help improve the delensing efficiency. 

We show in Fig.~\ref{Fig:window} the hit-count maps used for our simulation, which are one of the possible scan strategies being investigated for SO. Although the scan strategy is still to be finalized, we adopt the hit-count maps shown in the figure for this work. The disparity between the LAT and SAT hit-count maps is intentional: most of the science to be done with LAT maps is sample-variance limited at SO noise levels, even if the largest observable sky area is surveyed, and so calls for a wide survey with roughly uniform coverage; for the SAT, $B$-mode observations will likely be foreground, lensing, and noise limited, which leads one to concentrate integration time in a smaller region of low Galactic foreground emission.
As we show below, we achieve performance of delensing and constraints on $r$ similar to that obtained in idealized forecasts, i.e., the dissimilarity of the LAT and SAT hit-count maps does not significantly impact delensing and constraints on $r$.

Figure~\ref{Fig:cmb-aps} shows the noise angular power spectra for $\T\T$, $EE$, and $BB$ measured from simulations for the baseline and goal noise levels introduced in Ref.~\cite{SO:2018:forecast}. 
Pixels are weighted by the square root of the number of hits when computing these spectra. Note that, with such weighting, the spectra for inhomogeneous white noise is the same as if the hits were distributed uniformly, i.e., an isotropic survey.
The harmonic-space maps at each frequency are co-added with weighting given by the inverse variance computed from the noise spectra at each frequency. Note that we do not compute the temperature power spectrum for the SAT since we only use the SAT polarization in this work. 
In temperature, the atmospheric noise becomes significant below $\l\simeq 1000$. 
The LAT $B$-mode power spectrum is dominated by noise on all scales, but the SAT spectrum is signal dominated (by lensing) on large scales.

For external tracers, we generate Gaussian realizations of matter tracers that are appropriately correlated with a reference realization of the CMB lensing convergence map. To do this, we use the method and code presented in Appendix F of Ref.~\cite{ref:cib_delensing_biases}. 
Note that we do not include non-Gaussianity from the non-linear growth of density fluctuations in our simulations. It is known that non-Gaussianity of the CMB lensing convergence has a negligible impact on the power spectrum of the lensing $B$-mode template and covariance of the delensed $B$-modes \cite{Namikawa:2018:nldelens}. As shown in Ref.~\cite{Namikawa:2018:nldelens}, the same would be true when combining with matter tracers, despite potential complications from these typically being at lower redshift and from non-linear biasing. This is because the delensing utilizes mass tracers at high redshifts where the tracers are well correlated with the CMB lensing mass map and the non-linear growth is much less important. Additionally, the lensing $B$-modes on large angular scales ($\l\leq 100$) are most efficiently produced by intermediate scales of lensing mass ($300\alt L\alt 400$) \cite{Sherwin:2015,Simard:2015} where the non-linear growth is not significant.

\subsection{Lensing reconstruction} 

%<><><><><><><><><><><><><><><><><><><><><><><><><><><><><><><><><><><><><><><><><><><><><><><><><><>%
\begin{figure}[t] \bc \includegraphics[width=9.0cm,clip]{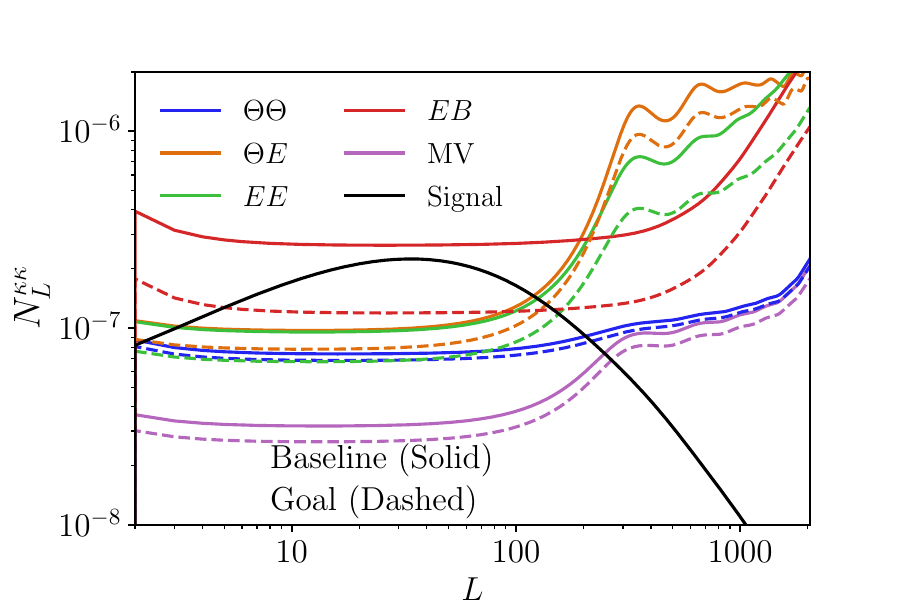} \caption{Analytic estimate of the lensing reconstruction noise using \eq{Eq:Rec:N0} for each individual quadratic estimator using the baseline (solid) and goal (dashed) instrumental noise spectra from simulations. 
Note that \eq{Eq:Rec:N0} is actually for the normalization and usually used for noise spectrum in a forecast, but in general differs from the actual noise spectrum. Given the neglect of $TE$ in the inverse-variance filtering, the ``analytic noise'' given here is considered as a rough estimate of the noise power for the quadratic estimator. 
For reference, we also show the analytic noise curve for the MV estimator of Ref.~\cite{OkamotoHu:quad}. The solid black curve shows the lensing convergence power spectrum.} \label{Fig:lens-noise} \ec \end{figure}
%<><><><><><><><><><><><><><><><><><><><><><><><><><><><><><><><><><><><><><><><><><><><><><><><><><>%

%<><><><><><><><><><><><><><><><><><><><><><><><><><><><><><><><><><><><><><><><><><><><><><><><><><>%
\begin{figure*}[t]
\bc
\includegraphics[width=8.5cm,clip]{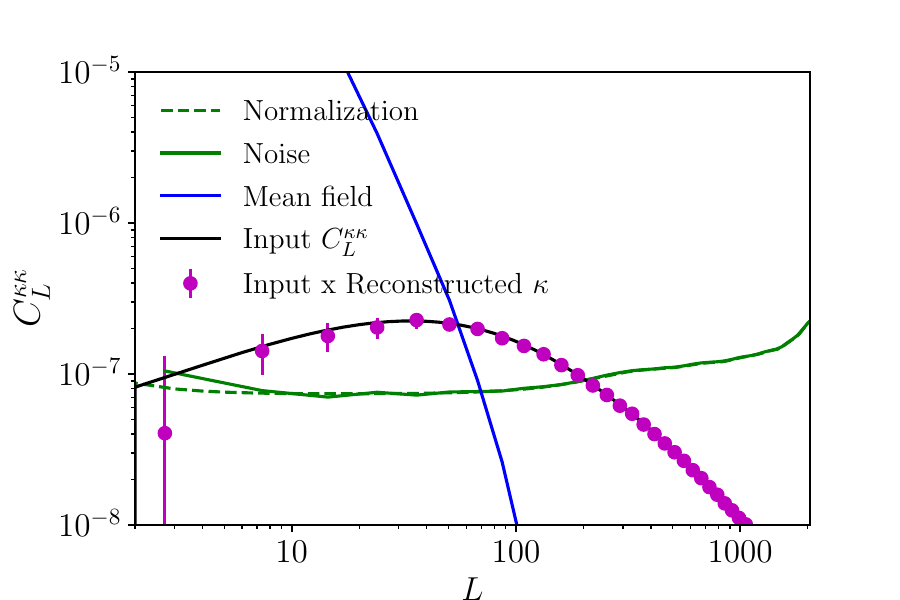}
\includegraphics[width=8.5cm,clip]{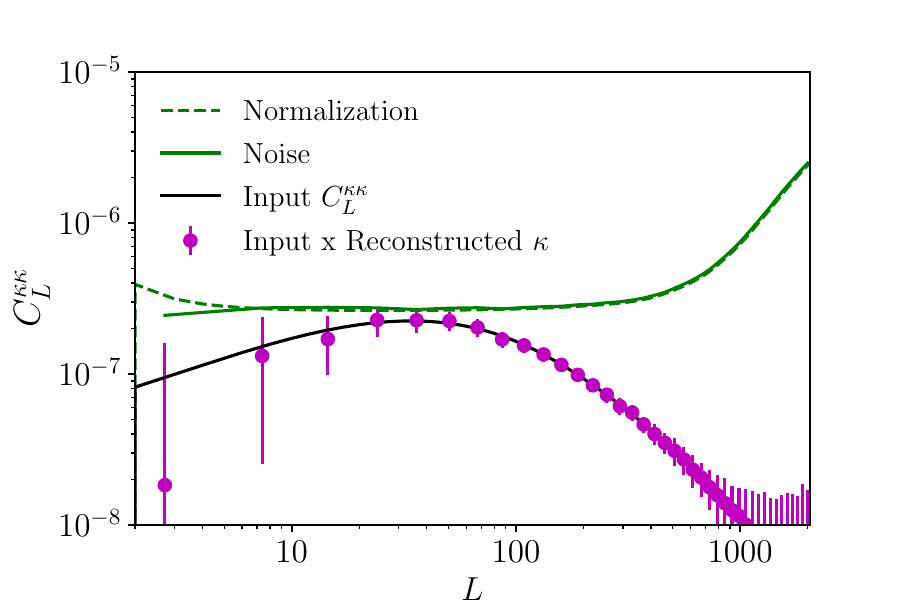}
\caption{
Reconstructed lensing map cross-correlated with the input lensing map using $\T\T$ (left) and $EB$ (right) quadratic estimators (magenta points). We also show the normalization (green dashed), noise (green solid), input lensing spectrum (black solid), and mean-field bias (blue). We use the baseline noise simulation. 
The mean-field bias in the $EB$ estimator is very small and not shown in the plot. The error bars denote representative of scatter in one simulation.
}
\label{Fig:lens-aps}
\ec
\end{figure*}
%<><><><><><><><><><><><><><><><><><><><><><><><><><><><><><><><><><><><><><><><><><><><><><><><><><>%

We first show the results of the internal lensing reconstruction from the CMB. 
To avoid the delensing bias on the scales of interest (see, e.g., Refs.~\cite{Teng:2011xc,Sehgal:2016,Namikawa:2017:delens,ref:baleato_20}), only the multipoles between $\l=301$ and $4096$ are taken into account in the calculation of the lensing reconstruction. 
For temperature, we further remove $\l\leq 500$ to suppress contamination from atmospheric noise without significant loss of signal-to-noise~\cite{Namikawa:2014:patchwork} and $\l\geq 3000$ to avoid expected contribution from the extragalactic foregrounds \cite{vanEngelen:2013rla}. 

Figure~\ref{Fig:lens-noise} shows the analytic estimates of the $\kappa$ noise power of the internal CMB lensing reconstructions for the two noise cases computed from the CMB noise spectra. The CMB instrumental noise power spectra are obtained from the simulations as shown in Fig.~\ref{Fig:cmb-aps}. 
Most of the signal-to-noise of the reconstructed lensing map comes from the $\T\T$, $\T E$ and $EE$ quadratic estimators for the baseline noise case. For the goal noise level, the $EB$ estimator becomes also important to improve the signal-to-noise of the lensing. 

Figure~\ref{Fig:lens-aps} shows the reconstructed lensing map cross-correlated with the input lensing map. We divide the spectra by $W_2\equiv\int {\rm d}^2\hatn W^2(\hatn)$ to account for the mis-normalization by the survey window \cite{Namikawa:2012:bhe}.
In the case of the $\T\T$ quadratic estimator, the power spectrum of the mean field becomes larger than the signal at $L\leq 60$. 
On the other hand, the reconstructed lensing map with the $EB$ estimator does not have a significant mean-field bias on all scales because $\langle EB\rangle =0$ by parity symmetry~\cite{Namikawa:2013:bhepol}. 
The difference between the input $\kappa$ spectrum and the cross-spectrum between input and reconstruction is within $5$\% at $L\agt 20$ for all of the quadratic estimators. This difference becomes larger than $10$\% on large scales, $L\alt 10$, which could be due to the presence of mode mixing induced by the survey mask, which is not corrected on large scales with our simple prescription for accounting for the survey mask, and higher-order lensing corrections~\cite{Hanson:2010:N2,Lewis:2011fk}. 
The bias is usually corrected by a Monte Carlo simulation.
Delensing, however, does not require the large-scale modes and we only use $L\geq 20$.

\subsection{Lensing template construction} 

%<><><><><><><><><><><><><><><><><><><><><><><><><><><><><><><><><><><><><><><><><><><><><><><><><><>%
\begin{figure*}[t]
\bc
\includegraphics[width=8.5cm,clip]{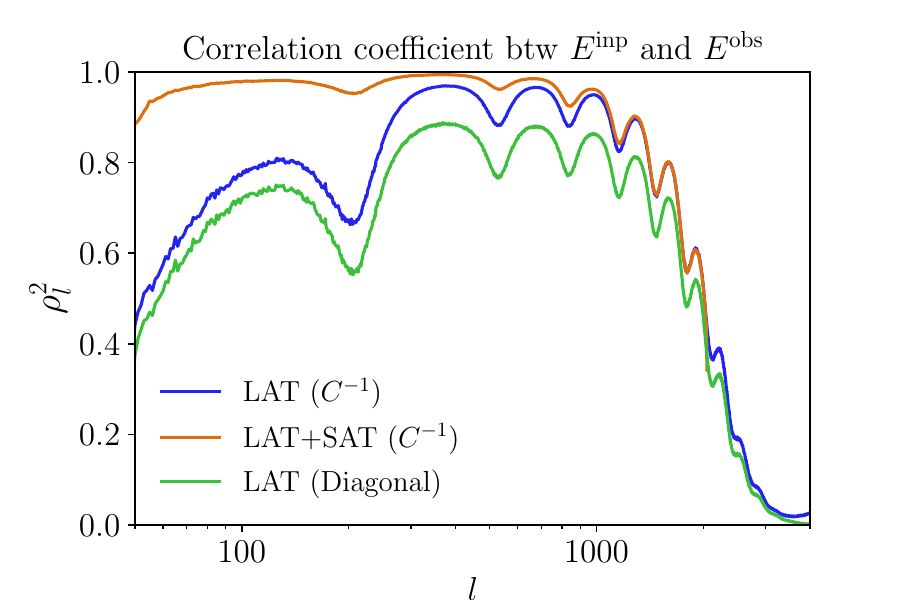}
\includegraphics[width=8.5cm,clip]{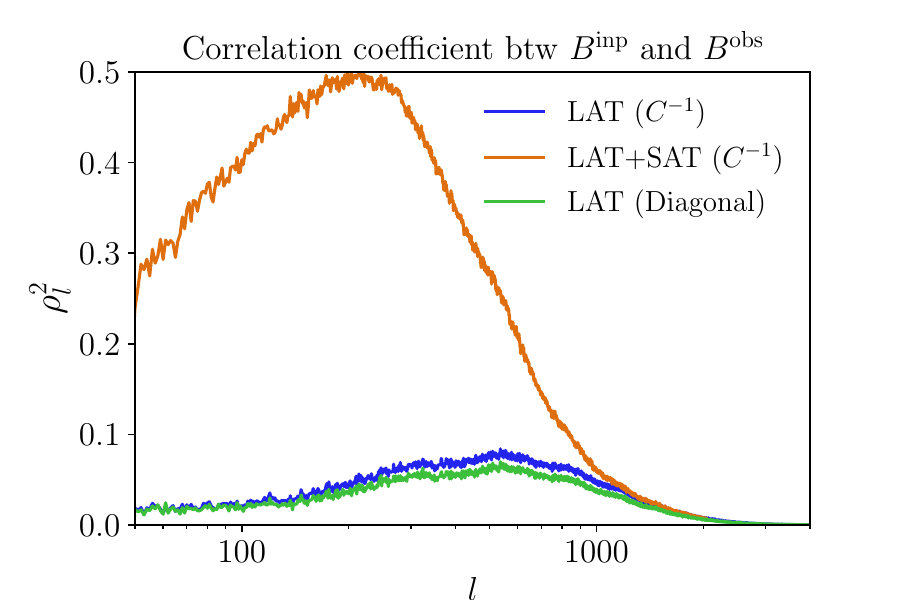}
\caption{
{\it Left}: Square of the correlation coefficient, $\rho_l^2$, between the Wiener-filtered $E$-modes and the input $E$-mode-only map, computed over the region of overlap of the SAT and LAT surveys, for the baseline noise case. Full ($C^{-1}$) filtering using only the LAT data is shown in blue; simple diagonal (in harmonic space) filtering of the LAT data in green; and full $C^{-1}$ filtering of the LAT and SAT data in orange. Note that ``LAT+SAT'' is only computed up to $\l=2048$ since it is used only for the $E$-modes in the lensing template construction. The correlation coefficient is close to unity for $\l\alt1000$ by applying LAT + SAT $C^{-1}$ filtering, meaning that it is close to optimal (i.e., signal-dominated) for $\l\alt1000$. 
{\it Right}: Same as the left panel, but for $B$-modes. 
}
\label{Fig:cinv}
\ec
\end{figure*}
%<><><><><><><><><><><><><><><><><><><><><><><><><><><><><><><><><><><><><><><><><><><><><><><><><><>%

%<><><><><><><><><><><><><><><><><><><><><><><><><><><><><><><><><><><><><><><><><><><><><><><><><><>%
\begin{figure*}[t]
\bc
\includegraphics[width=180mm,clip]{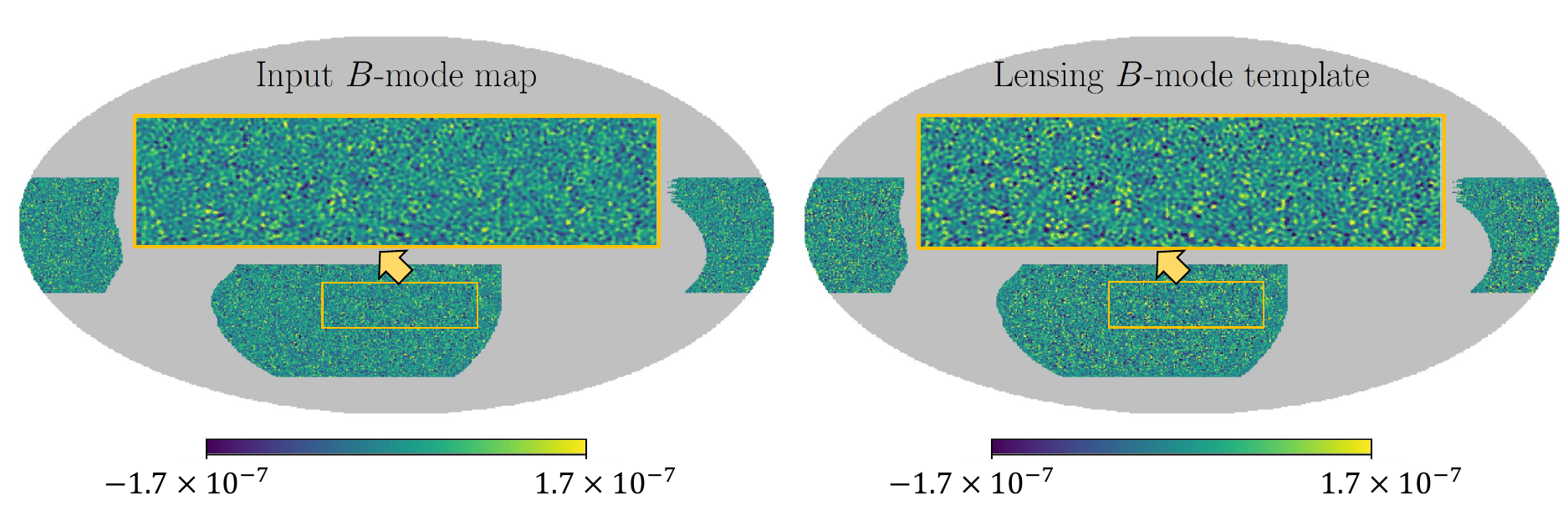}
\caption{
{\it Left}: Lensing $B$-mode template (scalar) map projected onto the region of overlap of the SAT and LAT surveys. The multipoles between $20\leq\l\leq128$ are included. A zoomed map is also plotted in the section. {\it Right}: Same as the left panel, but for the input $B$-mode map multiplied by $0.7$ which corresponds to the fraction of lensing $B$-modes removed by delensing (see text). One can visually see the correlation of these two maps. 
}
\label{Fig:bmap}
\ec
\end{figure*}
%<><><><><><><><><><><><><><><><><><><><><><><><><><><><><><><><><><><><><><><><><><><><><><><><><><>%

%<><><><><><><><><><><><><><><><><><><><><><><><><><><><><><><><><><><><><><><><><><><><><><><><><><>%
\begin{figure}[t]
\bc
\includegraphics[width=8.5cm,clip]{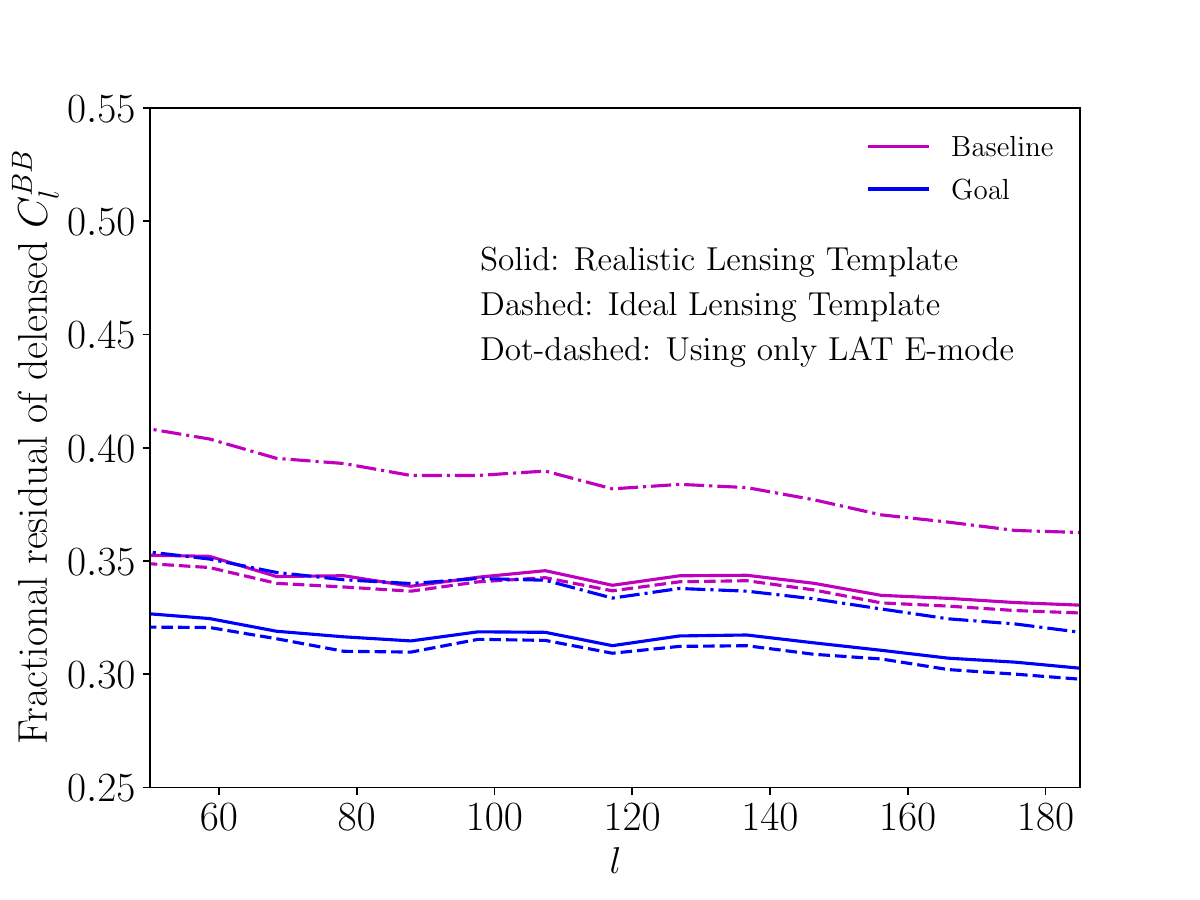}
\caption{
Fraction of lensing $B$-mode power left over after delensing in the region of overlap between the SAT and the LAT.
%which is given by $1-(C^{BB,{\rm cross}}_\l)^2/(\tCBB_\l C^{BB,{\rm temp}}_\l)$. 
The dashed lines show the ideal case where the CMB is observed over the full sky with isotropic noise. The dot-dashed lines show the case if we only use the LAT $E$-modes in the template construction. For the baseline (magenta) and goal (blue) noise cases, approximately $65$\% and $70$\% of lensing $B$-mode power is able to be removed, respectively. 
}
\label{Fig:bb-corr}
\ec
\end{figure}
%<><><><><><><><><><><><><><><><><><><><><><><><><><><><><><><><><><><><><><><><><><><><><><><><><><>%

%<><><><><><><><><><><><><><><><><><><><><><><><><><><><><><><><><><><><><><><><><><><><><><><><><><>%
\begin{figure}[t]
\bc
\includegraphics[width=8.5cm,clip]{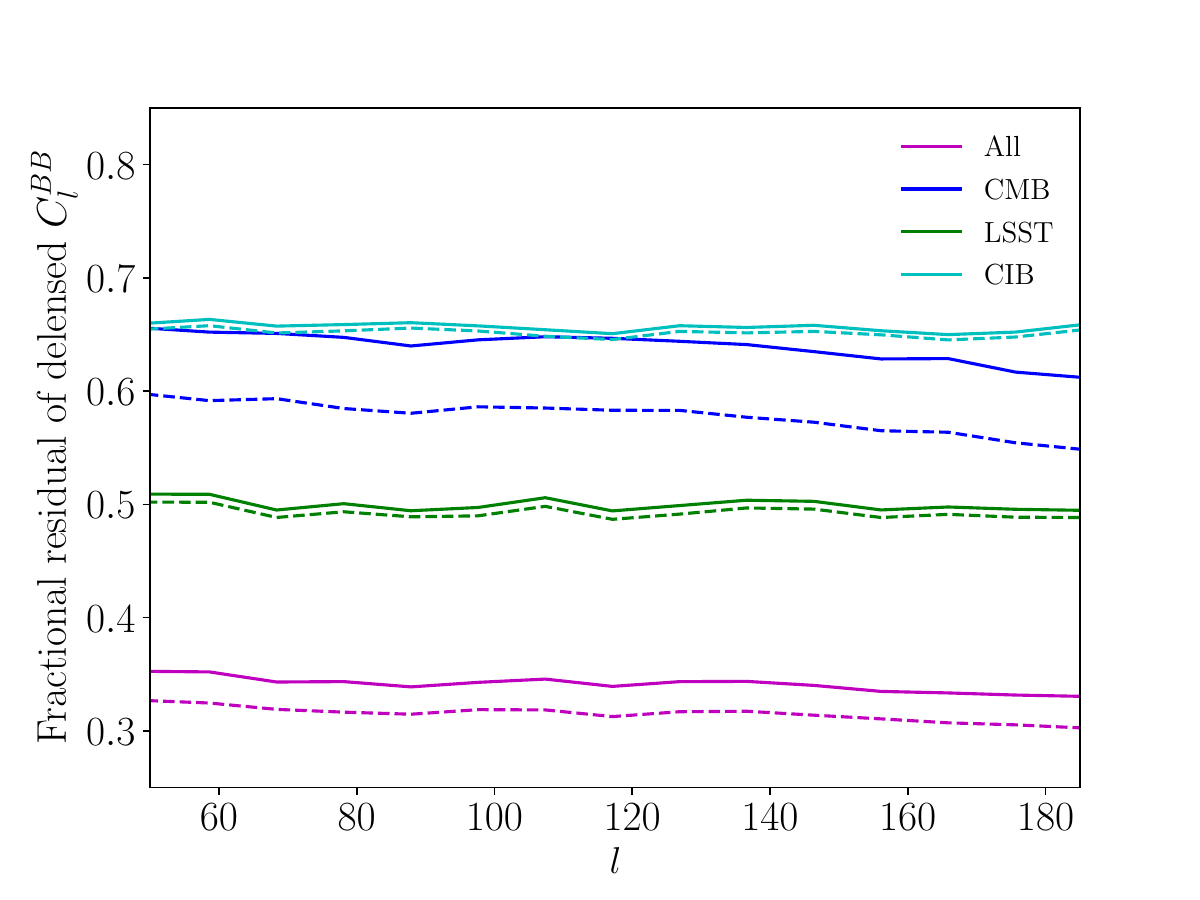}
\caption{
Same as Fig.~\ref{Fig:bb-corr} but with delensing using only the CMB lensing map (blue), galaxies from LSST (green) or the CIB (cyan), or their optimal combination (magenta), for the baseline (solid) and goal (dashed) noise levels. The LSST galaxies make the most significant contribution to delensing.
}
\label{Fig:bb-corr:each}
\ec
\end{figure}
%<><><><><><><><><><><><><><><><><><><><><><><><><><><><><><><><><><><><><><><><><><><><><><><><><><>%

Next, we show how significantly the SO inhomogeneous noise and survey geometry impacts the construction of the lensing $B$-mode template. In this section, the lensing template is built using the reconstructed lensing map at $20\leq L\leq 2048$ and $E$-modes at $50\leq \l\leq 2048$. 

Figure~\ref{Fig:cinv} shows the correlation coefficients between the input $E$/$B$-modes and Wiener-filtered ``observed'' $E$/$B$-modes, both of which are projected onto the region of overlap between the SAT and LAT surveys. Several options for the Wiener filtering are compared.
We see that the full Wiener-filtered LAT $E$-modes have better correlation with the input by around 5\%--10\% than if the simpler diagonal filtering is used. Optimally combining $E$-modes from the SAT and LAT polarization maps further improves the correlation with the input $E$-mode map, which is close to unity for $\l\alt1000$. The improvement of the correlation at $l\alt 500$ is important for the optimal lensing template because a significant fraction of the large-scale lensing $B$-modes are produced by $E$-modes at these scales. On the other hand, the LAT $B$-modes are dominated by noise and optimal filtering does not significantly recover the correlation with the input $B$-modes. 

Figure~\ref{Fig:bmap} shows two $B$-mode maps: the lensing $B$-mode template map; and the input $B$-mode map. Both maps are projected onto the region of overlap of the SAT and LAT surveys. We only include the multipoles between $20\leq\l\leq128$. 
The lensing $B$-modes in the template are suppressed due to the Wiener filtering process. 
We quantify this suppression by the cross-correlation of the template and input lensing $B$-modes divided by the input $B$-mode auto power spectrum. The ratio becomes $\sim 0.7$ using only large scales ($50\leq l\leq 190$). 
Thus, the input $B$-mode map is further multiplied by $0.7$ to correspond to the expected lensing $B$-mode signal in the template. 
The correlation between the two maps in the figure can be seen by eye.

Figure~\ref{Fig:bb-corr} shows the fraction of power left over after delensing $B$-modes in the region where the SAT and LAT surveys overlap. 
We consider the following delensing procedure:
\al{
    B^{\rm del}_{\l m} \equiv \hB_{\l m} - \alpha_\l B^{\rm temp}_{\l m} \,,
}
where $\alpha_\l$ is determined so that the variance of $B^{\rm del}$ is minimized: 
\al{
    \alpha_\l \equiv \frac{C^{BB,{\rm cross}}_\l}{C_\l^{BB,{\rm temp}}}
    \,. 
}
Here, $C^{BB,{\rm cross}}_\l$ is the cross-power spectrum of the template and the input lensing $B$-modes, and $C^{BB,{\rm temp}}_\l$ and $\hCBB_\l$ are the power spectra of the template and the observed $B$-modes, respectively. We ignore the dependence of $\alpha$ on $m$. In an idealistic case, $\alpha_\l=1$. The angular power spectrum of $B^{\rm del}_{\l m}$ is then given by:
\al{
    C^{BB,\rm del}_\l = \hC^{BB}_\l\left[1-\frac{(C^{BB,{\rm cross}}_\l)^2}{\hC^{BB}_\l C^{BB,{\rm temp}}_\l}\right]
    \,. 
}
The $B$-mode spectra, $\hCBB_\l$, $C^{BB,{\rm cross}}_\l$ and $C^{BB,{\rm temp}}_\l$, are computed over the region of overlap as follows. First, we construct the mask for the overlap region by simply multiplying the binary masks of each survey. We then multiply the Stokes parameters constructed from lensing template $B$-mode map and the input lensing $B$-mode map by this (binary) survey mask with a $5^\circ$ apodization 
and compute spectra from these masked maps. We do not apply any purification since we use the $B$-mode-only maps. 
We do not include noise in $\hB$ and $\hCBB$ is simply the lensing $B$-mode spectrum.
Figure \ref{Fig:bb-corr} shows the following three cases for either the baseline (magenta) or the goal (blue) noise scenarios:
(1) the realistic case using the SO map-based simulations (solid lines); (2) a relatively idealistic case in which all maps are full sky and the instrumental noise is isotropic with power spectrum equivalent to that obtained from the simulations (see Fig.~\ref{Fig:cmb-aps}), but the spectra are still computed over the overlap region (dashed lines); and (3) the case in which the template is constructed using only LAT $E$-modes (dot-dashed lines). In the realistic case (solid), the fraction of the lensing $B$-mode power left over after delensing is roughly $30$\%--$35$\% depending on angular scales. The results imply that our pipeline gives reconstructed lensing $B$-modes that are almost as correlated with the actual lensing $B$-modes as in the case of an isotropic survey. In particular, the amount of delensing is close to that given in Ref.~\cite{SO:2018:forecast} ($30$\% residual power for the goal noise levels).
Comparing cases (1) (solid) and (2) (dashed), the impact of the realistic inhomogeneous instrumental noise is small. The figure also indicates that adding $E$-modes from the SAT further reduces the lensing contribution by more than $5$\% and its benefit is not completely negligible. 

Figure~\ref{Fig:bb-corr:each} shows the individual contributions to the residual $B$-mode power after delensing with different tracers. LSST galaxies contribute most to delensing, while the reconstructed CMB lensing map and CIB have similar contributions. 

\subsection{Constraining IGWs} 

\begin{figure}
\includegraphics[width=8.5cm,clip]{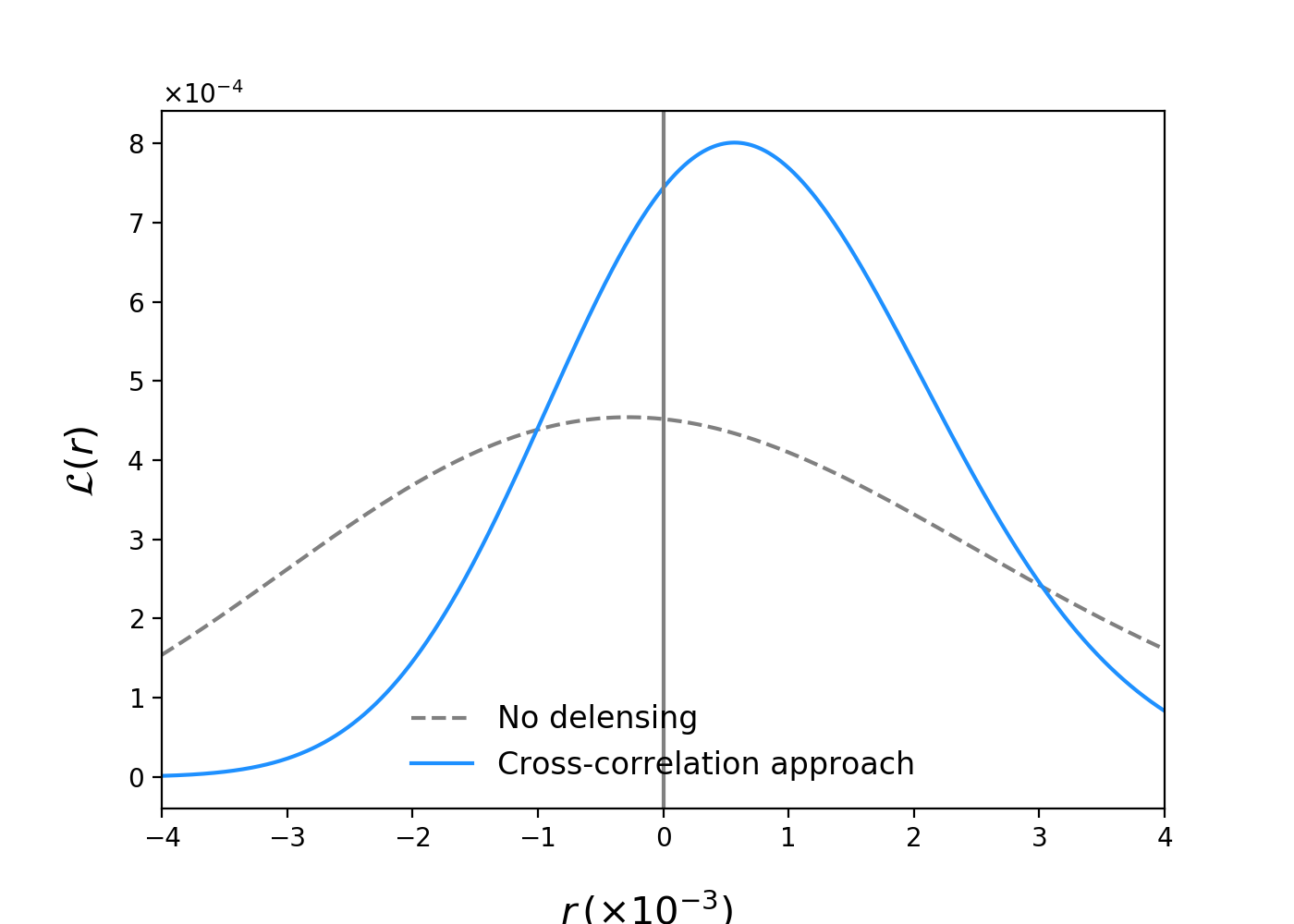}
\caption{
Demonstration of constraining the tensor-to-scalar ratio, $r$, with the cross-spectral (or ``cross-correlation'') approach in which all of the auto- and cross-power spectra between the observed $B$-modes and the lensing $B$-mode template are used in the likelihood analysis (solid blue line; see Sec.~\ref{sec:likelihood}). For comparison, we also show (in dashed) the case without the lensing $B$-mode template, i.e., no delensing. Note that we extend the likelihood into the unphysical region $r<0$ for illustration in the figure. The constraints on $r$ are $\sigma(r)=0.003$ for the no-delensing case and $\sigma(r)=0.0015$ with the lensing $B$-mode template, respectively. 
}
\label{fig:l_compare}
\end{figure}

Using the lensing $B$-mode template, we perform a likelihood analysis to determine the expected constraint on the tensor-to-scalar ratio, $r$, using the MBS simulations. The results are shown in Fig.~\ref{fig:l_compare}. 
As described in Sec.~\ref{sec:likelihood}, we compute the auto- and cross-power spectra of $B$-modes obtained from the SAT region and from the lensing template using the pure-$B$-mode formalism~\cite{Smith:2005:chi-estimator}, and use these in an approximate likelihood. 
We use $B$-mode multipoles between $\l=50$ and $200$. 
Since our purpose is to see the impact of practical effects in the construction of the lensing template on the constraint on $r$, we only consider one parameter, $r$, for simplicity, and ignore the Galactic foreground complexity. 
%Note that we extend the likelihood into the unphysical region $r<0$ for illustration in the figure.

We show two cases, with and without the lensing template in the likelihood. The $1\,\sigma$ constraint on $r$ with delensing is $\sigma(r)=0.0015$, which is close to the expectation from the ideal (isotropic) case and is nearly a factor of two improvement from the no-delensing case. 
This indicates that the non-idealities from non-white noise and masking do not significantly degrade the delensing performance which is enough to reproduce the constraint on $r$ expected from the idealized forecast, up to possible Galactic foreground non-idealities.

\section{Uncertainties in external tracer spectra}
\label{sec:systematics}

In this section, we study the impact of uncertainties in the spectra of external mass tracers on our efforts to constrain primordial $B$-modes. For a more thorough study of possible systematic effects arising when the CIB is used as the matter proxy for delensing, we refer the reader to Refs.~\cite{ref:bkspt_cib_delensing,ref:cib_delensing_biases}.

Evaluating the model spectra of Sec.~\ref{ref:model_spectra} requires knowledge of the auto-spectra and cross-spectra with CMB lensing of each of the tracers involved. In practical applications, the tracer spectra will likely be determined by fitting a smooth curve to measurements, and hence will be uncertain to some degree. It is important, then, to quantify accurately this uncertainty, as otherwise we run the risk of mistaking nontrivially shaped lensing residuals for a primordial signal, and thus biasing constraints on $r$~\cite{Sherwin:2015}. In this section, we explore this possibility quantitatively.

Before proceeding further, we note that this issue will also mean that, in principle, the weighting scheme summarized in Eq.~\eqref{eqn:multitracer_weights} will be sub-optimal whenever the fiducial spectra deviate from the truth. However, we ignore this effect because the corrections are second order in the error of the weight function and are therefore small~\cite{Sherwin:2015}.

For a quantitative analysis in this section, we first derive basic equations for the relevant $B$-mode power spectra. We do this in the flat-sky approximation as it has been shown that, on the angular scales relevant for SO, the approximation is in very good agreement with the exact curved-sky result (to within around $1$\%)~\cite{Challinor:2005jy}. 
First of all, we model the cross-correlation of observed $B$-modes with a leading-order lensing $B$-mode template formed from Wiener-filtered $E$-modes and a co-added mass map that involves both internally and externally estimated mass tracers.
In the flat-sky approximation, the $E$ and $B$-modes are given as the spin-$2$ Fourier transform of the Stokes $Q$ and $U$ maps \cite{HuOkamoto:2001}:
%----------------------------------------------------------------------------------------------------%
\al{
	E_{\bl} \pm \iu B_{\bl} = \Int{2}{\hatn}{} \E^{-\iu\hatn\cdot\bl} [Q\pm \iu U](\hatn) \E^{\mp2\iu\psi_{\bl}}
	\,, \label{Eq:EB-def:flat}
}
%----------------------------------------------------------------------------------------------------% 
where $\psi_{\bl}$ is the angle that $\bl$ makes with the axis defining positive Stokes $Q$.
Proceeding analogously to the full-sky case, we expand the lensed $E$ and $B$-modes to first order in $\kappa$ to obtain~\cite{HuOkamoto:2001}: 
%----------------------------------------------------------------------------------------------------%
\al{
	\Blens_{\bl} &= \Int{2}{\bl'}{(2\pi)^2} W(\bl,\bl') E_{\bl'} \kappa_{\bl-\bl'} 
	\label{Eq:Lensed-B:flat} 
	\,, 
} 
%----------------------------------------------------------------------------------------------------%
where $\kappa_{\bl}$ are the Fourier modes of the lensing convergence map and
%----------------------------------------------------------------------------------------------------%
\beq
    W(\bl,\bl') \equiv 2\,\frac{\bl'\cdot(\bl-\bl')}{|\bl-\bl'|^2} 
    \sin{2\left(\psi_{\bl}-\psi_{\bl'}\right)} 
    \,. 
\eeq
%----------------------------------------------------------------------------------------------------%
Equation~\eqref{Eq:Lensed-B:flat} is the flat-sky analogue of \eq{Eq:Lensed-B}. We compute the lensing $B$-mode template by replacing the true $E_{\bl}$ and $\kappa_{\bl}$ with the Wiener-filtered, measured $E$-modes, $\wE_{\bl}$, and the optimally-combined matter tracer map, $\estk^{\rm comb}_{\bl}$, in \eq{Eq:Lensed-B:flat}: 
%----------------------------------------------------------------------------------------------------%
\al{
	B^{\mathrm{temp}}_{\bl} &= \Int{2}{\bl'}{(2\pi)^2} W(\bl,\bl') 
	\wE_{\bl'} \hat{\kappa}^{\rm comb}_{\bl-\bl'} 
	\label{Eq:Lensed-B-est:flat} 
	\,. 
} 
%----------------------------------------------------------------------------------------------------%
We assume that the unlensed $E$-modes and lensing convergence are Gaussian distributed and uncorrelated with each other, and the Wiener-filtering is diagonal in $l$, i.e.,
\begin{equation}
    \mathcal{W}_\l^E \equiv \frac{\tCEE_\l}{\tCEE_\l+N_\l^{EE, \mathrm{fid}}} 
    \,,
\end{equation}
where $N_\l^{EE,\text{fid}}$ is a fiducial $E$-mode noise spectrum. Then, to $O(\kappa^2)$, the cross-spectrum is 
\begin{align}
    C_\l^{BB,{\rm cross}} 
    &= \Int{2}{\bl'}{(2\pi)^2}\, W^2(\bl,\bl') \mathcal{W}^E_{\l'}\CEE_{\l'} 
    C^{\kappa\hat{\kappa}^{\rm comb}}_{|\bl'-\bl|}
    \nonumber \\
    & = \Int{2}{\bl'}{(2\pi)^2}\, W^2(\bl,\bl') \mathcal{W}^E_{\l'}\CEE_{\l'}
    C^{\kappa \kappa}_{|\bl'-\bl|} \rho^{2}_{|\bl'-\bl|} 
    \,,
\end{align}
where we consider the terms up to $O(\kappa^2)$ and use the {\it unlensed} $E$-mode power spectrum to describe the cross-power spectrum. The correlation coefficient is given by:
\begin{equation}
    \rho_L \equiv \frac{C_L^{\kappa \hat{\kappa}^{\text{comb}}}}{\sqrt{C_L^{\kappa\kappa}C_L^{\hat{\kappa}^{\text{comb}}\hat{\kappa}^{\text{comb}}}}} 
    = \sqrt{\frac{C_L^{\kappa \hat{\kappa}^{\text{comb}}}}{C_L^{\kappa\kappa}}} 
    \,. \label{eq:corrcoeff}
\end{equation}
The second equality here follows from $\hat{\kappa}_{LM}^{\text{comb}}$ involving the Wiener-filtered combination of tracers (see Sec.~\ref{subsec:tracer}).
Note that evaluating the cross-power spectrum with the {\it lensed} $E$-mode power spectrum instead of the {\it unlensed} $E$-mode power spectrum makes very little difference as the acoustic peaks are smoothed out in the convolution integral. Note also that the contributions at the fourth order of $\kappa$ are significantly suppressed in the template delensing method due to a cancellation of terms, and the expressions here are quite accurate (see Ref.~\cite{BaleatoLizancos:2020b} for details). 
On the other hand, under the same set of assumptions, the auto-spectrum of the template can be modeled, to $O(\kappa^2)$, as
\begin{align}
    C_\l^{BB,{\rm temp}} & = \Int{2}{\bl'}{(2\pi)^2}\, 
    W^2(\bl,\bl') C_{\l'}^{\wE\wE}
    C^{\hat{\kappa}^{\rm comb}\hat{\kappa}^{\rm comb}}_{|\bl'-\bl|}
    \nonumber \\ 
    & = \Int{2}{\bl'}{(2\pi)^2}\, W^2(\bl,\bl') {\cal W}^E_{\l'}\CEE_{\l'}
    C^{\kappa \kappa}_{|\bl'-\bl|} \rho^{2}_{|\bl'-\bl|} 
    \,, 
\end{align}
which equals $C_\l^{BB,{\rm cross}}$.

Next, consider the angular power spectrum of residual lensing $B$-modes after delensing (i.e., subtracting the template from the observed $B$-modes) with the co-added tracer, $\hat{\kappa}_{LM}^{\text{comb}}$.
We choose this as our case study because the insights we gather from this simpler analysis should ultimately be applicable to one 
that combines all the individual auto- and cross-spectra of templates and observations, in the way of the BKSPT analysis followed earlier in this paper. 

To leading order in lensing, this is
\begin{align}\label{eqn:marginalising_over_alens_eq1}
    C_\l^{BB,{\rm res}} &= \Int{2}{\bl'}{(2\pi)^2} 
    W^2\left(\bl,\bl'\right) 
    \bigg[\CEE_{\l'}C_{|\bl-\bl'|}^{\kappa\kappa} 
    \nonumber \\
    & - 2\,\CEE_{\l'} \mathcal{W}_{\l'}^{E} 
    \sum_i c^{i}_{|\bl-\bl'|} C_{|\bl-\bl'|}^{\kappa \hat{\kappa}^i} 
    \nonumber \\
    & + \left( \CEE_{\l'} + N_{\l'}^{EE}\right) 
    \left(\mathcal{W}_{\l'}^{E} \right)^2 
    \nonumber \\
    & \times \sum_i \sum_j c^{i}_{|\bl-\bl'|} 
    c^{j}_{|\bl-\bl'|} C_{|\bl-\bl'|}^{ \hat{\kappa}^i\hat{\kappa}^j} 
    \bigg]
    \,, 
\end{align}
where the weights, $c_\l^i$, are calculated using fiducial spectra and we have not simplified further to allow for the case where the fiducial spectra differ from the truth.

In the case where the true spectra deviate from the fiducial model, we parametrize the true spectra as follows:
\begin{align}
    C_\l^{\kappa \hat{\kappa}^i} &= C_\l^{\kappa \hat{\kappa}^i, \mathrm{fid}} + \Delta C_\l^{\kappa \hat{\kappa}^i} 
    \,, \\
    C_\l^{\hat{\kappa}^i\hat{\kappa}^j} &= C_\l^{\hat{\kappa}^i \hat{\kappa}^j, \mathrm{fid}} + \Delta C_\l^{\hat{\kappa}^i \hat{\kappa}^j} 
    \,,
\end{align}

We allow for errors in the cross- and auto-spectra of the external tracers, and in the cross-spectra of the \emph{external} tracers with the true $\kappa$ and with the internally reconstructed $\kappa$, which we assume to be equal as the fiducial cross-spectra are calibrated on the same empirical spectra. We thus have
\begin{equation}
    \Delta C_l^{\kappa \hat{\kappa}^i} = \Delta C_l^{\hat{\kappa}\hat{\kappa}^i} \quad (\hat{\kappa}^i \neq \hat{\kappa}) 
    \,,
\end{equation}
where $\hat{\kappa}$ is the internal reconstruction. On the other hand, we assume the fiducial model is correct for the auto-spectrum of the internal reconstruction and its cross-spectrum with the true $\kappa$, since these can be predicted to high accuracy from first principles; hence,
\begin{equation}
    \Delta C_l^{\kappa \hat{\kappa}} = \Delta C_l^{\hat{\kappa}\hat{\kappa}} =0 
    \,.
\end{equation}
For the case of $n$ external tracers, we sample the $n(n+3)/2$ distinct deviations, $\Delta C_l^{\hat{\kappa}\hat{\kappa}^i}$ and $\Delta C_l^{\hat{\kappa}^i\hat{\kappa}^j}$ for $\hat{\kappa}^i$ and $\hat{\kappa}^j$ not equal to $\hat{\kappa}$, as zero-mean, Gaussian variables drawn from the appropriate covariance matrix. We model this with the covariances of the relevant empirical band powers. Using bins of width $\Delta l$ and a fraction $f_{\text{sky}}$ of the sky, the band power covariances are
\begin{multline}\label{eqn:covmat}
    \langle \Delta C_l^{\hat{\kappa}^i \hat{\kappa}^j} \Delta C_l^{\hat{\kappa}^m \hat{\kappa}^n} \rangle = \frac{1}{(2l+1)\Delta l f_{\text{sky}}}
    \\
    \times \left( C_l^{\hat{\kappa}^i \hat{\kappa}^m}
    C_l^{\hat{\kappa}^j \hat{\kappa}^n} + C_l^{\hat{\kappa}^i \hat{\kappa}^n} C_l^{\hat{\kappa}^j \hat{\kappa}^m} \right) 
    \,.
\end{multline}
For spectra involving the CIB, we use measurements from Planck; for those involving internal reconstructions, we assume SO \emph{goal} noise levels; and for the galaxies, we adopt the noise levels forecasted for the LSST \emph{gold} sample. When calculating elements of the covariance matrix not involving lensing, we set $f_{\mathrm{sky}}=0.05$; on the other hand, for elements featuring the cross-spectra of external tracers with lensing, we assume a larger footprint with $f_{\mathrm{sky}}=0.4$. We also choose $\Delta l=1$, and consider multipoles ranging approximately between $60<l<1500$.

%<><><><><><><><><><><><><><><><><><><><><><><><><><><><><><><><><><><><><><><><><><><><><><><><><><>%
\begin{figure*}[t]
\bc
\includegraphics[width=8.5cm,clip]{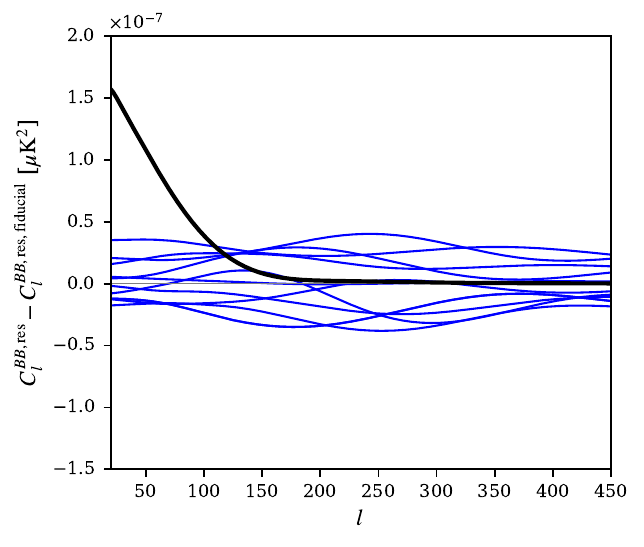}
\includegraphics[width=8.5cm,clip]{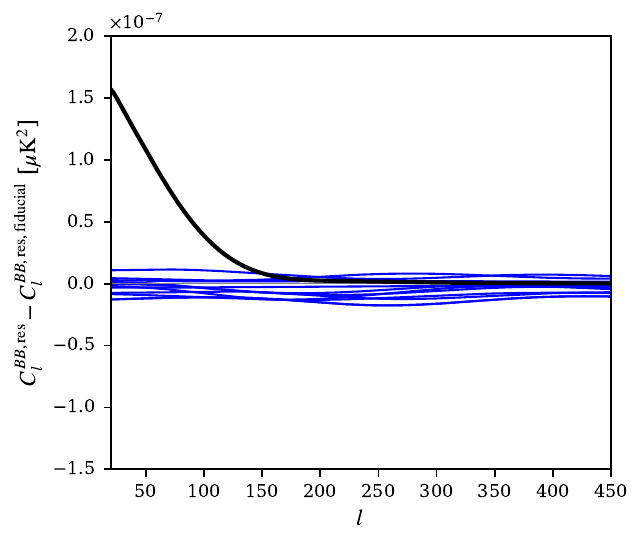}
\caption{
Impact of uncertainties in the tracer spectra on the power spectrum of $B$-mode lensing residuals after delensing. We quantify this by perturbing the true
auto-spectra of the external tracers and their cross-spectra with lensing and with each of the other tracers about the fiducial spectra, with errors drawn from a Gaussian distribution consistent with the covariance matrix described in the text.
\textit{Left:} the case where delensing is performed using only CIB maps, with spectra as measured by \textit{Planck} GNILC. \textit{Right:} the \textit{Planck} CIB maps are co-added with LSST galaxies (gold sample) and SO internal reconstructions. It is readily apparent that co-adding external tracers with internal reconstructions mitigates possible shapes in the residuals that might be confused with a primordial component. For comparison, a primordial signal with $r=0.001$ is shown in black. 
}
\label{fig:impact_of_tracer_spec_error}
\ec
\end{figure*}

\begin{figure*}[t]
\bc
\includegraphics[width=16cm]{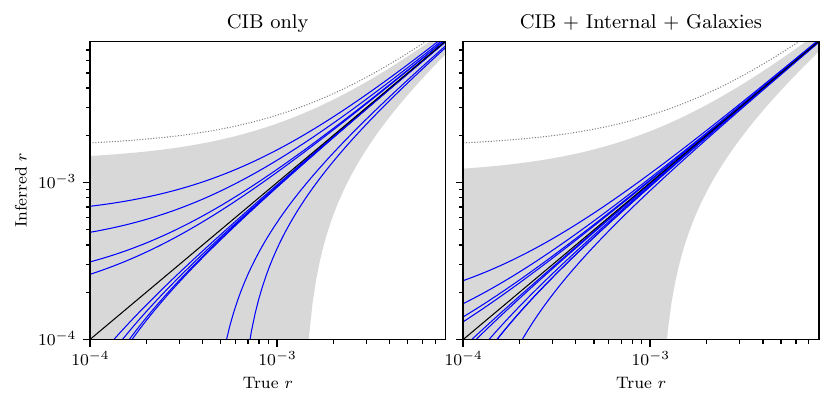}
\caption{Impact on the inferred tensor-to-scalar ratio $r$, as a function of input signal, of deviations from the fiducial model of the spectra of the external tracers used for delensing. We show results for ten different, random fluctuations of the tracer spectra consistent with measurement errors (see text) in the case of delensing with only the CIB (left) or the multi-tracer approach with CIB, an internal reconstruction and galaxies (right). (Note the overlap of some of the lines).
We constrain $r$ using scales $l_{\mathrm{min}}=50$ and $l_{\mathrm{max}}=200$. In general, the effect of modeling errors is small compared to the statistical uncertainty of SO. To show this, we plot as the shaded, gray region the $\pm1\,\sigma$ uncertainty for $r=0$ of an experiment covering $5$\% of the sky with the noise level of the SO SAT's 93\,GHz channel, no foregrounds and delensing as allowed by each of the tracer combinations, in the case where $r$ is constrained over the multipole range described above. For comparison also, the dotted lines show the size of the bias on $r$ if residual dust $B$-modes in the SO SAT maps (as forecasted by Ref.~\cite{SO:2018:forecast}) are not modeled in the $BB$ power spectrum.
}
\label{fig:propagating_bias_to_r}
\ec
\end{figure*}

\begin{figure}[t]
\bc
\includegraphics[width=8.5cm,clip]{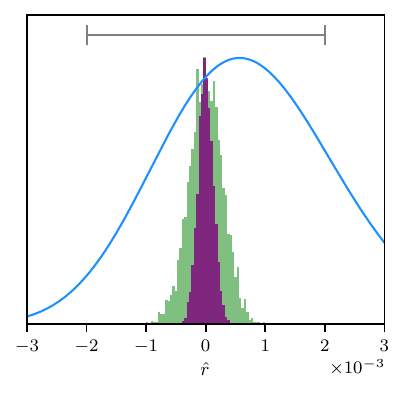}
\caption{Distribution of the deviations in the inferred tensor-to-scalar ratio, $\hat{r}$, from 5\,000 realizations (with input $r=0$), arising from uncertainties in the spectra of the external tracers used to delens $B$-modes, in the case where \emph{Planck} CIB maps alone are used (green histogram), or when these CIB maps are co-added with SO internal reconstructions and LSST galaxies (purple histogram). The bias is small compared to the precision of a typical likelihood curve when the inference is carried out in the presence of residual lensing and experimental noise (blue curve, which is the same as in Fig.~\ref{fig:l_compare}), and even smaller compared to the standard deviation on $r$ expected after foreground-cleaning and delensing SO observations (gray interval). The histograms and likelihood curve are all scaled to have similar heights to facilitate comparison.
}
\label{fig:bias_on_r_null_scenario}
\ec
\end{figure}
%<><><><><><><><><><><><><><><><><><><><><><><><><><><><><><><><><><><><><><><><><><><><><><><><><><>%

We can use Eq.~(\ref{eqn:marginalising_over_alens_eq1}), with $ C_\l^{\kappa \hat{\kappa}^{i}}$ replaced by $\Delta C_\l^{\kappa \hat{\kappa}^{i}}$  and $ C_\l^{\hat{\kappa}^{i} \hat{\kappa}^{j}}$ by $\Delta C_\l^{\hat{\kappa}^{i} \hat{\kappa}^{j}}$, to study possible deviations of the true lensing $B$-mode residual from a model constructed around the fiducial tracer spectra (that is, the same one used to calculate the weights). Several realizations of such deviations, consistent with the estimated measurement errors, are shown in Fig.~\ref{fig:impact_of_tracer_spec_error}. We see that the combination of external tracers with internal reconstructions (for which the correlation with lensing on large-scale scales is known very accurately) leads to residuals that are rather flat, significantly more so than in the case where external tracers alone are used. This behavior arises since the contribution of deviations in the tracer power spectra on small scales combine with the small-scale $E$-mode power to produce $B$-mode power that behaves as white noise on large scales. This is not the case for spectral deviations on large scales, but these are suppressed in the optimal combination with an internal lensing reconstruction and so contribute little.
This suggests that uncertainties in tracer spectra can be integrated into our constraints on $r$ by means of a simple marginalization procedure involving a single parameter governing the amplitude of a white-noise residual. Reference~\cite{Sherwin:2015} studied this procedure in the case where the CIB is the only tracer, finding that the uncertainty grows only moderately. Given that the residuals we see arise when co-adding multiple tracers are significantly flatter than when the CIB is used by itself, we expect the degradation in constraining power to be even smaller.

The residuals due to improper modeling shown in Fig.~\ref{fig:impact_of_tracer_spec_error} can be propagated to errors in estimates of $r$ using the relation (e.g., Ref.~\cite{ref:roy_et_al_21}) 
\begin{align}
    \Delta \hat{r} = & \left( \sum_{l=l_{\mathrm{min}}}^{l_{\mathrm{max}}} \frac{\left[C_l^{BB,\mathrm{prim}}(r=1)\right]^2}{ \mathrm{Var}\left(C_l^{BB,\mathrm{del}}\right)}\right)^{-1} \nonumber \\
    &\times \sum_{l=l_{\mathrm{min}}}^{l_{\mathrm{max}}} \frac{C_l^{BB,\mathrm{unmodeled}} C_l^{BB,\mathrm{prim}}(r=1)}{ \mathrm{Var}\left(C_l^{BB,\mathrm{del}}\right)}\,.
\end{align}
Here, 
$C_l^{BB,\mathrm{prim}}(r=1)$ is the angular power spectrum of primordial $B$-modes with $r=1$, $\mathrm{Var}(C_l^{BB,\mathrm{del}})$ is the variance of the power spectrum of delensed $B$-modes (which we assume to be Gaussian, free of foregrounds, and to feature a level of experimental noise appropriate for the 93\,GHz channel of the SO SATs in the \emph{goal} scenario) and $C_l^{BB,\mathrm{unmodeled}}$ is the part of the delensed $B$-mode spectrum that we have not modeled -- in this case, due to incorrect modeling of the external tracer spectra. In Fig.~\ref{fig:propagating_bias_to_r}, we compare the estimated shifts for ten random realizations to the standard deviation (assuming $r=0$) expected of an experiment covering $5$\% of the sky, with the noise levels of the SO SATs, and in the limit of no foreground $BB$ power and a removal of lensing as appropriate for delensing with the CIB alone or in the multi-tracer approach described above. We use $l_{\mathrm{min}}=50$ and $l_{\mathrm{max}}=200$. We see that the shifts induced by uncertainty in the tracer spectra are typically small compared to the statistical precision afforded by SO. 
Figure~\ref{fig:bias_on_r_null_scenario} illustrates this further by comparing the distribution of shifts in $r$ to the SO statistical uncertainty for the case of $r=0$. We find that the additional uncertainties on estimating $r$ are $\sigma(r)\lesssim 2\times 10^{-4}$.

%//////////////////////////////////////////////////////////////////////////////////////////%
\section{Summary and Conclusion}
\label{sec:summary}
%//////////////////////////////////////////////////////////////////////////////////////////%

% summary
We have developed a delensing methodology and pipeline for SO and tested its performance in the presence of the realistic survey effects.
We showed that, even in the presence of survey boundaries, inhomogeneous instrumental and atmospheric noise, the delensing method we developed produces
a statistical error on the tensor-to-scalar ratio, $\sigma(r)$, which is close to the ideal case of an isotropic survey of the same duration.
We also discussed potential errors associated with uncertainties in the spectral modeling of external mass tracers, by extending the study of Ref.~\cite{Sherwin:2015}. We showed that when combining an internal lensing reconstruction with external tracers, the impact of these tracer uncertainties is nearly flat residuals in the modeled delensed $B$-mode power spectrum, and can be captured with a single nuisance parameter. Marginalizing over such a parameter, with a prior informed by plausible errors in the modeling of the spectra of the external tracers, leads to additional uncertainties on $r$ as $\sigma(r)\lesssim 2\times 10^{-4}$ and should have negligible effect on the $r$ constraint from SO. 

% future direction
We generated our simulation realizations from a map-based approach and did not include any instrumental systematic effects. Although Ref.~\cite{Nagata:2021} explored the response of the residual $B$-mode power spectrum to observational systematic effects in a simple experimental setup, the impact of the instrumental systematics is not yet quantified accurately in the case of SO as doing so would require more realistic simulations at the level of the time-ordered data. 
In our study, we did not consider the point-source masks in CMB maps which could lead to a large mean-field bias in the reconstructed lensing map and a large reconstruction noise around the masks if we use the isotropic filtering to CMB. These bias would be, however, significantly mitigated by applying the optimal filtering (see e.g. Refs.~\cite{Namikawa:2012:bhe,Namikawa:2013:bhepol,Lembo:2021kxc}). 
We used idealized, full-sky external mass-tracer simulations, which we projected onto the LAT region. In practice, reality may be more complicated. For example, residual foregrounds in maps of the CIB will vary across the sky and could lead to a bias in delensing \cite{ref:cib_delensing_biases}, as may the depth of galaxy surveys. A more quantitative study requires realistic simulations of each mass tracer, which will be addressed in future work.

This paper focuses on application of multi-tracer delensing for SO. This approach is, however, expected also to be important for LiteBIRD \cite{LiteBIRD} and CMB-S4 \cite{CMBS4:r-forecast} to enhance the sensitivity to IGWs.
Therefore, the delensing methodology we presented in this paper may be also useful for delensing in these future CMB experiments.

%//////////////////////////////////////////////////////////////////////////////////////////%
% BACK MATTER 
%//////////////////////////////////////////////////////////////////////////////////////////%

% Acknowledgments %
\begin{acknowledgments}
Some of the results in this paper have been derived using public software: healpy \cite{healpy}; HEALPix \cite{Gorski:2004by}; and CAMB \cite{Lewis:1999bs}. 
For numerical calculations, this paper used resources of the National Energy Research Scientific Computing Center (NERSC), a U.S.\ Department of Energy Office of Science User Facility operated under Contract No. DE-AC02-05CH11231.
TN acknowledges support from JSPS KAKENHI Grant No. JP20H05859 and World Premier International Research Center Initiative (WPI), MEXT, Japan.
ABL acknowledges support from an Isaac Newton studentship at the University of Cambridge and the UK Science and Technology Facilities Council (STFC).
BDS acknowledges support from an Isaac Newton Trust Early Career Grant and from the European Research Council (ERC) under the European Union’s Horizon 2020 research and innovation programme (Grant agreement No. 851274) and an STFC Ernest Rutherford Fellowship. 
AC acknowledges support from the STFC Grant No. ST/S000623/1). 
DA is supported by the Science and Technology Facilities Council through an Ernest Rutherford Fellowship, grant reference ST/P004474
CB acknowledges support from the RADIOFOREGROUNDS grant of the European Unions Horizon 2020 research and innovation programme (COMPET-05-2015, Grant agreement No. 687312) as well as by the INDARK INFN Initiative and the COSMOS and LiteBIRD networks of the Italian Space Agency (cosmosnet.it). 
EC acknowledges support from the STFC Ernest Rutherford Fellowship ST/M004856/2, STFC Consolidated Grant ST/S00033X/1 and from the European Research Council (ERC) under the European Union’s Horizon 2020 research and innovation programme (Grant agreement No. 849169). 
JC acknowledges support from a SNSF Eccellenza Professorial Fellowship (No. 186879).
YC acknowledges the support from the JSPS KAKENHI Grant No. 18K13558, 19H00674, 21K03585.
JC and MR were supported by the ERC Consolidator Grant {\it CMBSPEC} (No.~725456) and The Royal Society (URF\textbackslash R\textbackslash 191023). 
GC is supported by the European Research Council under the Marie Sklodowska Curie actions through the Individual European Fellowship No. 892174 PROTOCALC. 
GF acknowledges the support of the European Research Council under the Marie Sk\l{}odowska Curie actions through the Individual Global Fellowship No.~892401 PiCOGAMBAS. 
AL is supported by the STFC ST/T000473/1. 
MM is supported in part by the Government of Canada through the Department of Innovation, Science and Industry Canada and by the Province of Ontario through the Ministry of Colleges and Universities. 
PDM and GO acknowledge support from the Netherlands organisation for scientific research (NWO) VIDI grant (dossier 639.042.730)
JM is supported by the US Department of Energy Office of Science under grant no. DE-SC0010129.
NS acknowledges support from NSF Grant No. AST-1907657.
OT acknowledges support from JSPS KAKENHI JP17H06134. 
ZX is supported by the Gordon and Betty Moore Foundation through Grant No. GBMF5215 to the Massachusetts Institute of Technology.
\end{acknowledgments}

% Appendix %
\onecolumngrid
\appendix

\section{Analytic Power Spectrum Covariances} \label{appendix:covariances}

In this section, we provide analytic models for the covariances of the different combinations of spectra involved in the inference of Sec.~\ref{sec:likelihood}. These serve as a complement and cross-check of simulated covariances. 

\subsection{Delensed $B$-mode power spectrum covariance}

In order to calculate the power spectrum covariance of delensed $B$-modes, we employ the following covariance \cite{Namikawa:2015:cov} which is an extension of the lensing $B$-mode covariance by \cite{BenoitLevy:2012va}
\begin{equation}\label{eqn:full_covariance}
    \mathrm{Cov}\left({C}_l^{BB,\mathrm{del}},C_{l'}^{BB,\mathrm{del}}\right) = \frac{2}{2l+1}\delta_{ll'}\left(C_l^{BB,\mathrm{del}}\right)^2 + \mathrm{Cov}_{\mathrm{NG}}\left({C}_l^{BB,\mathrm{del}},{C}_{l'}^{BB,\mathrm{del}}\right) 
    \,,
\end{equation}
where
\begin{align}\label{eqn:toshiyas_expression}
    \mathrm{Cov}_{\mathrm{NG}}\left({C}_l^{BB,\mathrm{del}},C_{l'}^{BB,\mathrm{del}}\right) = \sum_L \frac{2}{2L+1}\bigg[\frac{\partial C_l^{BB,\mathrm{del}}}{\partial C_L^{EE}} \left(C_L^{EE}\right)^2 \frac{\partial C_{l'}^{BB,\mathrm{del}}}{\partial C_L^{EE}} 
    + \frac{\partial C_l^{BB,\mathrm{del}}}{\partial C_L^{\kappa\kappa}} \left(C_L^{\kappa\kappa}\right)^2 \frac{\partial C_{l'}^{BB,\mathrm{del}}}{\partial C_L^{\kappa\kappa}}\bigg]
    \,,
\end{align}
is the non-Gaussian part of the covariance. This expression assumes (i) that the $E$-modes are cosmic-variance limited, (ii) that the noise in the matter tracer is uncorrelated with the lensing convergence and the CMB, and (iii) that $C_l^{EE}, C_l^{\kappa\kappa}$ and the noise spectrum all have Gaussian covariance.

Now, under the assumption that the $E$-modes are limited by cosmic variance, the power spectrum of delensed $B$-modes is~\cite{Sherwin:2015}
\begin{equation}
    C_l^{BB,\mathrm{del}} =  \frac{1}{2l+1} \sum_{\l'\l''} \left(p^- F^{(2)}_{\l\l''\l'}\right)^2 C_{\l'}^{EE}C^{\kappa\kappa}_{\l''} \left(1-\rho^2_{\l''}\right)
    \,, 
\end{equation}
where $\rho_L$ is the cross-correlation coefficients of our co-added tracer with the true CMB lensing. When taking the derivatives of Eq.~\eqref{eqn:toshiyas_expression}, $\rho_L$ can be regarded as a constant, which means that
\begin{equation}
    \PD{C_l^{BB,\mathrm{del}}}{C_L^{\kappa\kappa}} = \left(1-\rho^2_L\right)\PD{\tCBB_l}{C_L^{\kappa\kappa}}
    \,, 
\end{equation}
and
\begin{equation}
    \PD{C_l^{BB,\mathrm{del}}}{C_L^{EE}} = \left.\PD{\tCBB_l}{C_L^{EE}}\right|_{C^{\kappa\kappa}=(1-\rho^2)C^{\kappa\kappa}}
    \,.
\end{equation}
In order to evaluate this last expression, we use an expression in the style of Eq.~(27) of Ref.~\cite{ref:Schmittfull_13}.

With these insights in hand, the covariance of Eq.~\eqref{eqn:full_covariance} can be evaluated by modifying existing codes such as \texttt{LensCov}~\cite{ref:peloton_17} which compute the power spectrum covariance of lensed CMB fields.

\subsection{Cross-spectral approach}

In this section, we calculate the covariance of all auto- and cross-spectra of the observed and template $B$-modes. The covariance is needed when writing down the cross-spectral approach detailed in Sec.~\ref{sec:likelihood}. Although our analysis uses the analytic covariance, we also cross-check the results using the analytic covariance described in this section. 

\subsubsection{Covariance of cross-spectrum}

In Sec.~\ref{ref:model_spectra}, we saw that, to leading order, the cross-spectrum between lensing and template $B$-modes can be modeled as
\begin{equation}
    C_l^{BB,{\rm cross}} = \frac{1}{2l+1} \sum_{\l'\l''} \left(p^- F^{(2)}_{\l\l''\l'}\right)^2 C_{\l'}^{EE}C^{\kappa\kappa}_{\l''} \rho^2_{\l''} 
    \,.
\end{equation}
Proceeding analogously to how the lensing $B$-mode power spectrum covariance is calculated, we approximate the non-Gaussian part of the covariance as
\begin{align}
    \mathrm{Cov}_{\mathrm{NG}}\left(C_l^{BB,{\rm cross}},C_{l'}^{BB,{\rm cross}}\right) \approx \sum_L \frac{1}{2L+1}\bigg[\PD{C_l^{BB,{\rm cross}}}{C_L^{EE}} 2\left(C_L^{EE}\right)^2 \PD{C_{l'}^{BB,{\rm cross}}}{C_L^{EE}} + \PD{\tCBB_l}{C_L^{\kappa\kappa}}\mathrm{Var}\left(C_L^{\kappa\kappa^{\mathrm{WF}}}\right) \PD{\tCBB_{l'}}{C_L^{\kappa\kappa}}\bigg]
    \,. 
\end{align}
where $\kappa^{\mathrm{WF}}$ is the Wiener-filtered tracer map. As determined in Ref.~\cite{Sherwin:2015}, for a single tracer, this takes the form $\kappa^{\mathrm{WF}}_l=(C_l^{\kappa I}/C_l^{II}) I_l$, where $I$ is the tracer itself, so, 
\begin{equation}\label{eqn:kappa_kappa_comb_def}
    C_l^{\kappa^{\mathrm{WF}}\kappa^{\mathrm{WF}}} = \left(\frac{C_l^{\kappa I}}{C_l^{II}}\right)^2 C_l^{II} =  \left(\frac{C_l^{\kappa I}}{C_l^{II}}\right) C_l^{\kappa I} = C_l^{\kappa\kappa^{\mathrm{WF}}}.
\end{equation}
We can now use the correlation between the tracer and lensing, defined as $\rho_l=C_l^{\kappa I}/\sqrt{C_l^{\kappa \kappa} C_l^{II}}$, to rewrite $C_l^{\kappa^{\mathrm{WF}}\kappa^{\mathrm{WF}}} = C_l^{\kappa \kappa^{\mathrm{WF}}} = \rho_l^2 C_l^{\kappa\kappa}$. Under the assumption of Gaussianity of the lens power spectrum, the variance we are after can be computed using the usual prescription for Gaussian covariance:
\begin{equation}
    \mathrm{Cov}_{ll'}^{G_a G_b, G_c G_d}=\frac{\delta_{ll'}}{2l+1}\left[C_l^{G_a G_c}C_l^{G_b G_d} + C_l^{G_a G_d}C_l^{G_b G_c}\right]
    \,,
\end{equation}
yielding
\begin{align}
    \mathrm{Var}\left(C_L^{\kappa\kappa^{\mathrm{WF}}}\right) &= \frac{1}{2l+1}\left[C_L^{\kappa \kappa}C_L^{\kappa^{\mathrm{WF}} \kappa^{\mathrm{WF}}} + C_L^{\kappa \kappa^{\mathrm{WF}}}C_L^{\kappa \kappa^{\mathrm{WF}}}\right]\\
    & =\frac{1}{2l+1}\left(\rho^2_L+\rho^4_L\right)(C_L^{\kappa \kappa})^2
    \,.
\end{align}
Finally,
\begin{align}
    \mathrm{Cov}_{\mathrm{NG}}\left(C_l^{BB,{\rm cross}},C_{l'}^{BB,{\rm cross}}\right) &\approx \sum_L \frac{1}{2L+1}\bigg[\left.\PD{\tCBB_l}{C_L^{EE}}\right|_{C^{\kappa\kappa}=\rho^2 C^{\kappa\kappa}} 2\left(C_L^{EE}\right)^2 \left.\PD{\tCBB_{l'}}{C_L^{EE}}\right|_{C^{\kappa\kappa}=\rho^2 C^{\kappa\kappa}} \nonumber \\
    &\hphantom{\approx \sum_L \frac{2}{2L+1}\bigg[}+ \PD{\tCBB_l}{C_L^{\kappa\kappa}} \left(\rho^2_L + \rho^4_L\right)\left(C_L^{\kappa\kappa}\right)^2 \PD{\tCBB_{l'}}{C_L^{\kappa\kappa}}\bigg]
    \,.
\end{align}
In addition to this, the full covariance receives a purely Gaussian contribution. Including it, we obtain
\begin{align}
    \mathrm{Cov}\left(C_l^{BB,{\rm cross}},C_{l'}^{BB,{\rm cross}}\right) &= \frac{1}{2l+1}\delta_{ll'}\left[\tCBB_lC_l^{BB,{\rm temp}} + \left(C_l^{BB,{\rm cross}}\right)^2\right] +\mathrm{Cov}_{\mathrm{NG}}\left(C_l^{BB,{\rm cross}},C_{l'}^{BB,{\rm cross}}\right)
    \,.
\end{align}

\subsubsection{Covariance of template auto-spectrum}

To leading order, the auto- and cross-spectra are equal, $C_l^{BB,{\rm temp}} = C_l^{BB,{\rm cross}}$. This time, the non-Gaussian part of the covariance can be approximated as
\begin{align}
    \mathrm{Cov}_{\mathrm{NG}}\left(C_l^{BB,{\rm temp}},C_{l'}^{BB,{\rm temp}}\right) &\approx \sum_L \frac{1}{2L+1}\bigg[\PD{C_l^{BB,{\rm temp}}}{C_L^{EE}} 2\left(C_L^{EE}\right)^2 \PD{C_{l'}^{BB,{\rm temp}}}{C_L^{EE}} 
    \nonumber \\
    &\hphantom{\approx \sum_L \frac{2}{2L+1}\bigg[}+ \PD{\tCBB_l}{C_L^{\kappa\kappa}} \mathrm{Var}\left(C_L^{\kappa^{\mathrm{WF}}\kappa^{\mathrm{WF}}}\right) \PD{\tCBB_{l'}}{C_L^{\kappa\kappa}}\bigg] 
    \\
    &\approx \sum_L \frac{2}{2L+1}\bigg[\PD{C_l^{BB,{\rm temp}}}{C_L^{EE}} \left(C_L^{EE}\right)^2 \PD{C_{l'}^{BB,{\rm temp}}}{C_L^{EE}} 
    \nonumber \\
    &\hphantom{\approx \sum_L \frac{2}{2L+1}\bigg[}+ \PD{\tCBB_l}{C_L^{\kappa\kappa}} \left(\rho^2_L C_L^{\kappa\kappa}\right)^2 \PD{\tCBB_{l'}}{C_L^{\kappa\kappa}}\bigg]
    \,,
\end{align}
and the full covariance is
\begin{align}
    \mathrm{Cov}\left(C_l^{BB,{\rm temp}},C_{l'}^{BB,{\rm temp}}\right) = \frac{2}{2l+1}\left( C_l^{BB,{\rm temp}} \right)^2 +\mathrm{Cov}_{\mathrm{NG}}\left({C}_l^{BB,{\rm temp}},{C}_{l'}^{BB,{\rm temp}}\right)
    \,.
\end{align}

\subsubsection{Cross-covariance of lensing and template auto-spectra}

The non-Gaussian part of the covariance can be approximated as
\begin{align}
    \mathrm{Cov}_{\mathrm{NG}}\left(\tCBB_l,C_{l'}^{BB,{\rm temp}}\right) &\approx \sum_L \frac{1}{2L+1}\bigg[\PD{\tCBB_l}{C_L^{EE}} 2\left(C_L^{EE}\right)^2 \left.\PD{\tCBB_{l'}}{C_L^{EE}}\right|_{C^{\kappa\kappa}=\rho^2 C^{\kappa\kappa}} 
    \nonumber \\
    &\hphantom{\approx \sum_L \frac{2}{2L+1}\bigg[}+\PD{\tCBB_l}{C_L^{\kappa\kappa}} \mathrm{Cov}\left(C_L^{\kappa\kappa},C_L^{\kappa^{\mathrm{WF}}\kappa^{\mathrm{WF}}}\right) \PD{\tCBB_{l'}}{C_L^{\kappa\kappa}}\bigg] \\
    &\approx \sum_L \frac{2}{2L+1}\bigg[\PD{\tCBB_l}{C_L^{EE}} \left(C_L^{EE}\right)^2 \left.\PD{\tCBB_{l'}}{C_L^{EE}}\right|_{C^{\kappa\kappa}=\rho^2 C^{\kappa\kappa}} 
    \nonumber \\
    &\hphantom{\approx \sum_L \frac{2}{2L+1}\bigg[}+\PD{\tCBB_l}{C_L^{\kappa\kappa}} \left(\rho_L^2 C_L^{\kappa\kappa}\right)^2 \PD{\tCBB_{l'}}{C_L^{\kappa\kappa}}\bigg]
    \,.
\end{align}
and the full covariance is
\begin{align}
    \mathrm{Cov}\left(\tCBB_l,C_{l'}^{BB,{\rm temp}}\right) &= \frac{2}{2l+1}\delta_{ll'} \left({C}_l^{BB,{\rm cross}}\right)^2 + \mathrm{Cov}_{\mathrm{NG}}\left(\tCBB_l,C_{l'}^{BB,{\rm temp}}\right)
    \,.
\end{align}

\subsubsection{Cross-covariance of lensing auto- and template cross-spectra}

The non-Gaussian part of the covariance can be approximated as:
\begin{align}
    \mathrm{Cov}_{\mathrm{NG}}\left(\tCBB_l,C_{l'}^{BB,{\rm cross}}\right) &\approx \mathrm{Cov}_{\mathrm{NG}}\left(\tCBB_l,C_{l'}^{BB,{\rm temp}}\right) 
    \,,
\end{align}
and the full covariance is
\begin{align}
    \mathrm{Cov}\left(\tCBB_l,C_{l'}^{BB,{\rm temp}}\right) &= \frac{2}{2l+1}\delta_{ll'} \tCBB_l C_l^{BB,{\rm cross}} + \mathrm{Cov}_{\mathrm{NG}}\left(\tCBB_l,C_{l'}^{BB,{\rm temp}}\right)
    \,.
\end{align}

\subsubsection{Cross-covariance of template auto- and cross-spectra}

The non-Gaussian part of the covariance can be approximated as
\begin{align}
    \mathrm{Cov}_{\mathrm{NG}}\left(C_l^{BB,{\rm temp}},C_{l'}^{BB,{\rm cross}}\right) &\approx\mathrm{Cov}_{\mathrm{NG}}\left(C_l^{BB,{\rm temp}},C_{l'}^{BB,{\rm temp}}\right)
    \,,
\end{align}
and the full covariance is
\begin{align}
    \mathrm{Cov}\left(C_l^{BB,{\rm temp}},C_{l'}^{BB,{\rm cross}}\right) &= \frac{2}{2l+1}\delta_{ll'} \tilde{C}_l^{BB,{\rm temp}}{C}_l^{BB,{\rm cross}} + \mathrm{Cov}_{\mathrm{NG}}\left(C_l^{BB,{\rm temp}},C_{l'}^{BB,{\rm temp}}\right) \\
    &= \mathrm{Cov}\left(C_l^{BB,{\rm temp}},C_{l'}^{BB,{\rm temp}}\right)
    \,.
\end{align}

\twocolumngrid

% References %
\bibliographystyle{apsrev}
\bibliography{cite}

\begin{thebibliography}{76}
\expandafter\ifx\csname natexlab\endcsname\relax\def\natexlab#1{#1}\fi
\expandafter\ifx\csname bibnamefont\endcsname\relax
  \def\bibnamefont#1{#1}\fi
\expandafter\ifx\csname bibfnamefont\endcsname\relax
  \def\bibfnamefont#1{#1}\fi
\expandafter\ifx\csname citenamefont\endcsname\relax
  \def\citenamefont#1{#1}\fi
\expandafter\ifx\csname url\endcsname\relax
  \def\url#1{\texttt{#1}}\fi
\expandafter\ifx\csname urlprefix\endcsname\relax\def\urlprefix{URL }\fi
\providecommand{\bibinfo}[2]{#2}
\providecommand{\eprint}[2][]{\url{#2}}

\bibitem[{\citenamefont{Polnarev}(1985)}]{Polnarev:1985}
\bibinfo{author}{\bibfnamefont{A.~G.} \bibnamefont{Polnarev}},
  \bibinfo{journal}{Sov. Astron.} \textbf{\bibinfo{volume}{29}},
  \bibinfo{pages}{607} (\bibinfo{year}{1985}).

\bibitem[{\citenamefont{Kamionkowski et~al.}(1997)\citenamefont{Kamionkowski,
  Kosowsky, and Stebbins}}]{Kamionkowski:1996:GW}
\bibinfo{author}{\bibfnamefont{M.}~\bibnamefont{Kamionkowski}},
  \bibinfo{author}{\bibfnamefont{A.}~\bibnamefont{Kosowsky}}, \bibnamefont{and}
  \bibinfo{author}{\bibfnamefont{A.}~\bibnamefont{Stebbins}},
  \bibinfo{journal}{\prl} \textbf{\bibinfo{volume}{78}}, \bibinfo{pages}{2058}
  (\bibinfo{year}{1997}), \eprint{astro-ph/9609132}.

\bibitem[{\citenamefont{Seljak and Zaldarriaga}(1997)}]{Seljak:1996:GW}
\bibinfo{author}{\bibfnamefont{U.}~\bibnamefont{Seljak}} \bibnamefont{and}
  \bibinfo{author}{\bibfnamefont{M.}~\bibnamefont{Zaldarriaga}},
  \bibinfo{journal}{\prl} \textbf{\bibinfo{volume}{78}}, \bibinfo{pages}{2054}
  (\bibinfo{year}{1997}), \eprint{astro-ph/9609169}.

\bibitem[{\citenamefont{{BICEP/Keck Collaboration: P. A. R. Ade}
  et~al.}(2021)\citenamefont{{BICEP/Keck Collaboration: P. A. R. Ade}, {Ahmed},
  {Amiri}, {Barkats}, {Basu Thakur}, {Beck}, {Bischoff}, {Bock}, {Boenish},
  {Bullock} et~al.}}]{BK13}
\bibinfo{author}{\bibnamefont{{BICEP/Keck Collaboration: P. A. R. Ade}}},
  \bibinfo{author}{\bibfnamefont{Z.}~\bibnamefont{{Ahmed}}},
  \bibinfo{author}{\bibfnamefont{M.}~\bibnamefont{{Amiri}}},
  \bibinfo{author}{\bibfnamefont{D.}~\bibnamefont{{Barkats}}},
  \bibinfo{author}{\bibfnamefont{R.}~\bibnamefont{{Basu Thakur}}},
  \bibinfo{author}{\bibfnamefont{D.}~\bibnamefont{{Beck}}},
  \bibinfo{author}{\bibfnamefont{C.}~\bibnamefont{{Bischoff}}},
  \bibinfo{author}{\bibfnamefont{J.~J.} \bibnamefont{{Bock}}},
  \bibinfo{author}{\bibfnamefont{H.}~\bibnamefont{{Boenish}}},
  \bibinfo{author}{\bibfnamefont{E.}~\bibnamefont{{Bullock}}},
  \bibnamefont{et~al.}, \bibinfo{journal}{\prl} \textbf{\bibinfo{volume}{127}},
  \bibinfo{pages}{151301} (\bibinfo{year}{2021}), \eprint{2110.00483}.

\bibitem[{\citenamefont{{BICEP/Keck Collaboration: P.~A.~R. Ade}
  et~al.}(2021)\citenamefont{{BICEP/Keck Collaboration: P.~A.~R. Ade}, {Ahmed},
  {Amiri}, {Barkats}, {Basu Thakur}, {Beck}, {Bischoff}, {Bock}, {Boenish},
  {Bullock} et~al.}}]{BK15:data}
\bibinfo{author}{\bibnamefont{{BICEP/Keck Collaboration: P.~A.~R. Ade}}},
  \bibinfo{author}{\bibfnamefont{Z.}~\bibnamefont{{Ahmed}}},
  \bibinfo{author}{\bibfnamefont{M.}~\bibnamefont{{Amiri}}},
  \bibinfo{author}{\bibfnamefont{D.}~\bibnamefont{{Barkats}}},
  \bibinfo{author}{\bibfnamefont{R.}~\bibnamefont{{Basu Thakur}}},
  \bibinfo{author}{\bibfnamefont{D.}~\bibnamefont{{Beck}}},
  \bibinfo{author}{\bibfnamefont{C.}~\bibnamefont{{Bischoff}}},
  \bibinfo{author}{\bibfnamefont{J.~J.} \bibnamefont{{Bock}}},
  \bibinfo{author}{\bibfnamefont{H.}~\bibnamefont{{Boenish}}},
  \bibinfo{author}{\bibfnamefont{E.}~\bibnamefont{{Bullock}}},
  \bibnamefont{et~al.} (\bibinfo{year}{2021}), \eprint{2110.00482}.

\bibitem[{\citenamefont{{Hui} et~al.}(2018)\citenamefont{{Hui}, {Ade}, {Ahmed},
  {Aikin}, {Alexander}, {Barkats}, {Benton}, {Bischoff}, {Bock}, {Bowens-Rubin}
  et~al.}}]{BICEPArray}
\bibinfo{author}{\bibfnamefont{H.}~\bibnamefont{{Hui}}},
  \bibinfo{author}{\bibfnamefont{P.~A.~R.} \bibnamefont{{Ade}}},
  \bibinfo{author}{\bibfnamefont{Z.}~\bibnamefont{{Ahmed}}},
  \bibinfo{author}{\bibfnamefont{R.~W.} \bibnamefont{{Aikin}}},
  \bibinfo{author}{\bibfnamefont{K.~D.} \bibnamefont{{Alexander}}},
  \bibinfo{author}{\bibfnamefont{D.}~\bibnamefont{{Barkats}}},
  \bibinfo{author}{\bibfnamefont{S.~J.} \bibnamefont{{Benton}}},
  \bibinfo{author}{\bibfnamefont{C.~A.} \bibnamefont{{Bischoff}}},
  \bibinfo{author}{\bibfnamefont{J.~J.} \bibnamefont{{Bock}}},
  \bibinfo{author}{\bibfnamefont{R.}~\bibnamefont{{Bowens-Rubin}}},
  \bibnamefont{et~al.}, \bibinfo{journal}{Proc. SPIE Int. Soc. Opt. Eng.}
  \textbf{\bibinfo{volume}{10708}}, \bibinfo{pages}{1070807}
  (\bibinfo{year}{2018}), \eprint{1808.00568}.

\bibitem[{\citenamefont{{Suzuki} et~al.}(2016)\citenamefont{{Suzuki}, {Ade},
  {Akiba}, {Aleman}, {Arnold}, {Baccigalupi}, {Barch}, {Barron}, {Bender},
  {Boettger} et~al.}}]{SimonsArray}
\bibinfo{author}{\bibfnamefont{A.}~\bibnamefont{{Suzuki}}},
  \bibinfo{author}{\bibfnamefont{P.}~\bibnamefont{{Ade}}},
  \bibinfo{author}{\bibfnamefont{Y.}~\bibnamefont{{Akiba}}},
  \bibinfo{author}{\bibfnamefont{C.}~\bibnamefont{{Aleman}}},
  \bibinfo{author}{\bibfnamefont{K.}~\bibnamefont{{Arnold}}},
  \bibinfo{author}{\bibfnamefont{C.}~\bibnamefont{{Baccigalupi}}},
  \bibinfo{author}{\bibfnamefont{B.}~\bibnamefont{{Barch}}},
  \bibinfo{author}{\bibfnamefont{D.}~\bibnamefont{{Barron}}},
  \bibinfo{author}{\bibfnamefont{A.}~\bibnamefont{{Bender}}},
  \bibinfo{author}{\bibfnamefont{D.}~\bibnamefont{{Boettger}}},
  \bibnamefont{et~al.}, \bibinfo{journal}{\jltp}
  \textbf{\bibinfo{volume}{184}}, \bibinfo{pages}{805} (\bibinfo{year}{2016}),
  \eprint{1512.07299}.

\bibitem[{\citenamefont{{Lee} et~al.}(2019)\citenamefont{{Lee}, {Abitbol},
  {Adachi}, {Ade}, {Aguirre}, {Ahmed}, {Aiola}, {Ali}, {Alonso}, {Alvarez}
  et~al.}}]{SimonsObservatory}
\bibinfo{author}{\bibfnamefont{A.}~\bibnamefont{{Lee}}},
  \bibinfo{author}{\bibfnamefont{M.~H.} \bibnamefont{{Abitbol}}},
  \bibinfo{author}{\bibfnamefont{S.}~\bibnamefont{{Adachi}}},
  \bibinfo{author}{\bibfnamefont{P.}~\bibnamefont{{Ade}}},
  \bibinfo{author}{\bibfnamefont{J.}~\bibnamefont{{Aguirre}}},
  \bibinfo{author}{\bibfnamefont{Z.}~\bibnamefont{{Ahmed}}},
  \bibinfo{author}{\bibfnamefont{S.}~\bibnamefont{{Aiola}}},
  \bibinfo{author}{\bibfnamefont{A.}~\bibnamefont{{Ali}}},
  \bibinfo{author}{\bibfnamefont{D.}~\bibnamefont{{Alonso}}},
  \bibinfo{author}{\bibfnamefont{M.~A.} \bibnamefont{{Alvarez}}},
  \bibnamefont{et~al.}, \bibinfo{journal}{Bull. Am. Astron. Soc.}
  \textbf{\bibinfo{volume}{51}}, \bibinfo{pages}{147} (\bibinfo{year}{2019}),
  \eprint{1907.08284}.

\bibitem[{\citenamefont{{Hazumi} et~al.}(2019)\citenamefont{{Hazumi}, {Ade},
  {Akiba}, {Alonso}, {Arnold}, {Aumont}, {Baccigalupi}, {Barron}, {Basak},
  {Beckman} et~al.}}]{LiteBIRD}
\bibinfo{author}{\bibfnamefont{M.}~\bibnamefont{{Hazumi}}},
  \bibinfo{author}{\bibfnamefont{P.~A.~R.} \bibnamefont{{Ade}}},
  \bibinfo{author}{\bibfnamefont{Y.}~\bibnamefont{{Akiba}}},
  \bibinfo{author}{\bibfnamefont{D.}~\bibnamefont{{Alonso}}},
  \bibinfo{author}{\bibfnamefont{K.}~\bibnamefont{{Arnold}}},
  \bibinfo{author}{\bibfnamefont{J.}~\bibnamefont{{Aumont}}},
  \bibinfo{author}{\bibfnamefont{C.}~\bibnamefont{{Baccigalupi}}},
  \bibinfo{author}{\bibfnamefont{D.}~\bibnamefont{{Barron}}},
  \bibinfo{author}{\bibfnamefont{S.}~\bibnamefont{{Basak}}},
  \bibinfo{author}{\bibfnamefont{S.}~\bibnamefont{{Beckman}}},
  \bibnamefont{et~al.}, \bibinfo{journal}{J. Low. Temp. Phys.}
  \textbf{\bibinfo{volume}{194}}, \bibinfo{pages}{443} (\bibinfo{year}{2019}).

\bibitem[{\citenamefont{{CMB-S4 Collaboration: K. Abazajian}
  et~al.}(2019)\citenamefont{{CMB-S4 Collaboration: K. Abazajian}, {Addison},
  {Adshead}, {Ahmed}, {Allen}, {Alonso}, {Alvarez}, {Anderson}, {Arnold},
  {Baccigalupi} et~al.}}]{CMBS4}
\bibinfo{author}{\bibnamefont{{CMB-S4 Collaboration: K. Abazajian}}},
  \bibinfo{author}{\bibfnamefont{G.}~\bibnamefont{{Addison}}},
  \bibinfo{author}{\bibfnamefont{P.}~\bibnamefont{{Adshead}}},
  \bibinfo{author}{\bibfnamefont{Z.}~\bibnamefont{{Ahmed}}},
  \bibinfo{author}{\bibfnamefont{S.~W.} \bibnamefont{{Allen}}},
  \bibinfo{author}{\bibfnamefont{D.}~\bibnamefont{{Alonso}}},
  \bibinfo{author}{\bibfnamefont{M.}~\bibnamefont{{Alvarez}}},
  \bibinfo{author}{\bibfnamefont{A.}~\bibnamefont{{Anderson}}},
  \bibinfo{author}{\bibfnamefont{K.~S.} \bibnamefont{{Arnold}}},
  \bibinfo{author}{\bibfnamefont{C.}~\bibnamefont{{Baccigalupi}}},
  \bibnamefont{et~al.} (\bibinfo{year}{2019}), \eprint{1907.04473}.

\bibitem[{\citenamefont{{CMB-S4 Collaboration: K. Abazajian}
  et~al.}(2022)\citenamefont{{CMB-S4 Collaboration: K. Abazajian}, {Addison},
  {Adshead}, {Ahmed}, {Akerib}, {Ali}, {Allen}, {Alonso}, {Alvarez}, {Amin}
  et~al.}}]{CMBS4:r-forecast}
\bibinfo{author}{\bibnamefont{{CMB-S4 Collaboration: K. Abazajian}}},
  \bibinfo{author}{\bibfnamefont{G.~E.} \bibnamefont{{Addison}}},
  \bibinfo{author}{\bibfnamefont{P.}~\bibnamefont{{Adshead}}},
  \bibinfo{author}{\bibfnamefont{Z.}~\bibnamefont{{Ahmed}}},
  \bibinfo{author}{\bibfnamefont{D.}~\bibnamefont{{Akerib}}},
  \bibinfo{author}{\bibfnamefont{A.}~\bibnamefont{{Ali}}},
  \bibinfo{author}{\bibfnamefont{S.~W.} \bibnamefont{{Allen}}},
  \bibinfo{author}{\bibfnamefont{D.}~\bibnamefont{{Alonso}}},
  \bibinfo{author}{\bibfnamefont{M.}~\bibnamefont{{Alvarez}}},
  \bibinfo{author}{\bibfnamefont{M.~A.} \bibnamefont{{Amin}}},
  \bibnamefont{et~al.}, \bibinfo{journal}{\apj} \textbf{\bibinfo{volume}{926}},
  \bibinfo{pages}{54} (\bibinfo{year}{2022}), \eprint{2008.12619}.

\bibitem[{\citenamefont{{BICEP2 Collaboration: P.~A.~R. Ade}
  et~al.}(2014)\citenamefont{{BICEP2 Collaboration: P.~A.~R. Ade}, {Aikin},
  {Barkats}, {Benton}, {Bischoff}, {Bock}, {Brevik}, {Buder}, {Bullock},
  {Dowell} et~al.}}]{B2I}
\bibinfo{author}{\bibnamefont{{BICEP2 Collaboration: P.~A.~R. Ade}}},
  \bibinfo{author}{\bibfnamefont{R.~W.} \bibnamefont{{Aikin}}},
  \bibinfo{author}{\bibfnamefont{D.}~\bibnamefont{{Barkats}}},
  \bibinfo{author}{\bibfnamefont{S.~J.} \bibnamefont{{Benton}}},
  \bibinfo{author}{\bibfnamefont{C.~A.} \bibnamefont{{Bischoff}}},
  \bibinfo{author}{\bibfnamefont{J.~J.} \bibnamefont{{Bock}}},
  \bibinfo{author}{\bibfnamefont{J.~A.} \bibnamefont{{Brevik}}},
  \bibinfo{author}{\bibfnamefont{I.}~\bibnamefont{{Buder}}},
  \bibinfo{author}{\bibfnamefont{E.}~\bibnamefont{{Bullock}}},
  \bibinfo{author}{\bibfnamefont{C.~D.} \bibnamefont{{Dowell}}},
  \bibnamefont{et~al.}, \bibinfo{journal}{\prl} \textbf{\bibinfo{volume}{112}},
  \bibinfo{pages}{241101} (\bibinfo{year}{2014}), \eprint{1403.3985}.

\bibitem[{\citenamefont{Zaldarriaga and Seljak}(1998)}]{Zaldarriaga:1998:LensB}
\bibinfo{author}{\bibfnamefont{M.}~\bibnamefont{Zaldarriaga}} \bibnamefont{and}
  \bibinfo{author}{\bibfnamefont{U.}~\bibnamefont{Seljak}},
  \bibinfo{journal}{\prd} \textbf{\bibinfo{volume}{58}},
  \bibinfo{pages}{023003} (\bibinfo{year}{1998}), \eprint{astro-ph/9803150}.

\bibitem[{\citenamefont{Kesden et~al.}(2002)\citenamefont{Kesden, Cooray, and
  Kamionkowski}}]{Kesden:2002ku}
\bibinfo{author}{\bibfnamefont{M.}~\bibnamefont{Kesden}},
  \bibinfo{author}{\bibfnamefont{A.}~\bibnamefont{Cooray}}, \bibnamefont{and}
  \bibinfo{author}{\bibfnamefont{M.}~\bibnamefont{Kamionkowski}},
  \bibinfo{journal}{\prl} \textbf{\bibinfo{volume}{89}},
  \bibinfo{pages}{011304} (\bibinfo{year}{2002}), \eprint{astro-ph/0202434}.

\bibitem[{\citenamefont{Seljak and Hirata}(2004)}]{Seljak:2003pn}
\bibinfo{author}{\bibfnamefont{U.}~\bibnamefont{Seljak}} \bibnamefont{and}
  \bibinfo{author}{\bibfnamefont{C.~M.} \bibnamefont{Hirata}},
  \bibinfo{journal}{\prd} \textbf{\bibinfo{volume}{69}},
  \bibinfo{pages}{043005} (\bibinfo{year}{2004}), \eprint{astro-ph/0310163}.

\bibitem[{\citenamefont{Teng et~al.}(2011)\citenamefont{Teng, Kuo, and
  Wu}}]{Teng:2011xc}
\bibinfo{author}{\bibfnamefont{W.-H.} \bibnamefont{Teng}},
  \bibinfo{author}{\bibfnamefont{C.-L.} \bibnamefont{Kuo}}, \bibnamefont{and}
  \bibinfo{author}{\bibfnamefont{J.-H.~P.} \bibnamefont{Wu}}
  (\bibinfo{year}{2011}), \eprint{1102.5729}.

\bibitem[{\citenamefont{Namikawa and Nagata}(2014)}]{Namikawa:2014:patchwork}
\bibinfo{author}{\bibfnamefont{T.}~\bibnamefont{Namikawa}} \bibnamefont{and}
  \bibinfo{author}{\bibfnamefont{R.}~\bibnamefont{Nagata}},
  \bibinfo{journal}{\jcap} \textbf{\bibinfo{volume}{09}}, \bibinfo{pages}{009}
  (\bibinfo{year}{2014}), \eprint{1405.6568}.

\bibitem[{\citenamefont{Sehgal et~al.}(2017)\citenamefont{Sehgal,
  Madhavacheril, Sherwin, and {van Engelen}}}]{Sehgal:2016}
\bibinfo{author}{\bibfnamefont{N.}~\bibnamefont{Sehgal}},
  \bibinfo{author}{\bibfnamefont{M.~S.} \bibnamefont{Madhavacheril}},
  \bibinfo{author}{\bibfnamefont{B.}~\bibnamefont{Sherwin}}, \bibnamefont{and}
  \bibinfo{author}{\bibfnamefont{A.}~\bibnamefont{{van Engelen}}},
  \bibinfo{journal}{\prd} \textbf{\bibinfo{volume}{95}}, \bibinfo{eid}{103512}
  (\bibinfo{year}{2017}), \eprint{1612.03898}.

\bibitem[{\citenamefont{Namikawa}(2017)}]{Namikawa:2017:delens}
\bibinfo{author}{\bibfnamefont{T.}~\bibnamefont{Namikawa}},
  \bibinfo{journal}{\prd} \textbf{\bibinfo{volume}{95}},
  \bibinfo{pages}{103514} (\bibinfo{year}{2017}), \eprint{1703.00169}.

\bibitem[{\citenamefont{Carron et~al.}(2017)\citenamefont{Carron, Lewis, and
  Challinor}}]{Carron:2017}
\bibinfo{author}{\bibfnamefont{J.}~\bibnamefont{Carron}},
  \bibinfo{author}{\bibfnamefont{A.}~\bibnamefont{Lewis}}, \bibnamefont{and}
  \bibinfo{author}{\bibfnamefont{A.}~\bibnamefont{Challinor}},
  \bibinfo{journal}{\jcap} \textbf{\bibinfo{volume}{05}}, \bibinfo{pages}{035}
  (\bibinfo{year}{2017}), \eprint{1701.01712}.

\bibitem[{\citenamefont{{{POLARBEAR Collaboration: S. Adachi}}
  et~al.}(2020)\citenamefont{{{POLARBEAR Collaboration: S. Adachi}}, {Aguilar
  Fa{\'u}ndez}, {Akiba}, {Ali}, {Arnold}, {Baccigalupi}, {Barron}, {Beck},
  {Bianchini}, {Borrill} et~al.}}]{PB:2019:delens}
\bibinfo{author}{\bibnamefont{{{POLARBEAR Collaboration: S. Adachi}}}},
  \bibinfo{author}{\bibfnamefont{M.~A.~O.} \bibnamefont{{Aguilar
  Fa{\'u}ndez}}}, \bibinfo{author}{\bibfnamefont{Y.}~\bibnamefont{{Akiba}}},
  \bibinfo{author}{\bibfnamefont{A.}~\bibnamefont{{Ali}}},
  \bibinfo{author}{\bibfnamefont{K.}~\bibnamefont{{Arnold}}},
  \bibinfo{author}{\bibfnamefont{C.}~\bibnamefont{{Baccigalupi}}},
  \bibinfo{author}{\bibfnamefont{D.}~\bibnamefont{{Barron}}},
  \bibinfo{author}{\bibfnamefont{D.}~\bibnamefont{{Beck}}},
  \bibinfo{author}{\bibfnamefont{F.}~\bibnamefont{{Bianchini}}},
  \bibinfo{author}{\bibfnamefont{J.}~\bibnamefont{{Borrill}}},
  \bibnamefont{et~al.}, \bibinfo{journal}{\prl} \textbf{\bibinfo{volume}{124}},
  \bibinfo{pages}{131301} (\bibinfo{year}{2020}), \eprint{1909.13832}.

\bibitem[{\citenamefont{{Baleato Lizancos}
  et~al.}(2021{\natexlab{a}})\citenamefont{{Baleato Lizancos}, {Challinor}, and
  {Carron}}}]{ref:baleato_20}
\bibinfo{author}{\bibfnamefont{A.}~\bibnamefont{{Baleato Lizancos}}},
  \bibinfo{author}{\bibfnamefont{A.}~\bibnamefont{{Challinor}}},
  \bibnamefont{and} \bibinfo{author}{\bibfnamefont{J.}~\bibnamefont{{Carron}}},
  \bibinfo{journal}{\jcap} \textbf{\bibinfo{volume}{2021}}, \bibinfo{eid}{016}
  (\bibinfo{year}{2021}{\natexlab{a}}), \eprint{2007.01622}.

\bibitem[{\citenamefont{Smith et~al.}(2012)\citenamefont{Smith, Hanson,
  LoVerde, Hirata, and Zahn}}]{Smith:2010gu}
\bibinfo{author}{\bibfnamefont{K.~M.} \bibnamefont{Smith}},
  \bibinfo{author}{\bibfnamefont{D.}~\bibnamefont{Hanson}},
  \bibinfo{author}{\bibfnamefont{M.}~\bibnamefont{LoVerde}},
  \bibinfo{author}{\bibfnamefont{C.~M.} \bibnamefont{Hirata}},
  \bibnamefont{and} \bibinfo{author}{\bibfnamefont{O.}~\bibnamefont{Zahn}},
  \bibinfo{journal}{\jcap} \textbf{\bibinfo{volume}{06}}, \bibinfo{pages}{014}
  (\bibinfo{year}{2012}), \eprint{1010.0048}.

\bibitem[{\citenamefont{Carron}(2019)}]{Carron:2018:GW}
\bibinfo{author}{\bibfnamefont{J.}~\bibnamefont{Carron}},
  \bibinfo{journal}{\prd} \textbf{\bibinfo{volume}{99}},
  \bibinfo{pages}{043518} (\bibinfo{year}{2019}), \eprint{1808.10349}.

\bibitem[{\citenamefont{Sherwin and Schmittfull}(2015)}]{Sherwin:2015}
\bibinfo{author}{\bibfnamefont{B.~D.} \bibnamefont{Sherwin}} \bibnamefont{and}
  \bibinfo{author}{\bibfnamefont{M.}~\bibnamefont{Schmittfull}},
  \bibinfo{journal}{\prd} \textbf{\bibinfo{volume}{92}},
  \bibinfo{pages}{043005} (\bibinfo{year}{2015}), \eprint{1502.05356}.

\bibitem[{\citenamefont{Simard et~al.}(2015)\citenamefont{Simard, Hanson, and
  Holder}}]{Simard:2015}
\bibinfo{author}{\bibfnamefont{G.}~\bibnamefont{Simard}},
  \bibinfo{author}{\bibfnamefont{D.}~\bibnamefont{Hanson}}, \bibnamefont{and}
  \bibinfo{author}{\bibfnamefont{G.}~\bibnamefont{Holder}},
  \bibinfo{journal}{\apj} \textbf{\bibinfo{volume}{807}}, \bibinfo{pages}{166}
  (\bibinfo{year}{2015}), \eprint{1410.0691}.

\bibitem[{\citenamefont{Namikawa et~al.}(2016)\citenamefont{Namikawa, Yamauchi,
  Sherwin, and Nagata}}]{Namikawa:2015:delens}
\bibinfo{author}{\bibfnamefont{T.}~\bibnamefont{Namikawa}},
  \bibinfo{author}{\bibfnamefont{D.}~\bibnamefont{Yamauchi}},
  \bibinfo{author}{\bibfnamefont{D.}~\bibnamefont{Sherwin}}, \bibnamefont{and}
  \bibinfo{author}{\bibfnamefont{R.}~\bibnamefont{Nagata}},
  \bibinfo{journal}{\prd} \textbf{\bibinfo{volume}{93}},
  \bibinfo{pages}{043527} (\bibinfo{year}{2016}), \eprint{1511.04653}.

\bibitem[{\citenamefont{Manzotti}(2018)}]{Manzotti:2018}
\bibinfo{author}{\bibfnamefont{A.}~\bibnamefont{Manzotti}},
  \bibinfo{journal}{\prd} \textbf{\bibinfo{volume}{97}},
  \bibinfo{pages}{043527} (\bibinfo{year}{2018}), \eprint{1710.11038}.

\bibitem[{\citenamefont{Marian and Bernstein}(2007)}]{Marian:2007}
\bibinfo{author}{\bibfnamefont{L.}~\bibnamefont{Marian}} \bibnamefont{and}
  \bibinfo{author}{\bibfnamefont{G.~M.} \bibnamefont{Bernstein}},
  \bibinfo{journal}{\prd} \textbf{\bibinfo{volume}{76}},
  \bibinfo{pages}{123009} (\bibinfo{year}{2007}), \eprint{0710.2538}.

\bibitem[{\citenamefont{Sigurdson and Cooray}(2005)}]{Sigurdson:2005cp}
\bibinfo{author}{\bibfnamefont{K.}~\bibnamefont{Sigurdson}} \bibnamefont{and}
  \bibinfo{author}{\bibfnamefont{A.}~\bibnamefont{Cooray}},
  \bibinfo{journal}{\prl} \textbf{\bibinfo{volume}{95}},
  \bibinfo{pages}{211303} (\bibinfo{year}{2005}), \eprint{astro-ph/0502549}.

\bibitem[{\citenamefont{Karkare}(2019)}]{Karkare:2019:delens}
\bibinfo{author}{\bibfnamefont{K.~S.} \bibnamefont{Karkare}},
  \bibinfo{journal}{Phys. Rev. D} \textbf{\bibinfo{volume}{100}},
  \bibinfo{pages}{043529} (\bibinfo{year}{2019}), \eprint{1908.08128}.

\bibitem[{\citenamefont{Larsen et~al.}(2016)\citenamefont{Larsen, Challinor,
  Sherwin, and Mak}}]{Larsen:2016}
\bibinfo{author}{\bibfnamefont{P.}~\bibnamefont{Larsen}},
  \bibinfo{author}{\bibfnamefont{A.}~\bibnamefont{Challinor}},
  \bibinfo{author}{\bibfnamefont{B.~D.} \bibnamefont{Sherwin}},
  \bibnamefont{and} \bibinfo{author}{\bibfnamefont{D.}~\bibnamefont{Mak}},
  \bibinfo{journal}{\prl} \textbf{\bibinfo{volume}{117}},
  \bibinfo{pages}{151102} (\bibinfo{year}{2016}), \eprint{1607.05733}.

\bibitem[{\citenamefont{{A. Manzotti et al. (SPTpol
  Collaboration)}}(2017)}]{SPTpol:delens}
\bibinfo{author}{\bibnamefont{{A. Manzotti et al. (SPTpol Collaboration)}}},
  \bibinfo{journal}{\apj} \textbf{\bibinfo{volume}{846}}, \bibinfo{pages}{45}
  (\bibinfo{year}{2017}), \eprint{1701.04396}.

\bibitem[{\citenamefont{{Han} et~al.}(2021)\citenamefont{{Han}, {Sehgal},
  {MacInnis}, {van Engelen}, {Sherwin}, {Madhavacheril}, {Aiola}, {Battaglia},
  {Beall}, {Becker} et~al.}}]{DW:2020:delens}
\bibinfo{author}{\bibfnamefont{D.}~\bibnamefont{{Han}}},
  \bibinfo{author}{\bibfnamefont{N.}~\bibnamefont{{Sehgal}}},
  \bibinfo{author}{\bibfnamefont{A.}~\bibnamefont{{MacInnis}}},
  \bibinfo{author}{\bibfnamefont{A.}~\bibnamefont{{van Engelen}}},
  \bibinfo{author}{\bibfnamefont{B.~D.} \bibnamefont{{Sherwin}}},
  \bibinfo{author}{\bibfnamefont{M.~S.} \bibnamefont{{Madhavacheril}}},
  \bibinfo{author}{\bibfnamefont{S.}~\bibnamefont{{Aiola}}},
  \bibinfo{author}{\bibfnamefont{N.}~\bibnamefont{{Battaglia}}},
  \bibinfo{author}{\bibfnamefont{J.~A.} \bibnamefont{{Beall}}},
  \bibinfo{author}{\bibfnamefont{D.~T.} \bibnamefont{{Becker}}},
  \bibnamefont{et~al.}, \bibinfo{journal}{\jcap}
  \textbf{\bibinfo{volume}{2021}}, \bibinfo{eid}{031} (\bibinfo{year}{2021}),
  \eprint{2007.14405}.

\bibitem[{\citenamefont{{BICEP/Keck and SPTpol Collaborations: P.~A.~R. Ade}
  et~al.}(2021)\citenamefont{{BICEP/Keck and SPTpol Collaborations: P.~A.~R.
  Ade}, {Ahmed}, {Amiri}, {Anderson}, {Austermann}, {Avva}, {Barkats},
  {Thakur}, {Beall}, {Bender} et~al.}}]{ref:bkspt_cib_delensing}
\bibinfo{author}{\bibnamefont{{BICEP/Keck and SPTpol Collaborations: P.~A.~R.
  Ade}}}, \bibinfo{author}{\bibfnamefont{Z.}~\bibnamefont{{Ahmed}}},
  \bibinfo{author}{\bibfnamefont{M.}~\bibnamefont{{Amiri}}},
  \bibinfo{author}{\bibfnamefont{A.~J.} \bibnamefont{{Anderson}}},
  \bibinfo{author}{\bibfnamefont{J.~E.} \bibnamefont{{Austermann}}},
  \bibinfo{author}{\bibfnamefont{J.~S.} \bibnamefont{{Avva}}},
  \bibinfo{author}{\bibfnamefont{D.}~\bibnamefont{{Barkats}}},
  \bibinfo{author}{\bibfnamefont{R.~B.} \bibnamefont{{Thakur}}},
  \bibinfo{author}{\bibfnamefont{J.~A.} \bibnamefont{{Beall}}},
  \bibinfo{author}{\bibfnamefont{A.~N.} \bibnamefont{{Bender}}},
  \bibnamefont{et~al.}, \bibinfo{journal}{\prd} \textbf{\bibinfo{volume}{103}},
  \bibinfo{eid}{022004} (\bibinfo{year}{2021}), \eprint{2011.08163}.

\bibitem[{\citenamefont{{The Simons Observatory
  Collaboration}}(2019)}]{SO:2018:forecast}
\bibinfo{author}{\bibnamefont{{The Simons Observatory Collaboration}}},
  \bibinfo{journal}{\jcap} \textbf{\bibinfo{volume}{02}}, \bibinfo{pages}{056}
  (\bibinfo{year}{2019}), \eprint{1808.07445}.

\bibitem[{\citenamefont{Lewis and Challinor}(2006)}]{Lewis:2006fu}
\bibinfo{author}{\bibfnamefont{A.}~\bibnamefont{Lewis}} \bibnamefont{and}
  \bibinfo{author}{\bibfnamefont{A.}~\bibnamefont{Challinor}},
  \bibinfo{journal}{Phys. Rep.} \textbf{\bibinfo{volume}{429}},
  \bibinfo{pages}{1} (\bibinfo{year}{2006}), \eprint{astro-ph/0601594}.

\bibitem[{\citenamefont{Challinor and Chon}(2002)}]{Challinor:2002cd}
\bibinfo{author}{\bibfnamefont{A.}~\bibnamefont{Challinor}} \bibnamefont{and}
  \bibinfo{author}{\bibfnamefont{G.}~\bibnamefont{Chon}},
  \bibinfo{journal}{\prd} \textbf{\bibinfo{volume}{66}},
  \bibinfo{pages}{127301} (\bibinfo{year}{2002}), \eprint{astro-ph/0301064}.

\bibitem[{\citenamefont{Okamoto and Hu}(2003)}]{OkamotoHu:quad}
\bibinfo{author}{\bibfnamefont{T.}~\bibnamefont{Okamoto}} \bibnamefont{and}
  \bibinfo{author}{\bibfnamefont{W.}~\bibnamefont{Hu}}, \bibinfo{journal}{\prd}
  \textbf{\bibinfo{volume}{67}}, \bibinfo{pages}{083002}
  (\bibinfo{year}{2003}), \eprint{astro-ph/0301031}.

\bibitem[{\citenamefont{Challinor and Lewis}(2005)}]{Challinor:2005jy}
\bibinfo{author}{\bibfnamefont{A.}~\bibnamefont{Challinor}} \bibnamefont{and}
  \bibinfo{author}{\bibfnamefont{A.}~\bibnamefont{Lewis}},
  \bibinfo{journal}{\prd} \textbf{\bibinfo{volume}{71}},
  \bibinfo{pages}{103010} (\bibinfo{year}{2005}), \eprint{astro-ph/0502425}.

\bibitem[{\citenamefont{{Baleato Lizancos}
  et~al.}(2021{\natexlab{b}})\citenamefont{{Baleato Lizancos}, {Challinor}, and
  {Carron}}}]{BaleatoLizancos:2020b}
\bibinfo{author}{\bibfnamefont{A.}~\bibnamefont{{Baleato Lizancos}}},
  \bibinfo{author}{\bibfnamefont{A.}~\bibnamefont{{Challinor}}},
  \bibnamefont{and} \bibinfo{author}{\bibfnamefont{J.}~\bibnamefont{{Carron}}},
  \bibinfo{journal}{\prd} \textbf{\bibinfo{volume}{103}}, \bibinfo{eid}{023518}
  (\bibinfo{year}{2021}{\natexlab{b}}), \eprint{2010.14286}.

\bibitem[{\citenamefont{Carron and Lewis}(2017)}]{Carron:2017mqf}
\bibinfo{author}{\bibfnamefont{J.}~\bibnamefont{Carron}} \bibnamefont{and}
  \bibinfo{author}{\bibfnamefont{A.}~\bibnamefont{Lewis}},
  \bibinfo{journal}{\prd} \textbf{\bibinfo{volume}{96}},
  \bibinfo{pages}{063510} (\bibinfo{year}{2017}), \eprint{1704.08230}.

\bibitem[{\citenamefont{Lewis et~al.}(2011)\citenamefont{Lewis, Challinor, and
  Hanson}}]{Lewis:2011fk}
\bibinfo{author}{\bibfnamefont{A.}~\bibnamefont{Lewis}},
  \bibinfo{author}{\bibfnamefont{A.}~\bibnamefont{Challinor}},
  \bibnamefont{and} \bibinfo{author}{\bibfnamefont{D.}~\bibnamefont{Hanson}},
  \bibinfo{journal}{\jcap} \textbf{\bibinfo{volume}{03}}, \bibinfo{pages}{018}
  (\bibinfo{year}{2011}), \eprint{1101.2234}.

\bibitem[{\citenamefont{Hanson et~al.}(2011)\citenamefont{Hanson, Challinor,
  Efstathiou, and Bielewicz}}]{Hanson:2010:N2}
\bibinfo{author}{\bibfnamefont{D.}~\bibnamefont{Hanson}},
  \bibinfo{author}{\bibfnamefont{A.}~\bibnamefont{Challinor}},
  \bibinfo{author}{\bibfnamefont{G.}~\bibnamefont{Efstathiou}},
  \bibnamefont{and}
  \bibinfo{author}{\bibfnamefont{P.}~\bibnamefont{Bielewicz}},
  \bibinfo{journal}{\prd} \textbf{\bibinfo{volume}{83}},
  \bibinfo{pages}{043005} (\bibinfo{year}{2011}), \eprint{1008.4403}.

\bibitem[{\citenamefont{Namikawa et~al.}(2013)\citenamefont{Namikawa, Hanson,
  and Takahashi}}]{Namikawa:2012:bhe}
\bibinfo{author}{\bibfnamefont{T.}~\bibnamefont{Namikawa}},
  \bibinfo{author}{\bibfnamefont{D.}~\bibnamefont{Hanson}}, \bibnamefont{and}
  \bibinfo{author}{\bibfnamefont{R.}~\bibnamefont{Takahashi}},
  \bibinfo{journal}{\mnras} \textbf{\bibinfo{volume}{431}},
  \bibinfo{pages}{609} (\bibinfo{year}{2013}), \eprint{1209.0091}.

\bibitem[{\citenamefont{Namikawa and Takahashi}(2014)}]{Namikawa:2013:bhepol}
\bibinfo{author}{\bibfnamefont{T.}~\bibnamefont{Namikawa}} \bibnamefont{and}
  \bibinfo{author}{\bibfnamefont{R.}~\bibnamefont{Takahashi}},
  \bibinfo{journal}{\mnras} \textbf{\bibinfo{volume}{438}},
  \bibinfo{pages}{1507} (\bibinfo{year}{2014}), \eprint{1310.2372}.

\bibitem[{\citenamefont{Maniyar et~al.}(2021)\citenamefont{Maniyar,
  Ali-Ha\"\i{}moud, Carron, Lewis, and Madhavacheril}}]{Maniyar:2021:GMV}
\bibinfo{author}{\bibfnamefont{A.~S.} \bibnamefont{Maniyar}},
  \bibinfo{author}{\bibfnamefont{Y.}~\bibnamefont{Ali-Ha\"\i{}moud}},
  \bibinfo{author}{\bibfnamefont{J.}~\bibnamefont{Carron}},
  \bibinfo{author}{\bibfnamefont{A.}~\bibnamefont{Lewis}}, \bibnamefont{and}
  \bibinfo{author}{\bibfnamefont{M.~S.} \bibnamefont{Madhavacheril}},
  \bibinfo{journal}{\prd} \textbf{\bibinfo{volume}{103}}, \bibinfo{eid}{083524}
  (\bibinfo{year}{2021}), \eprint{2101.12193}.

\bibitem[{\citenamefont{Hirata and Seljak}(2003)}]{Hirata:2003:mle:pol}
\bibinfo{author}{\bibfnamefont{C.~M.} \bibnamefont{Hirata}} \bibnamefont{and}
  \bibinfo{author}{\bibfnamefont{U.}~\bibnamefont{Seljak}},
  \bibinfo{journal}{\prd} \textbf{\bibinfo{volume}{68}},
  \bibinfo{pages}{083002} (\bibinfo{year}{2003}), \eprint{astro-ph/0306354}.

\bibitem[{\citenamefont{{Yu} et~al.}(2017)\citenamefont{{Yu}, {Hill}, and
  {Sherwin}}}]{ref:yu_17}
\bibinfo{author}{\bibfnamefont{B.}~\bibnamefont{{Yu}}},
  \bibinfo{author}{\bibfnamefont{J.~C.} \bibnamefont{{Hill}}},
  \bibnamefont{and} \bibinfo{author}{\bibfnamefont{B.~D.}
  \bibnamefont{{Sherwin}}}, \bibinfo{journal}{\prd}
  \textbf{\bibinfo{volume}{96}}, \bibinfo{eid}{123511} (\bibinfo{year}{2017}),
  \eprint{1705.02332}.

\bibitem[{\citenamefont{{Remazeilles} et~al.}(2011)\citenamefont{{Remazeilles},
  {Delabrouille}, and {Cardoso}}}]{Remazeilles:2011:GNILC}
\bibinfo{author}{\bibfnamefont{M.}~\bibnamefont{{Remazeilles}}},
  \bibinfo{author}{\bibfnamefont{J.}~\bibnamefont{{Delabrouille}}},
  \bibnamefont{and} \bibinfo{author}{\bibfnamefont{J.-F.}
  \bibnamefont{{Cardoso}}}, \bibinfo{journal}{\mnras}
  \textbf{\bibinfo{volume}{418}}, \bibinfo{pages}{467} (\bibinfo{year}{2011}),
  \eprint{1103.1166}.

\bibitem[{\citenamefont{{Planck Collaboration}}(2016)}]{ref:gnilc}
\bibinfo{author}{\bibnamefont{{Planck Collaboration}}}, \bibinfo{journal}{\aap}
  \textbf{\bibinfo{volume}{596}}, \bibinfo{eid}{A109} (\bibinfo{year}{2016}),
  \eprint{1605.09387}.

\bibitem[{\citenamefont{{Ivezi{\'c}} et~al.}(2019)\citenamefont{{Ivezi{\'c}},
  {Kahn}, {Tyson}, {Abel}, {Acosta}, {Allsman}, {Alonso}, {AlSayyad},
  {Anderson}, {Andrew} et~al.}}]{ref:lsst}
\bibinfo{author}{\bibfnamefont{{\v Z}.}~\bibnamefont{{Ivezi{\'c}}}},
  \bibinfo{author}{\bibfnamefont{S.~M.} \bibnamefont{{Kahn}}},
  \bibinfo{author}{\bibfnamefont{J.~A.} \bibnamefont{{Tyson}}},
  \bibinfo{author}{\bibfnamefont{B.}~\bibnamefont{{Abel}}},
  \bibinfo{author}{\bibfnamefont{E.}~\bibnamefont{{Acosta}}},
  \bibinfo{author}{\bibfnamefont{R.}~\bibnamefont{{Allsman}}},
  \bibinfo{author}{\bibfnamefont{D.}~\bibnamefont{{Alonso}}},
  \bibinfo{author}{\bibfnamefont{Y.}~\bibnamefont{{AlSayyad}}},
  \bibinfo{author}{\bibfnamefont{S.~F.} \bibnamefont{{Anderson}}},
  \bibinfo{author}{\bibfnamefont{J.}~\bibnamefont{{Andrew}}},
  \bibnamefont{et~al.}, \bibinfo{journal}{\apj} \textbf{\bibinfo{volume}{873}},
  \bibinfo{eid}{111} (\bibinfo{year}{2019}), \eprint{0805.2366}.

\bibitem[{\citenamefont{Smith}(2009)}]{Smith:2011we}
\bibinfo{author}{\bibfnamefont{K.~M.} \bibnamefont{Smith}},
  \bibinfo{journal}{ASP Conf. Ser.} \textbf{\bibinfo{volume}{432}},
  \bibinfo{pages}{147} (\bibinfo{year}{2009}), \eprint{1111.1783}.

\bibitem[{\citenamefont{{\textit{Planck} Collaboration}
  et~al.}(2020)\citenamefont{{\textit{Planck} Collaboration}, {Aghanim},
  {Akrami}, {Ashdown}, {Aumont}, {Baccigalupi}, {Ballardini}, {Banday},
  {Barreiro}, {Bartolo} et~al.}}]{P18:phi}
\bibinfo{author}{\bibnamefont{{\textit{Planck} Collaboration}}},
  \bibinfo{author}{\bibfnamefont{N.}~\bibnamefont{{Aghanim}}},
  \bibinfo{author}{\bibfnamefont{Y.}~\bibnamefont{{Akrami}}},
  \bibinfo{author}{\bibfnamefont{M.}~\bibnamefont{{Ashdown}}},
  \bibinfo{author}{\bibfnamefont{J.}~\bibnamefont{{Aumont}}},
  \bibinfo{author}{\bibfnamefont{C.}~\bibnamefont{{Baccigalupi}}},
  \bibinfo{author}{\bibfnamefont{M.}~\bibnamefont{{Ballardini}}},
  \bibinfo{author}{\bibfnamefont{A.~J.} \bibnamefont{{Banday}}},
  \bibinfo{author}{\bibfnamefont{R.~B.} \bibnamefont{{Barreiro}}},
  \bibinfo{author}{\bibfnamefont{N.}~\bibnamefont{{Bartolo}}},
  \bibnamefont{et~al.}, \bibinfo{journal}{\aap} \textbf{\bibinfo{volume}{641}},
  \bibinfo{eid}{A8} (\bibinfo{year}{2020}), \eprint{1807.06210}.

\bibitem[{\citenamefont{Eriksen et~al.}(2004)\citenamefont{Eriksen, O'Dwyer,
  Jewell, Wandelt, Larson, Gorski, Levin, Banday, and
  Lilje}}]{Eriksen:2004:wiener}
\bibinfo{author}{\bibfnamefont{H.~K.} \bibnamefont{Eriksen}},
  \bibinfo{author}{\bibfnamefont{I.~J.} \bibnamefont{O'Dwyer}},
  \bibinfo{author}{\bibfnamefont{J.~B.} \bibnamefont{Jewell}},
  \bibinfo{author}{\bibfnamefont{B.~D.} \bibnamefont{Wandelt}},
  \bibinfo{author}{\bibfnamefont{D.~L.} \bibnamefont{Larson}},
  \bibinfo{author}{\bibfnamefont{K.~M.} \bibnamefont{Gorski}},
  \bibinfo{author}{\bibfnamefont{S.}~\bibnamefont{Levin}},
  \bibinfo{author}{\bibfnamefont{A.~J.} \bibnamefont{Banday}},
  \bibnamefont{and} \bibinfo{author}{\bibfnamefont{P.~B.} \bibnamefont{Lilje}},
  \bibinfo{journal}{Astrophys. J. Suppl.} \textbf{\bibinfo{volume}{155}},
  \bibinfo{pages}{227} (\bibinfo{year}{2004}), \eprint{astro-ph/0407028}.

\bibitem[{\citenamefont{Press et~al.}(1992)\citenamefont{Press, Teukolsky,
  Vetterling, and Flannery}}]{NumericalRecipes}
\bibinfo{author}{\bibfnamefont{W.~H.} \bibnamefont{Press}},
  \bibinfo{author}{\bibfnamefont{S.~A.} \bibnamefont{Teukolsky}},
  \bibinfo{author}{\bibfnamefont{W.~T.} \bibnamefont{Vetterling}},
  \bibnamefont{and} \bibinfo{author}{\bibfnamefont{B.~P.}
  \bibnamefont{Flannery}}, \emph{\bibinfo{title}{Numerical Recipes in C (2nd
  Ed.): The Art of Scientific Computing}} (\bibinfo{publisher}{Cambridge
  University Press}, \bibinfo{address}{USA}, \bibinfo{year}{1992}), ISBN
  \bibinfo{isbn}{0521431085}.

\bibitem[{\citenamefont{{\textsc{Bicep2} and {\it Planck}
  Collaborations}}(2015)}]{BKP}
\bibinfo{author}{\bibnamefont{{\textsc{Bicep2} and {\it Planck}
  Collaborations}}}, \bibinfo{journal}{\prl} \textbf{\bibinfo{volume}{114}},
  \bibinfo{pages}{101301} (\bibinfo{year}{2015}), \eprint{1502.00612}.

\bibitem[{\citenamefont{Azzoni et~al.}(2021)\citenamefont{Azzoni, Abitbol,
  Alonso, Gough, Katayama, and Matsumura}}]{Azzoni:2020hpw}
\bibinfo{author}{\bibfnamefont{S.}~\bibnamefont{Azzoni}},
  \bibinfo{author}{\bibfnamefont{M.~H.} \bibnamefont{Abitbol}},
  \bibinfo{author}{\bibfnamefont{D.}~\bibnamefont{Alonso}},
  \bibinfo{author}{\bibfnamefont{A.}~\bibnamefont{Gough}},
  \bibinfo{author}{\bibfnamefont{N.}~\bibnamefont{Katayama}}, \bibnamefont{and}
  \bibinfo{author}{\bibfnamefont{T.}~\bibnamefont{Matsumura}},
  \bibinfo{journal}{\jcap} \textbf{\bibinfo{volume}{05}}, \bibinfo{pages}{047}
  (\bibinfo{year}{2021}), \eprint{2011.11575}.

\bibitem[{\citenamefont{Smith}(2006)}]{Smith:2005:chi-estimator}
\bibinfo{author}{\bibfnamefont{K.~M.} \bibnamefont{Smith}},
  \bibinfo{journal}{\prd} \textbf{\bibinfo{volume}{74}},
  \bibinfo{pages}{083002} (\bibinfo{year}{2006}), \eprint{astro-ph/0511629}.

\bibitem[{\citenamefont{Ghosh et~al.}(2021)\citenamefont{Ghosh, Delabrouille,
  Zhao, and Santos}}]{Ghosh:2020}
\bibinfo{author}{\bibfnamefont{S.}~\bibnamefont{Ghosh}},
  \bibinfo{author}{\bibfnamefont{J.}~\bibnamefont{Delabrouille}},
  \bibinfo{author}{\bibfnamefont{W.}~\bibnamefont{Zhao}}, \bibnamefont{and}
  \bibinfo{author}{\bibfnamefont{L.}~\bibnamefont{Santos}},
  \bibinfo{journal}{\jcap} \textbf{\bibinfo{volume}{02}}, \bibinfo{pages}{036}
  (\bibinfo{year}{2021}), \eprint{2007.09928}.

\bibitem[{\citenamefont{Hamimeche and Lewis}(2008)}]{Hamimeche:2008}
\bibinfo{author}{\bibfnamefont{S.}~\bibnamefont{Hamimeche}} \bibnamefont{and}
  \bibinfo{author}{\bibfnamefont{A.}~\bibnamefont{Lewis}},
  \bibinfo{journal}{\prd} \textbf{\bibinfo{volume}{77}},
  \bibinfo{pages}{103013} (\bibinfo{year}{2008}), \eprint{0801.0554}.

\bibitem[{\citenamefont{{Smith} et~al.}(2004)\citenamefont{{Smith}, {Hu}, and
  {Kaplinghat}}}]{ref:smith_2004}
\bibinfo{author}{\bibfnamefont{K.~M.} \bibnamefont{{Smith}}},
  \bibinfo{author}{\bibfnamefont{W.}~\bibnamefont{{Hu}}}, \bibnamefont{and}
  \bibinfo{author}{\bibfnamefont{M.}~\bibnamefont{{Kaplinghat}}},
  \bibinfo{journal}{\prd} \textbf{\bibinfo{volume}{70}}, \bibinfo{eid}{043002}
  (\bibinfo{year}{2004}), \eprint{astro-ph/0402442}.

\bibitem[{\citenamefont{Benoit-Levy et~al.}(2012)\citenamefont{Benoit-Levy,
  Smith, and Hu}}]{BenoitLevy:2012va}
\bibinfo{author}{\bibfnamefont{A.}~\bibnamefont{Benoit-Levy}},
  \bibinfo{author}{\bibfnamefont{K.~M.} \bibnamefont{Smith}}, \bibnamefont{and}
  \bibinfo{author}{\bibfnamefont{W.}~\bibnamefont{Hu}}, \bibinfo{journal}{\prd}
  \textbf{\bibinfo{volume}{86}}, \bibinfo{pages}{123008}
  (\bibinfo{year}{2012}), \eprint{1205.0474}.

\bibitem[{\citenamefont{{Baleato Lizancos}
  et~al.}(2021{\natexlab{c}})\citenamefont{{Baleato Lizancos}, {Challinor},
  {Sherwin}, and {Namikawa}}}]{ref:cib_delensing_biases}
\bibinfo{author}{\bibfnamefont{A.}~\bibnamefont{{Baleato Lizancos}}},
  \bibinfo{author}{\bibfnamefont{A.}~\bibnamefont{{Challinor}}},
  \bibinfo{author}{\bibfnamefont{B.~D.} \bibnamefont{{Sherwin}}},
  \bibnamefont{and}
  \bibinfo{author}{\bibfnamefont{T.}~\bibnamefont{{Namikawa}}}
  (\bibinfo{year}{2021}{\natexlab{c}}), \eprint{2102.01045}.

\bibitem[{\citenamefont{Namikawa and Takahashi}(2019)}]{Namikawa:2018:nldelens}
\bibinfo{author}{\bibfnamefont{T.}~\bibnamefont{Namikawa}} \bibnamefont{and}
  \bibinfo{author}{\bibfnamefont{R.}~\bibnamefont{Takahashi}},
  \bibinfo{journal}{\prd} \textbf{\bibinfo{volume}{99}},
  \bibinfo{pages}{023530} (\bibinfo{year}{2019}), \eprint{1810.03346}.

\bibitem[{\citenamefont{van Engelen et~al.}(2014)\citenamefont{van Engelen,
  Bhattacharya, Sehgal, Holder, Zahn, and Nagai}}]{vanEngelen:2013rla}
\bibinfo{author}{\bibfnamefont{A.}~\bibnamefont{van Engelen}},
  \bibinfo{author}{\bibfnamefont{S.}~\bibnamefont{Bhattacharya}},
  \bibinfo{author}{\bibfnamefont{N.}~\bibnamefont{Sehgal}},
  \bibinfo{author}{\bibfnamefont{G.~P.} \bibnamefont{Holder}},
  \bibinfo{author}{\bibfnamefont{O.}~\bibnamefont{Zahn}}, \bibnamefont{and}
  \bibinfo{author}{\bibfnamefont{D.}~\bibnamefont{Nagai}},
  \bibinfo{journal}{\apj} \textbf{\bibinfo{volume}{786}}, \bibinfo{pages}{14}
  (\bibinfo{year}{2014}), \eprint{1310.7023}.

\bibitem[{\citenamefont{Hu and Okamoto}(2002)}]{HuOkamoto:2001}
\bibinfo{author}{\bibfnamefont{W.}~\bibnamefont{Hu}} \bibnamefont{and}
  \bibinfo{author}{\bibfnamefont{T.}~\bibnamefont{Okamoto}},
  \bibinfo{journal}{\apj} \textbf{\bibinfo{volume}{574}}, \bibinfo{pages}{566}
  (\bibinfo{year}{2002}), \eprint{astro-ph/0111606}.

\bibitem[{\citenamefont{{Roy} et~al.}(2021)\citenamefont{{Roy}, {Kulkarni},
  {Meerburg}, {Challinor}, {Baccigalupi}, {Lapi}, and
  {Haehnelt}}}]{ref:roy_et_al_21}
\bibinfo{author}{\bibfnamefont{A.}~\bibnamefont{{Roy}}},
  \bibinfo{author}{\bibfnamefont{G.}~\bibnamefont{{Kulkarni}}},
  \bibinfo{author}{\bibfnamefont{P.~D.} \bibnamefont{{Meerburg}}},
  \bibinfo{author}{\bibfnamefont{A.}~\bibnamefont{{Challinor}}},
  \bibinfo{author}{\bibfnamefont{C.}~\bibnamefont{{Baccigalupi}}},
  \bibinfo{author}{\bibfnamefont{A.}~\bibnamefont{{Lapi}}}, \bibnamefont{and}
  \bibinfo{author}{\bibfnamefont{M.~G.} \bibnamefont{{Haehnelt}}},
  \bibinfo{journal}{\jcap} \textbf{\bibinfo{volume}{2021}}, \bibinfo{eid}{003}
  (\bibinfo{year}{2021}), \eprint{2004.02927}.

\bibitem[{\citenamefont{Nagata and Namikawa}(2021)}]{Nagata:2021}
\bibinfo{author}{\bibfnamefont{R.}~\bibnamefont{Nagata}} \bibnamefont{and}
  \bibinfo{author}{\bibfnamefont{T.}~\bibnamefont{Namikawa}},
  \bibinfo{journal}{\ptep} \textbf{\bibinfo{volume}{2021}},
  \bibinfo{eid}{053E01} (\bibinfo{year}{2021}), \eprint{2102.00133}.

\bibitem[{\citenamefont{Lembo et~al.}(2021)\citenamefont{Lembo, Fabbian,
  Carron, and Lewis}}]{Lembo:2021kxc}
\bibinfo{author}{\bibfnamefont{M.}~\bibnamefont{Lembo}},
  \bibinfo{author}{\bibfnamefont{G.}~\bibnamefont{Fabbian}},
  \bibinfo{author}{\bibfnamefont{J.}~\bibnamefont{Carron}}, \bibnamefont{and}
  \bibinfo{author}{\bibfnamefont{A.}~\bibnamefont{Lewis}}
  (\bibinfo{year}{2021}), \eprint{2109.13911}.

\bibitem[{\citenamefont{Zonca et~al.}(2019)\citenamefont{Zonca, Singer, Lenz,
  Reinecke, Rosset, Hivon, and Gorski}}]{healpy}
\bibinfo{author}{\bibfnamefont{A.}~\bibnamefont{Zonca}},
  \bibinfo{author}{\bibfnamefont{L.}~\bibnamefont{Singer}},
  \bibinfo{author}{\bibfnamefont{D.}~\bibnamefont{Lenz}},
  \bibinfo{author}{\bibfnamefont{M.}~\bibnamefont{Reinecke}},
  \bibinfo{author}{\bibfnamefont{C.}~\bibnamefont{Rosset}},
  \bibinfo{author}{\bibfnamefont{H.}~\bibnamefont{Hivon}}, \bibnamefont{and}
  \bibinfo{author}{\bibfnamefont{K.}~\bibnamefont{Gorski}},
  \bibinfo{journal}{Journal of Open Source Software}
  \textbf{\bibinfo{volume}{4}}, \bibinfo{pages}{1298} (\bibinfo{year}{2019}).

\bibitem[{\citenamefont{{G{\'o}rski} et~al.}(2005)\citenamefont{{G{\'o}rski},
  {Hivon}, {Banday}, {Wandelt}, {Hansen}, {Reinecke}, and
  {Bartelmann}}}]{Gorski:2004by}
\bibinfo{author}{\bibfnamefont{K.~M.} \bibnamefont{{G{\'o}rski}}},
  \bibinfo{author}{\bibfnamefont{E.}~\bibnamefont{{Hivon}}},
  \bibinfo{author}{\bibfnamefont{A.~J.} \bibnamefont{{Banday}}},
  \bibinfo{author}{\bibfnamefont{B.~D.} \bibnamefont{{Wandelt}}},
  \bibinfo{author}{\bibfnamefont{F.~K.} \bibnamefont{{Hansen}}},
  \bibinfo{author}{\bibfnamefont{M.}~\bibnamefont{{Reinecke}}},
  \bibnamefont{and}
  \bibinfo{author}{\bibfnamefont{M.}~\bibnamefont{{Bartelmann}}},
  \bibinfo{journal}{\apj} \textbf{\bibinfo{volume}{622}}, \bibinfo{pages}{759}
  (\bibinfo{year}{2005}), \eprint{astro-ph/0409513}.

\bibitem[{\citenamefont{Lewis et~al.}(2000)\citenamefont{Lewis, Challinor, and
  Lasenby}}]{Lewis:1999bs}
\bibinfo{author}{\bibfnamefont{A.}~\bibnamefont{Lewis}},
  \bibinfo{author}{\bibfnamefont{A.}~\bibnamefont{Challinor}},
  \bibnamefont{and} \bibinfo{author}{\bibfnamefont{A.}~\bibnamefont{Lasenby}},
  \bibinfo{journal}{\apj} \textbf{\bibinfo{volume}{538}}, \bibinfo{pages}{473}
  (\bibinfo{year}{2000}), \eprint{astro-ph/9911177}.

\bibitem[{\citenamefont{Namikawa and Nagata}(2015)}]{Namikawa:2015:cov}
\bibinfo{author}{\bibfnamefont{T.}~\bibnamefont{Namikawa}} \bibnamefont{and}
  \bibinfo{author}{\bibfnamefont{R.}~\bibnamefont{Nagata}},
  \bibinfo{journal}{\jcap} \textbf{\bibinfo{volume}{10}}, \bibinfo{pages}{004}
  (\bibinfo{year}{2015}), \eprint{1506.09209}.

\bibitem[{\citenamefont{{Schmittfull} et~al.}(2013)\citenamefont{{Schmittfull},
  {Challinor}, {Hanson}, and {Lewis}}}]{ref:Schmittfull_13}
\bibinfo{author}{\bibfnamefont{M.~M.} \bibnamefont{{Schmittfull}}},
  \bibinfo{author}{\bibfnamefont{A.}~\bibnamefont{{Challinor}}},
  \bibinfo{author}{\bibfnamefont{D.}~\bibnamefont{{Hanson}}}, \bibnamefont{and}
  \bibinfo{author}{\bibfnamefont{A.}~\bibnamefont{{Lewis}}},
  \bibinfo{journal}{\prd} \textbf{\bibinfo{volume}{88}}, \bibinfo{eid}{063012}
  (\bibinfo{year}{2013}), \eprint{1308.0286}.

\bibitem[{\citenamefont{{Peloton} et~al.}(2017)\citenamefont{{Peloton},
  {Schmittfull}, {Lewis}, {Carron}, and {Zahn}}}]{ref:peloton_17}
\bibinfo{author}{\bibfnamefont{J.}~\bibnamefont{{Peloton}}},
  \bibinfo{author}{\bibfnamefont{M.}~\bibnamefont{{Schmittfull}}},
  \bibinfo{author}{\bibfnamefont{A.}~\bibnamefont{{Lewis}}},
  \bibinfo{author}{\bibfnamefont{J.}~\bibnamefont{{Carron}}}, \bibnamefont{and}
  \bibinfo{author}{\bibfnamefont{O.}~\bibnamefont{{Zahn}}},
  \bibinfo{journal}{\prd} \textbf{\bibinfo{volume}{95}}, \bibinfo{eid}{043508}
  (\bibinfo{year}{2017}), \eprint{1611.01446}.

\end{thebibliography}

\end{document}